\documentclass[12pt,preprint]{aastex}

\shorttitle{Mid-Infrared Spectral Variability Atlas of Young Stars}
\shortauthors{K\'osp\'al et al.}

\begin{document}


\title{Mid-Infrared Spectral Variability Atlas of Young Stellar
  Objects\footnote{This work is based on observations made with the
    Infrared Space Observatory (ISO) and with the Spitzer Space
    Telescope. ISO is an ESA project with instruments funded by ESA
    Member States (especially the PI countries: France, Germany, the
    Netherlands and the UK) and with the participation of ISAS and
    NASA. Spitzer is operated by the Jet Propulsion Laboratory,
  California Institute Technology under a contract with NASA.}}

\author{\'A. K\'osp\'al\altaffilmark{1,2}}
\email{akospal@rssd.esa.int}
\author{P. \'Abrah\'am\altaffilmark{3}}
\author{J. A. Acosta-Pulido\altaffilmark{4,5}}
\author{C. P. Dullemond\altaffilmark{6,7}}
\author{Th. Henning\altaffilmark{7}}
\author{M. Kun\altaffilmark{3}}
\author{Ch. Leinert\altaffilmark{7}}
\author{A. Mo\'or\altaffilmark{3}}
\author{N. J. Turner\altaffilmark{8}}

\altaffiltext{1}{Leiden Observatory, Leiden University, PO Box 9513,
  2300 RA, Leiden, The Netherlands}
\altaffiltext{2}{Research and Scientific Support Department, European
  Space Agency (ESA-ESTEC, SRE-SA), PO Box 299, 2200 AG, Noordwijk,
  The Netherlands}
\altaffiltext{3}{Konkoly Observatory, Research Centre for Astronomy and
     Earth Sciences, Hungarian Academy of Sciences, 
     PO Box 67, 1525 Budapest, Hungary}
\altaffiltext{4}{Instituto de Astrof\'\i{}sica de Canarias, Via
  L\'actea s/n, 38200 La Laguna, Tenerife, Spain}
\altaffiltext{5}{Departamento de Astrof\'\i{}sica, Universidad de La
  Laguna, 38205 La Laguna, Tenerife, Spain}
\altaffiltext{6}{Institut f\"ur Theoretische Astrophysik, Zentrum
  f\"ur Astronomie der Universit\"at Heidelberg,
  Albert-Ueberle-Str. 2, 69120 Heidelberg, Germany}
\altaffiltext{7}{Max-Planck-Institut f\"ur Astronomie, K\"onigstuhl
  17, 69117 Heidelberg, Germany}
\altaffiltext{8}{Jet Propulsion Laboratory, California Institute of
  Technology, Pasadena, CA 91109, USA}

\begin{abstract}
  Optical and near-infrared variability is a well-known property of
  young stellar objects. However, a growing number of recent studies
  claim that a considerable fraction of them also exhibit mid-infrared
  flux changes. With the aim of studying and interpreting variability
  on a decadal timescale, here we present a mid-infrared spectral
  atlas containing observations of 68 low- and intermediate mass young
  stellar objects. The atlas consists of 2.5--11.6$\,\mu$m
  low-resolution spectra obtained with the ISOPHOT-S instrument
  on-board the {\it Infrared Space Observatory (ISO)} between 1996 and
  1998, as well as 5.2--14.5$\,\mu$m low-resolution spectra obtained
  with the IRS instrument on-board the {\it Spitzer Space Telescope}
  between 2004 and 2007. The observations were retrieved from the ISO
  and Spitzer archives and were post-processed interactively by our
  own routines. For those 47 objects where multi-epoch spectra were
  available, we analyze mid-infrared spectral variability on annual
  and/or decadal timescales. We identify 37 variable candidate
  sources. Many stars show wavelength-independent flux changes,
  possibly due to variable accretion rate. In several systems, all
  exhibiting 10$\,\mu$m silicate emission, the variability of the
  6$-$8$\,\mu$m continuum and the silicate feature exhibit different
  amplitudes. A possible explanation is variable shadowing of the
  silicate emitting region by an inner disk structure of changing
  height or extra silicate emission from dust clouds in the disk
  atmosphere. Our results suggest that mid-infrared variability, in
  particular the wavelength-dependent changes, are more ubiquitous
  than was known before. Interpreting this variability is a new
  possibility to explore the structure of the disk and its dynamical
  processes.
\end{abstract}


\section{Introduction}

At the early phases of their formation, stars are intimately linked to
their circumstellar environment. Dust particles in the circumstellar
disk and envelope are heated both internally by viscous energy
released in the midplane of the accretion disk, and externally by
irradiation from the central star. The absorbed energy is re-radiated
as thermal emission at infrared (IR) wavelengths. The shape of the IR
spectral energy distribution (SED) reflects the temperature and
density structure of the circumstellar environment. Of particular
importance is the mid-IR domain (3--25$\,\mu$m), where the disk
emission originates from the region where planets are formed. The same
domain contains spectral features of PAH particles, small silicate
grains, and molecular ices. PAH features give information about the UV
radiation field and ionization balance in the circumstellar gas
\citep{tielens2008}; the shape and amplitude of the 10$\,\mu$m
silicate feature is partly related to the degree of crystallinity of
the silicate grains, indicating how far these grains evolved from the
interstellar medium \citep{henning2010}; while ice features are used
to study the organic chemistry and thermal history of the
circumstellar material \citep{vandishoeck2004}.

Although optical and near-IR variability is a well-known property of
young stellar objects (YSOs), a growing number of recent studies claim
that a considerable fraction of them also exhibit mid-IR photometric
changes \citep[e.g.][]{barsony2005, morales2009, morales2011}. In
young eruptive stars and some T\,Tauri stars, changing accretion is
thought to be responsible for mid-IR variability
\citep[e.g.][]{abraham2004, kospal2007, vanboekel2010}. There are
claims that in some young eruptive stars the temporarily increased
heat leads to structural changes in the innermost part of the system,
resulting in decreased extinction and consequently in IR brightness
changes (see, e.g., V1647\,Ori in \citealt{muzerolle2005}, or PV\,Cep
in \citealt{kun2011}). In some intermediate mass young stars,
structural changes in the inner disk were invoked to explain their
mid-IR flux changes \citep{juhasz2007, sitko2008}. A similar reason,
i.e.~changing inner disk height, may also explain the variability of
those low-mass objects whose disks contain holes and gaps
\citep{muzerolle2009, espaillat2011}. These indications of dynamical
processes suggest that disks are less steady-state than thought
before.

The wavelength-dependence of the mid-IR variability carries
information on the disk and envelope structure, because the emission
at different wavelengths corresponds to disk regions with different
temperatures and thus to different radial distances from the star. An
interesting aspect is the exploration of possible changes in the
strength and profile of IR spectral features. Several studies
published so far concentrate on the variability of the 10$\,\mu$m
silicate feature. Time-dependent shadowing of disk areas by the inner
gas disk could explain the spectral changes in the Herbig stars
HD\,31648 and HD\,163296 \citep{sitko2008}. In the highly accreting
T\,Tauri stars DG\,Tau and XZ\,Tau \citep{bary2009}, self-absorption
may play a role in the variability of the 10$\,\mu$m feature. Changes
in the mineralogical composition of the emitting dust particles were
also observed by \citet{abraham2009} in the most recent outburst of EX
Lupi, where they detected the appearance of crystalline silicates that
were not present in quiescence. However, more detailed investigations
are limited by the lack of multi-epoch mid-IR spectroscopic databases.

The timescale of mid-IR variability is related to the timescale on
which the heating of the emitting region can vary and on which the
dust particles in the disk can react to changes in the energy
input. For example, almost instantaneous (seconds to minutes)
responses to fluctuations in the stellar illumination may indicate
emission from an optically thin medium, while longer timescales
(years) may indicate an optically thick medium \citep{cg}. Dynamical
changes leading to rearrangement of the density structure can occur on
the timescale of an orbital period. Thus, the timescale of the
variability is an essential input to understand the physics behind the
flux changes. Recent studies carried out with the Spitzer Space
Telescope sampled MIR variability on several different
timescales. While hourly variations are rare and small in amplitude
\citep{morales2009}, variability on daily, weekly, and yearly
timescales is ubiquitous, and amplitudes range from a few tenths of
magnitudes to $>$1\,mag \citep{morales2009, luhman2010, morales2011,
  rebull2011}. These timescales, if caused by changes in the disk
structure, indicate that the variability is connected to the innermost
0.1--1\,AU of the disk. With Spitzer alone, however, one cannot
investigate whether variability on longer timescale exists. Such data
would provide information on dynamical changes at larger radial
distances from the star, typically several AU. Such study can only be
performed by comparing data from two IR space missions separated in
time by several years.

Motivated by the above mentioned considerations, here we present a new
atlas of 2.5--11.6$\,\mu$m low-resolution spectra obtained with the
ISOPHOT-S instrument on-board the {\it Infrared Space Observatory}
(ISO) between 1996 and 1998, as well as 5.2--14.5$\,\mu$m
low-resolution spectra obtained with the IRS instrument on-board the
{\it Spitzer Space Telescope} between 2004 and 2007. The overlapping
wavelength range and different epochs of ISO and Spitzer make it
possible to do a systematic mid-IR spectral variability study for a
large sample of well-known YSOs both on annual timescale (less than a
few years, i.e.~within the cryogenic lifetime of either telescope),
and of nearly a decade (by comparing the results of the two
telescopes). Although the beams of the two instruments significantly
differ, beam confusion is usually not an issue, because at mid-IR
wavelengths, the size of the emitting region (warm dust in the inner
few AUs) is smaller than either in the optical (scattered light) or in
the far-IR/mm (thermal emission from cold dust). We identify typical
variability patterns and hypothesize about their possible physical
origin. Our results on monitoring and interpreting variability may
provide a powerful ``extra dimension'' of information on the structure
of the circumstellar material. Our sample may also serve as a starting
point to plan follow-up mid-IR monitoring campaigns.

\section{Observations and data reduction}

\subsection{Sample selection}

We made a query of the ISO data archive for ISOPHOT-S observations
(AOT PHT40) and selected observations of low- and intermediate mass
YSOs by consulting the SIMBAD database and the literature. The query
resulted in 94 measurements of 68 YSOs; including 35 low-mass and 33
intermediate mass stars. This forms the target list of the present
study. Then we checked the Spitzer archive for 5.2--14.5$\,\mu$m
low-resolution IRS spectra. We found that such Spitzer spectra were
available for 51 of our targets. Table~\ref{Table1} presents the list
of our targets and the log of observations.

\subsection{ISO data reduction}
\label{sec:isodata}

ISOPHOT-S was a sub-instrument of ISOPHOT on-board the ISO. It
produced low-resolution mid-IR spectra of 128 data points covering the
2.5--4.9 (ISOPHOT-SS) and 5.8--11.6$\,\mu$m (ISOPHOT-SL) wavelength
ranges with a spectral resolution of $R\approx$100 during the active
period of ISO (1995--1998). The aperture was 24$''$$\times$24$''$. For
a detailed description of the ISOPHOT-S instrument and its calibration
strategies, see \citet{acosta2003}.

Most observations presented here consist of pairs of separate ON/OFF
measurements, but chopping between the source and background
position(s), as well as small scans/maps were also utilized. Edited
Raw Data (ERD), produced by the ISOPHOT Off-Line Processing software
(OLP v.~10.0, \citealt{isophothandbook}), were downloaded from the ISO
Data Archive for all of our targets. Standard processing to Auto
Analysis Product Data (AAP) level was performed with the ISOPHOT
Interactive Analysis Software Package (PIA v10.0,
\citealt{pia}). Further processing was done using self-developed IDL
routines consisting of the following steps: improved deglitching,
correction for memory effects due to preceding bright source
measurements, background subtraction in the lack of dedicated OFF
observation, correction for flux loss in off-center pointings (in case
the offset was larger than 1$''$), and an empirical photometric
correction. Several of these algorithms were aimed at correcting
artifacts that were discovered after the development of the ISO
offline processing software was finished. We found that in a number of
cases, the reprocessing significantly improved the data
quality. According to our systematic study of 28 normal star spectra,
we concluded that the uncertainty of the absolute calibration is
5-10\% for bright sources (brighter than a few Jy) and 30$-$100\,mJy
for fainter sources ($<$1\,Jy). Note that these figures are valid for
each independently calibrated pixel. A detailed explanation of the
correction steps and the error budget can be found in Appendix~A.

\subsection{Spitzer data reduction}
\label{sec:spitzerdata}

The Infrared Spectrograph (IRS) on-board the {\it Spitzer Space
  Telescope} had two low resolution and two high resolution modules,
covering the 5.2--38$\,\mu$m wavelength range. Since our aim is to
look for mid-IR variability, here we utilize the Short-Low channel
that covers the 5.2--14.5$\,\mu$m wavelength range, thus has the
largest wavelength overlap with ISOPHOT-S. The Short-Low channel has a
spectral resolution of $R\approx\,$60--127, comparable to that of
ISOPHOT-S. The width and length of the IRS Short-Low slit is
3$\farcs$6 and 57$''$, respectively. According to the IRS instrument
handbook (Version 4.0), the uncertainty of the absolute flux
calibration is 5-10\%, while the repeatability is 2-5\%.

We downloaded basic calibrated data (BCD) files from the Spitzer
archive, processed with the pipeline version 18.18.0. Measurements
were taken either in staring mode (two nod positions), or in spectral
mapping mode (small 2$\times$3 maps). In the latter case, we used the
central 2 positions, and considered them as normal staring
measurements. We first subtracted the two nod positions from each
other. Then we used the Spitzer IRS Custom Extraction software (SPICE)
to extract the positive signal from the two-dimensional dispersed
images. We extracted the spectra from a wavelength-dependent, tapered
aperture around the target, and we averaged the two spectra
corresponding to the two nod positions. In case the target was not
well-centered in the slit (i.e.~offset from the slit center
perpendicular to the slit was larger than 0.5$''$), we corrected the
spectra for flux loss using the measured IRS beam profiles (for
details, see \citealt{par21}).

\subsection{Comparison of the ISOPHOT-S and Spitzer/IRS calibration}

\begin{figure}
\epsscale{1.0}
\plotone{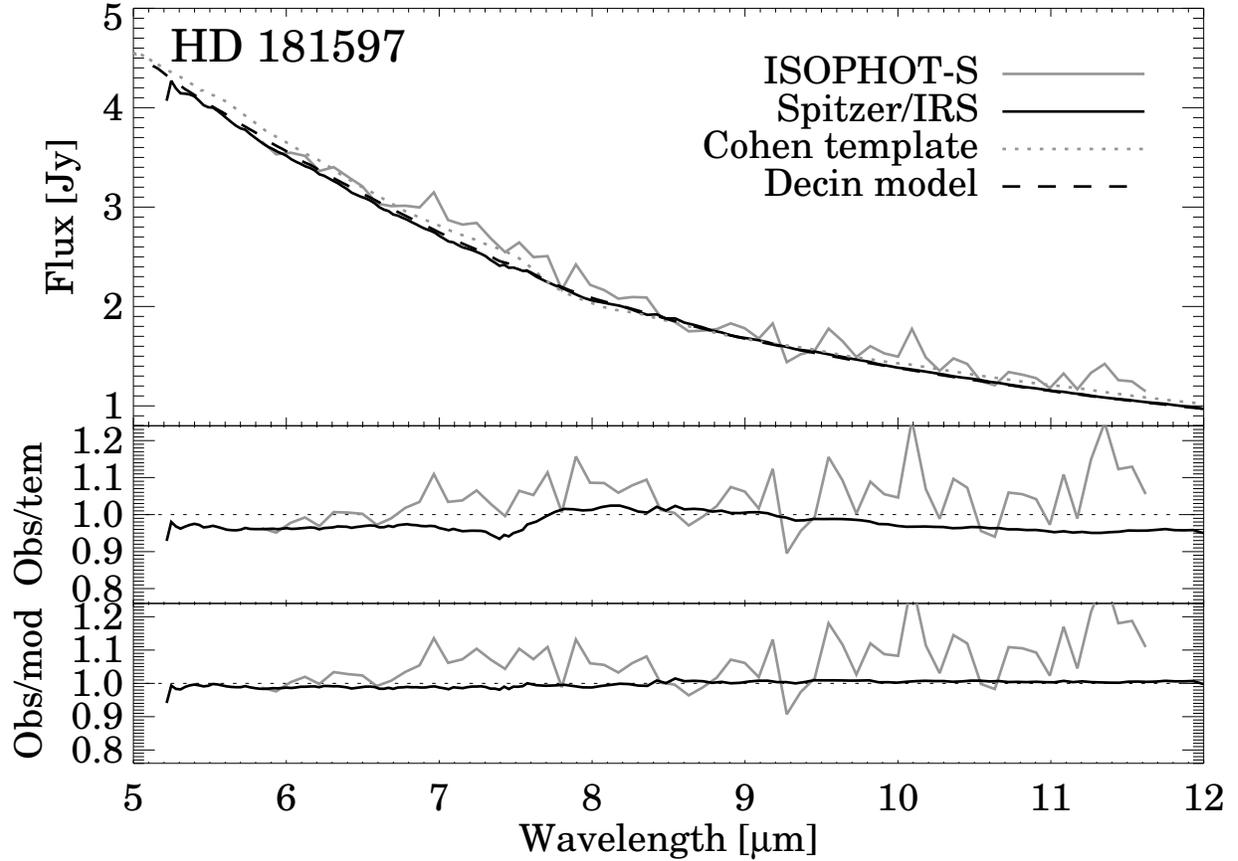}
\caption{Spectra of the spectrophotometric standard star HD\,181597
  (HR\,7341). In the upper panel, solid lines show the observations
  (TDT 53102202 for ISOPHOT-S, AOR 28709632 for Spitzer) , while
  dotted and dashed lines show the template and model spectra. The
  middle panel displays the ratio between the observed and the Cohen
  template spectrum, while the lower panel is the ratio between the
  observed and the Decin model spectrum.
\label{fig_x}}
\end{figure}

The absolute spectrophotometric calibration of ISOPHOT-S was based
both on \emph{empirical} spectral templates of standard stars derived
by \citet{cohen2003}, and on stellar atmosphere models provided by
\citet{hammersley2003} using infrared observations. Spitzer/IRS
spectra, on the other hand, were calibrated using standard stars for
which \emph{model} atmospheres are available in
\citet{decin2004}. Nevertheless, Cohen templates were used to verify
the calibration of IRS \citep{houck2004}. Thus, despite the
differences in calibration strategy, we do not expect significant
systematic discrepancies between the ISOPHOT-S and the IRS spectra. In
order to check this, we reduced the spectra of HD\,181597 (HR\,7341),
a K1\,III type star that was used as a primary low resolution standard
for Spitzer and was also used in the calibration of ISOPHOT-S. In
Fig.~\ref{fig_x} we plotted the observations along with the template
and model spectra, and we also calculated their ratio. The figure
shows that the template and model spectra are close to each other, and
the observations are within 10\% (for ISOPHOT-S) or 5\% (for
Spitzer/IRS) of the template/model spectra. We emphasize that no
wavelength-dependent trend can be seen in the ratios, suggesting that
the calibration uncertainties would not affect the observed colors of
the targets. Thus, we conclude that the accuracies of the absolute
calibration mentioned in Sec.~\ref{sec:isodata} and
\ref{sec:spitzerdata} are plausible and there is no need to increase
the error bars when comparing spectra obtained with the two different
telescopes.

\section{Catalog of observations and the spectral atlas}
\label{sec:catalog}

In Table~\ref{Table1} we present the catalog of spectroscopic
observations. The table contains the following columns:

\noindent {\it Column (1):} Name of the target as common in the
literature.

\noindent {\it Column (2):} L -- low-mass pre-main sequence stars (T
Tauri-type stars and embedded young stellar objects); I --
intermediate mass pre-main sequence stars (Herbig Ae/Be stars).  B, A,
and F stars are considered intermediate mass stars, while G, K, and M
stars are considered low-mass stars. Spectral types or stellar masses
for sources where SIMBAD does not provide spectral types were taken
from \citet{straizys2002} for IRAS\,03260+3111, \citet{myers1987} for
BARN\,5\,IRS\,1, \citet{brinch2007} for LDN\,1489\,IRS,
\citet{lim2006} for LDN\,1551\,IRS\,5, \citet{sandell2001} for
Reipurth\,50\,N\,IRS\,1, \citet{simon1995} for WL 16,
\citet{bontemps2001} for WL\,6, \citet{hodapp1996} for OO\,Ser, and
\citet{pontoppidan2004} for [SVS76]\,Ser\,4.

\noindent {\it Column (3):} Spectral types from the SIMBAD database.

\noindent {\it Column (4)-(5):} J2000 coordinates of the source from
the 2MASS All-Sky Catalog of Point Sources \citep{cutri2003}, except
where indicated in the table. Since the peak of the optical and IR
emission might not coincide, we decided to use coordinates from 2MASS
instead of optical positions. These coordinates are precise enough to
carry out the offset-correction if the source was off-center in the
ISOPHOT-S aperture or in the Spitzer/IRS slit.

\noindent {\it Column (6):} Unique 8-digit identifier of the on-source
ISOPHOT-S observation in the ISO Data Archive (TDT number).

\noindent {\it Column (7):} Corrections applied to the ISOPHOT-S
observations: M -- memory correction; B -- in the lack of dedicated
OFF measurement, zodiacal background was predicted and subtracted; O
-- position offset correction applied. An exclamation mark signals
spectra where the uncertainty of the correction is above 10\%, thus
they are not used for variability analysis (for details, see
Appendices~A and B).

\noindent {\it Column (8):} Date of the ISOPHOT-S observation.

\noindent {\it Column (9):} Unique identifier of the Spitzer/IRS
observation in the Spitzer Data Archive (AOR).

\noindent {\it Column (10):} Corrections applied to the Spitzer/IRS
observations: O -- position offset correction applied. 

\noindent {\it Column (11):} Date of the Spitzer/IRS observation.

\noindent {\it Column (12)-(14):} Inventory of clearly visible
spectral features. Si-O -- 10$\,\mu$m silicate feature; Ices -- H$_2$O
3.1$\,\mu$m, CO$_2$ 4.27$\,\mu$m, H$_2$O 6.0$\,\mu$m, CH$_3$OH
6.8$\,\mu$m, or CH$_4$ 7.7$\,\mu$m; PAH -- 3.3, 6.2, 7.7, 8.6, 11.2,
or 12.7$\,\mu$m bands of polycyclic aromatic hydrocarbons. em --
feature in emission; abs -- feature in absorption; \dots -- feature
not present.

\noindent {\it Column (15):} Type of the spectral shape: PAH dom. --
the spectrum is dominated by PAH emission features, no 10$\,\mu$m
silicate feature is discernible; sil.~em. -- the 10$\,\mu$m silicate
feature is in emission; sil.~abs. -- the 10$\,\mu$m silicate or ice
bands are visible in absorption; \dots -- undecided; (for details, see
Sec.~\ref{sec:results}).

\noindent {\it Column (16):} Variability: yes -- candidate variable
source; no -- constant source; \dots -- undecided.

The table and the spectra are also available electronically.

\section{Results}
\label{sec:results}

In Fig.~\ref{map_var} we present the ISOPHOT-S and Spitzer/IRS spectra
for each object. Where clearly detected, we marked with vertical
dashed lines the wavelengths of molecular ice bands (H$_2$O
3.1$\,\mu$m, CO$_2$ 4.27$\,\mu$m, H$_2$O 6.0$\,\mu$m, CH$_3$OH
6.8$\,\mu$m, CH$_4$ 7.7$\,\mu$m); and with vertical dotted lines the
wavelengths of PAH features (3.3, 6.2, 7.7, 8.6, 11.2,
12.7$\,\mu$m). The figures show both cases where the plotted spectra
agree, and sources where the measurements at different epochs seem to
differ. Some of the differences can be explained by serious
instrumental artifacts or source confusion resulting from the
different beam sizes of the ISOPHOT-S and IRS instruments. We studied
the literature of each object in order to establish whether it is part
of a multiple stellar system and whether there is any indication for
extended near or mid-IR emission in the vicinity. These information
are presented in Appendix~B for the affected objects. Based on this,
we could decide whether it is meaningful to compare the different
spectra in Fig.~\ref{map_var} to look for temporal variability. For
certain objects, Appendix~B also contains a non-exhaustive comparison
with other mid-IR spectra published in the literature.

In order to explore the significance of the observed flux changes and
check their wavelength-dependence, we computed the ratio of spectra
taken at different epochs. In Fig.~\ref{compare_var} we plotted these
flux ratios for sources where spectra at more than one epoch were
available and their comparison was meaningful (47 stars). Where
several spectra existed, we took the two most extremes. In case the
ratio of a Spitzer/IRS and an ISOPHOT-S spectrum is plotted, the
former was first resampled to the latter's wavelength
resolution. Error bars plotted in this figure represent the quadratic
sum of the errors of each spectrum. Where the ratio of two Spitzer
spectra is computed, instead of the 10\% absolute calibration
uncertainty, we used the 5\% uncertainty of the repeatability (BF\,Ori
and HD\,98800). We also overplotted with vertical dotted/dashed lines
the wavelengths of PAH/ice features for sources where these features
are present.

For the purpose of further analyses, we classified our sources into
three groups based on the shape of the mid-IR SED and the presence of
various spectral features as follows:
\begin{itemize}
\item PAH dominated sources: the spectrum is dominated by PAH bands,
  no apparent 10$\,\mu$m silicate emission band is visible (13
  objects);
\item Silicate emission sources: the spectrum is dominated by the
  10$\,\mu$m silicate emission band, otherwise it is rather
  featureless (36 objects). In a few sources, weak PAH emission is
  present;
\item Silicate absorption sources: the spectrum is rising towards
  longer wavelengths, it displays 10$\,\mu$m silicate absorption
  and/or several ice absorption bands (16 objects).
\end{itemize}
Our classification is presented in Column (15) of Table~\ref{Table1},
as well as in Fig.~\ref{map_var} in parentheses under the name of each
source. We could classify all sources except DG\,Tau,
Ced\,112\,IRS\,4, and CK\,1, which display an apparent combination of
silicate emission and absorption.

\subsection{Statistics and wavelength-dependence of mid-IR variability}
\label{sec:statistics}

Inspecting Figs.~\ref{map_var} and \ref{compare_var}, we could
conclude that the flux ratio in many sources differ from unity at a
level of more than 1$\sigma$, for part of or the whole wavelength
range. Note that 1$\sigma$ here refers to individual wavelength
channels, and averaging the spectrum for a wider wavelength range
suppresses the random, channel-to-channel noise. Thus, a 10-15\% flux
change for a continuous wavelength range is in fact already a strong
indication of variability. Out of the full sample of 68 stars,
variability could not be decided for 21 objects, for having only one
spectrum, or due to problems related to faintness, source confusion,
or instrumental artifacts. We found 10 stars (21\%) to be constant
between 6 and 11.5$\,\mu$m. The remaining 37 sources (78\%) were
classified as variable objects. These are marked in Column 16 of
Table~\ref{Table1}.

The wavelength-dependence of the observed variability can also be
characterized from our data. By looking at Figs.~\ref{map_var} and
\ref{compare_var}, we conclude that the wavelength-dependence of the
flux changes is mostly smooth over the 5--12$\,\mu$m spectral range,
and radical changes within a few channel distance are not
observed. Although changes in the strength of the 10$\,\mu$m silicate
emission feature are ubiquitous and will be quantified and discussed
later (see Sections~\ref{sec:trends} and \ref{sec:type2}), we found no
sources where the silicate or any other spectral feature emerged or
completely disappeared from one epoch to the other. Neither did we
find sources where the shape of the feature changed significantly,
implying that alteration of the size distribution or mineralogical
composition of the emitting dust grains (cf.~the crystallization
during the outburst of EX\,Lup, \citealt{abraham2009}) is not
widespread in our sample.

\subsection{Synthetic photometry and variability timescales}
\label{sec:timescales}

The smooth trends of the flux ratio curves in Fig.~\ref{compare_var}
suggest that the wavelength-dependence of the variability can be fully
described, with high signal-to-noise ratio, via integrating the
spectra over a few wavelength intervals and looking at the changes of
these integrals. Following this idea, we derived synthetic photometry
from the spectra. For objects with silicate emission or absorption, we
calculated synthetic photometry for each source by averaging the ratio
curves in Fig.~\ref{compare_var} between 6 and 8$\,\mu$m and between 8
and 11.5$\,\mu$m, and converting them to magnitude differences. The
former wavelength range is representative of the continuum, while the
latter one covers the 10$\,\mu$m silicate feature and is close to the
usual N photometric band. For PAH dominated objects, we defined a
``continuum band'' by averaging ratios at 6.5--7 and 9.5--10.5$\,\mu$m
and a ``PAH feature band'' containing the remaining parts of the
spectrum.

The synthetic magnitudes can be used to assess quantitatively a
fundamental question: whether the characteristics (frequency and
amplitude) of the variability are the same on annual timescales
(ISOPHOT-S vs.~ISOPHOT-S or Spitzer vs.~Spitzer) and on decadal
timescales (ISOPHOT-S vs.~Spitzer), or there is a hint for long-term
variability trends (higher frequency or larger amplitude) not
measurable by ISOPHOT-S alone or Spitzer alone. In
Fig.~\ref{timescales} we compared the distribution of magnitude
changes in the annual and decadal cases for the three objects types
defined above. Histograms are plotted separately for the ``continuum''
and the ``feature'' magnitude changes. For PAH dominated objects, no
systematic difference can be seen between the short-term and the
long-term distributions, either in the continuum or the feature
(Fig.~\ref{timescales}a, e). For objects with silicate emission, we
plotted separately the intermediate mass and the low-mass stars, but
even with this separation, no significant differences can be seen
between the two timescales (Fig.~\ref{timescales}b, c, f, and g). The
only weak hint for such long-term trends might be seen for silicate
absorption sources, where both in the continuum and the feature, the
decadal changes seem to be somewhat more common and exhibit somewhat
larger magnitude differences than on annual scale
(Fig.~\ref{timescales}d, h). Although the number statistics are
limited, we can conclude that the characteristic timescales of the
mid-IR variability we are observing is less than or equal to a few
years, and in the following we do not differentiate between annual and
decadal changes when studying the physical reasons of the variability.

\subsection{Trends in the mid-IR flux changes}
\label{sec:trends}

In Fig.~\ref{graph} we plot the obtained synthetic magnitude changes.
As in Fig.~\ref{compare_var}, we always considered the two most
extreme spectra. Sources with constant flux in the whole
6--11.5$\,\mu$m wavelength range (Fig.~\ref{compare_var}) are situated
close to the (0,0) point. Most of them fall into an area within
0.1\,mag, indicated by a dark gray circle. Moreover, we checked the
synthetic photometry for the calibration star HD\,181597, which was
observed by both ISOPHOT-S and Spitzer/IRS (Fig.~\ref{fig_x}), and
found $\Delta$mag(6--8$\,\mu$m)=0.067\,mag,
$\Delta$mag(8--11.5$\,\mu$m)=0.076\,mag. For these reasons, we
consider the value of 0.1\,mag as a practical threshold between
variable and constant sources. Objects with a wavelength independent
flux change (either brightening or fading) are distributed along the
45 degree diagonal line (for guiding the eye, this line, broadened by
a $\pm$0.1\,mag typical uncertainty, is plotted with light gray). A
similar, but vertical stripe is plotted for objects with silicate
emission, marking the location of those sources where the continuum
part was constant, but the 10$\,\mu$m silicate emission feature varied
significantly.

Several trends in the distribution of the points can be recognized in
Fig.~\ref{graph}.

\begin{itemize}
\item Objects whose spectra dominated by PAH emission are plotted in
  the upper left panel of Fig.~\ref{graph}. Out of these 8 objects, 5
  are marked as constant in column 15 of Tab.~\ref{Table1}. These are
  situated within 0.1\,mag of the (0,0) point. All the remaining three
  sources (RR\,Tau, CU\,Cha, and HD\,135344\,B) are distributed along
  the diagonal stripe with a magnitude difference of 0.2-0.3\,mag both
  in the continuum and in the PAH feature band. We suggest that these
  objects show real physical variability (RR\,Tau exhibited flux
  changes of 0.5\,mag also between 2.5--5\,$\mu$m, see
  Fig.~\ref{map_var}). The variability in the 6--11.5\,$\mu$m range
  seems to be wavelength independent, as indicated by the distribution
  of points along the diagonal stripe. This means that for each source
  the variation in the PAH features and in the continuum has similar
  amplitude.
\item Stars with silicate emission are plotted in the upper right
  (Herbig stars) and lower left (T\,Tauri stars) panels of
  Fig.~\ref{graph}. They exhibit the largest diversity in variability
  patterns within the whole sample. Three Herbig stars (HD\,95881,
  HD\,142527, and AK\,Sco) and no T\,Tauri stars are constant. 11
  stars show wavelength independent flux changes of $\ge$0.1\,mag,
  located along the diagonal stripe (Herbig stars: BF\,Ori,
  HD\,104237, HD\,144432, VV\,Ser, WW\,Vul; T\,Tauri stars: DR\,Tau,
  VZ\,Cha, WX\,Cha, WW\,Cha, EX\,Lup, S\,CrA). The amplitude can be as
  high as 0.7\,mag. Remarkably, another 11 sources change their
  brightness primarily in the 8--11.5$\,\mu$m band (amplitude is
  $\leq$0.6\,mag) while at shorter wavelengths, they are less variable
  or constant (Herbig stars: HD\,98800, HD\,139614, HD\,142666,
  51\,Oph; T\,Tauri stars: UZ\,Tau, VY\,Tau, CR\,Cha, CT\,Cha,
  VW\,Cha, Glass\,I, CV\,Cha). Interestingly, one T\,Tauri star does
  not follow any of the above described trends. XX\,Cha became
  brighter at shorter wavelengths and fainter at longer wavelengths,
  exhibiting an anti-correlation between the two synthetic photometric
  bands. Although the amplitude of the variability of this source is
  small, the phenomenon is probably real, since it is the slope of the
  mid-IR SED that changed, rather than the more uncertain absolute
  flux level only.
\item Stars with silicate and/or ice absorption (lower right panel in
  Fig.~\ref{graph}) show frequent variability (only one source out of
  10, Ced\,111\,IRS\,5, is constant). The observed variations mostly
  follow the diagonal stripe, indicating wavelength independent flux
  changes. The amplitude of the variability is at most
  0.4\,mag. Slight deviations from this trend can be seen in those
  objects where the 10$\,\mu$m silicate absorption feature is very
  deep, thus, their 8--11.5$\,\mu$m flux is close to zero and remains
  essentially constant (e.g.~Haro\,6-10, LDN\,1551\,IRS\,5, OO\,Ser).
\end{itemize}

There are objects where the detected flux changes are only marginal
($\gtrsim$0.1\,mag), and their association with the above mentioned
trends may not be unambiguous. However, for each trend, there are
clear examples with magnitude changes $>$0.3\,mag, which proves the
existence of these trends. The association of a certain source with a
variability trend might depend on which measurements are
compared. E.g.~in the case of WW\,Cha, we have one ISOPHOT-S and three
Spitzer/IRS spectra. The comparison of the Spitzer spectra indicate
wavelength-independent changes, while the ratio of the ISOPHOT-S
spectrum and the Spitzer/IRS spectrum obtained on 07-Mar-2006 provides
a clear example where the flux variations above and below
$\approx$9$\,\mu$m anti-correlate, similarly to XX\,Cha. VY\,Tau,
BF\,Ori, CR\,Cha, and Glass\,I are further examples for cases where
certain spectral ratios show wavelength independent flux changes while
others indicate changes only in the silicate feature or hints for
anti-correlation between the continuum and the feature. This suggests
that the trend of variability may not be a unique characteristic of a
given source, but may change in time.

For 15 sources where two or more ISOPHOT-S spectra were available, we
could also study variability in the 2.5--4.9$\,\mu$m wavelength
range. We found that five sources (UZ\,Tau, VY\,Tau, HD\,97300,
CV\,Cha, HD\,135344\,B) are constant between 2.5--4.9$\,\mu$m. Out of
these, UZ\,Tau, VY\,Tau, and CV\,Cha are slightly variable at longer
wavelengths, while HD\,98300 and HD\,135344\,B is constant throughout
the whole 2.5$-$11.6$\,\mu$m domain. The two variable PAH dominated
objects in this sample (RR\,Tau and CU\,Cha) have larger flux changes
at shorter wavelengths than at $\lambda\,{>}\,$5$\,\mu$m
(0.2--0.5\,mag). A number of silicate emission objects show
variability amplitudes comparable to those at longer wavelengths
(DR\,Tau, Glass\,I, HD\,104237, and EX\,Lup). One silicate emission
object, however, displays anti-correlation between the shorter and
longer wavelength regimes (CR\,Cha). This behavior is similar to what
we observed for WW\,Cha and XX\,Cha at longer wavelength, except that
the constant pivot point is at somewhat shorter wavelength. The three
variable silicate absorption objects (Haro\,6-10, HH\,100\,IRS, and
OO\,Ser) display slightly higher variability amplitude in the
2--5$\,\mu$m range than at longer wavelengths.

\section{Discussion}

In the previous sections we investigated the characteristics of
variability for a sample of well-known YSOs. Based on the magnitude
changes plotted in Figs.~\ref{timescales} and \ref{graph}, out of 47
objects where multi-epoch observations were available, we found 37
stars (79\%) to vary $\ge\,$0.1\,mag and 20 stars (43\%) to be
variable at $\ge\,$0.3\,mag level either in the continuum or the
feature. Our results can be compared with those of earlier
mid-infrared variability studies carried out on annual (typically 1--4
years) timescales. \citet{barsony2005} surveyed the $\rho$\,Oph cloud
and found that at least $\sim$20\% of the YSOs are variable at
10$\,\mu$m on a few years timescale. Based on multi-epoch Spitzer/IRAC
photometry, \citet{luhman2010} reported that 44\% of Class 0/I/II
sources in the Taurus, and 26\% in the Chamaeleon\,I star forming
regions exhibit at least 0.05\,mag flux changes. Recently, from the
analysis of multi-epoch Spitzer/IRS spectra, \citet{espaillat2011}
concluded that 86\% of the transitional and pre-transitional disks in
the Taurus and Chamaeleon star forming regions display larger than
10\% flux variability in the 5--38$\,\mu$m wavelength range. Our
results confirm the high incidence rates found in these studies and
demonstrate that in our -- rather heterogeneous -- sample of young
objects, annual/decadal mid-IR variability is ubiquitous.

Our data set also offers a possibility to explore the characteristic
timescales of the processes responsible for the observed
variability. These timescales could place constraints on the nature of
the underlying physics. In Section~\ref{sec:timescales} and
Fig.~\ref{timescales} we statistically proved that the distribution of
the variability amplitudes on annual and on decadal timescales do not
differ significantly. This new result implies that on decadal
timescales we observe the same variability process than on shorter
timescales, and no extra long-term physical mechanism is evident from
our data. This fact constrains the physical timescale of the
variability to less than a few years. There are hints that this
characteristic timescale may even be shorter. In our Spitzer/IRAC
monitoring program (PID 60167, PI: P.~\'Abrah\'am), we obtained
14-day-long light curves at 3.6 and 4.5$\,\mu$m of 38 selected low-
and intermediate mass YSOs and found that practically all targets were
variable with amplitudes from a few tens of millimagnitudes to
0.4\,mag. \citet{morales2009} observed 69 YSOs in the IC\,1396A dark
globule between 3.6--8.0$\,\mu$m with Spitzer/IRAC. The obtained
14-day-long light curves indicated that more than half of the observed
YSOs show daily variations, with amplitudes ranging from 0.05\,mag to
0.2\,mag. Finally, the initial results of the Spitzer YSOVAR program
\citep{rebull2011,morales2011}, based on the analysis of 40-day-long
light curves obtained for sources in the Orion Nebula Cluster and
other well-known star-forming regions, indicate that 70\% of the YSOs
with IR excess are variable with days to weeks timescales and with
amplitudes up to higher than 1\,mag. These papers suggest that the
incidence and the amplitude of flux changes we found on decadal
timescale seem to be present already on the much shorter timescale of
a few weeks. If confirmed, this result would strongly constrain the
characteristic timescale of the physical variability process, possibly
pointing to the innermost part of the system as the region of the true
cause of variability.

In order to examine flux changes, we integrated our spectra in
relatively broad synthetic photometric bands. Consequently, the
amplitude of variability within limited wavelength ranges can be even
higher. Moreover, since our conclusions are based on the comparison of
only two or three different spectra, the incidence numbers are lower
limits, and the fraction of mid-IR variable stars may very well be
larger. A dedicated mid-IR monitoring program would probably reveal
variability for many of those objects that we observed to be
constant. The amplitude of variability might also turn out to be
larger if more epochs are taken into account. Examples are HD\,169142
in Fig.~\ref{hd169142}, HD\,104237 in Fig.~\ref{hd104237} and
HD\,142527 in Fig.~\ref{hd142527}, where we plotted additional
mid-infrared spectra from \citet{sylvester1996} and
\citet{vanboekel2005}. Based on these results one might predict that a
very high fraction of all low- and intermediate mass young stars are
variable at a certain level in the mid-IR wavelength
regime. Nevertheless, our data also revealed significant differences
in the incidence of variability among objects with different SED
types. As an example, Figures~\ref{timescales} and \ref{graph} clearly
show that the fraction of variable stars is lower among the
PAH-dominated objects than among the T\,Tauri stars. In the following
we discuss the possible physical reason of variability for each object
type separately.

\subsection{PAH dominated objects}
\label{sec:type1}

In our sample, sources whose spectra are dominated by PAH emission are
all intermediate mass stars with spectral types ranging from B0 to F6,
except the G0-type T\,Tauri multiple system HD\,34700.  This is in
accordance with the observation that stars with spectral type later
than G8 display no PAH emission \citep{geers2006, furlan2006}. The PAH
emission, both spectral features and pseudo-continuum
\citep{desert1990}, originates from molecules transiently heated by
the UV photons of the central source. These species can be located at
relatively large distances from the star. \citet{habart2004,
  habart2006} resolved the PAH emission around several Herbig stars
(e.g.~HD\,169142, CU\,Cha, HD\,100453, and HD\,100546) and found that
PAH emission is originated from the illuminated surfaces of flared
disks. The 3.3$\,\mu$m feature is the least extended ($R\,{<}\,$30
AU), while the 6.2, 7.7, and 11.3$\,\mu$m features are emitted by the
outer disk region ($R\,{\sim}\,$100 AU). In other cases, the source of
PAH emission is an even more extended envelope (e.g.~HD\,97300,
\citealt{siebenmorgen1998, kospal2012}), or might be coming from the
surrounding cloud material (e.g.~IRAS 03260+3111 and MWC\,865,
\citealt{sloan1999, boersma2009}).

When explaining the spectral variability of PAH dominated objects, one
has to take into account that the spatial location of the regions
emitting the mid-IR continuum may differ from that of the PAH
emission. The bulk of the continuum emission at mid-IR wavelengths is
radiated by warm, large dust grains in thermal equilibrium. Depending
on the temperature and luminosity of the central star, these grains
are confined to the inner disk, within a few AUs from the star, much
closer than the source of the PAH emission. Our spectra do not exhibit
any discernible silicate emission around 10$\,\mu$m. In a few cases,
the PAH emission is so strong that a weak silicate emission might be
present but difficult to notice (IRAS\,03260+3111, CU\,Cha, HD\,97300,
WL\,16, MWC\,865). In other cases, the underlying continuum can be
clearly seen and the silicate emission is apparently absent
\citep[HD\,34700, RR\,Tau, HD\,100453, HD\,135344\,B, HD\,141569,
  HD\,169142, BD\,+40\,4124, and LkHa\,224, see
  also][]{juhasz2010herbig}. The lack of the silicate emission feature
suggests an unusually low density of submicron-sized silicate grains
in the surface layer of the disk. One possible reason for this is that
the small silicate grains settled to the midplane, do not receive any
stellar irradiation, thus became too cold to emit at 10$\,\mu$m
\citep{meeus2002, dullemond2007}. Alternatively, it may also happen
that the silicate grains are too large to emit at 10$\,\mu$m due to
grain growth \citep{collins2009}.

Figs.~\ref{compare_var} and \ref{graph} show that the observed flux
changes, if there are any, are within 30\% and are mostly
wavelength-independent: similar for continuum and PAH band
wavelengths. In seeking a physical explanation for the observed
variability, first we investigate whether the flux changes could be
caused by fluctuating obscuration along the line-of-sight. Such a
model is motivated by the fact that one of our targets, RR\,Tau, is
known to be UXor-type star. UXors, named after the prototype UX\,Ori,
show several magnitude deep optical minima due to variable
circumstellar obscuration along the line of sight in a nearly edge-on
system \citep{uxors}. Theoretically, one could imagine that the
observed mid-IR variability is caused by a ``super-UXor'' event when a
circumstellar dust cloud obscures not only the star but the whole
mid-IR emitting region. However, the observed magnitude of the mid-IR
changes would require very large variations in the visual extinction
(the extinction in the V band is $\approx$30 times that at 5$\,\mu$m),
which is not observed at optical wavelengths for our targets.
Moreover, the continuum and in particular the PAH emission originate
from a very extended circumstellar region (at least several AUs in
size), which is unlikely to be completely obscured. Therefore, we
conclude that the UXor-phenomenon cannot be the physical origin of the
mid-IR variations in PAH dominated objects. Any variability seen in
the mid-IR regime should reflect variability in the emission from PAH
or dust particles due to the changing illumination from the central
source.

Variable dust emission may occur when the stellar irradiation of the
disk is modulated by time dependent shadowing effects. Changing the
vertical structure of the inner disk can cast a shadow on a
significant portion of the disk, decreasing its total stellar
irradiation. However, this would only affect the continuum emitted by
large grains and not the PAH emission. Since the PAH particles are
strongly coupled to the gas and do not settle to the midplane,
shadowing them is difficult, because the shadow should extend to a
large solid angle \citep{dullemond2007}. 

The only remaining way to change the irradiation of the disk is by
varying the stellar luminosity. Herbig stars often show modest optical
variability (see long-term optical monitoring programs e.g.~in
\citealt{manfroid1991, herbst1999}), whose amplitude is similar to the
mid-IR changes we observed for the PAH dominated objects
(Fig.~\ref{timescales}). In some cases the optical variability is due
to line of sight effects (extinction, long-lived stellar spots), which
cannot change the luminosity of the star, consequently the integrated
mid-IR emission would stay constant. In a few cases, however, it is
possible that the central luminosity actually changes, for instance
due to variable accretion. Herbig stars are modest accretors:
accretion rates from Br$\gamma$ observations are below a few times
10$^{-7}$ M$_{\odot}$/yr. However, there are indications that in some
sources (e.g.~R\,CrA), the Br$\gamma$ equivalent width is variable,
pointing to variable accretion rate \citep{garcialopez2006}. The fact
that not all intermediate mass stars exhibit luminosity changes is
consistent with our finding that mid-IR variability is not widespread
among PAH dominated stars (three variables out of eight).

Our results suggest a circumstellar geometry where both the regions of
PAH and continuum emission are directly illuminated by the central
source. The most straightforward model to reproduce these results is a
simple flared disk geometry. In this model, the PAH particles and the
large dust grains in the disk surface layer react to the changing
illumination almost instantaneously. Even the optically thick disk
midplane would react on a few years timescale \citep{cg}, which is
less than the temporal baseline of our variability survey. No opaque
puffed-up inner rim is included in this model because no shadowing
effects were revealed by our data. Note that our simple geometrical
picture, deduced from the results of the variability measurements, is
similar to the scheme proposed by \citet{meeus2001} for their Group\,I
sources of Herbig stars. The lack of small silicate grains high above
the midplane of the disk, indicated by the weak or missing silicate
emission feature, may help to keep clear the way for the stellar UV
photons to reach the outer flared disk regions and excite the PAH
molecules. In summary, varying illumination of a flared disk, possibly
caused by variable accretion, is a promising idea to explain the
mid-IR variability of PAH dominated objects.

\subsection{Objects with silicate emission}
\label{sec:type2}

The group of young stars exhibiting the 10$\,\mu$m silicate feature in
emission contains both low- and intermediate mass objects: 20 Herbig
Ae/Be and 16 T\,Tauri stars belong here. Seven Herbig stars show PAH
emission as well, but the PAH features do not dominate the spectrum
(T\,Tauri stars show no PAH emission in our sample). It is unclear why
PAH emission is weak or absent in our intermediate mass objects with
silicate emission. The distribution of spectral types among PAH
dominated and silicate emission Herbig stars are similar, thus the
reason might be related to e.g.~intrinsically low PAH abundance in the
disk atmosphere; the destruction of PAHs in the disk atmosphere by
strong UV flux; or shadowing or softening the stellar irradiation by
the puffed-up inner rim \citep{keller2008}. Taking into account that
the PAH molecules are typically situated in the outer parts of a
flared disk (Section~\ref{sec:type1}), the last possibility invokes a
modestly flared or flat disk geometry, similar to the scheme proposed
by \citet{meeus2001} for their Group II Herbig stars. The fact that in
Fig.~\ref{graph} there is no obvious difference in the distribution of
data points of silicate emission Herbig stars with and without PAH
emission suggests that variability studies of the silicate feature and
continuum are not affected by the presence of PAH emission. This
conclusion is supported by the flux ratios in Fig.~\ref{compare_var}
which suggest that, similarly to PAH dominated objects, for silicate
emission Herbig stars the PAH features and the adjacent continuum show
identical variability, thus magnitude changes in the 6--8$\,\mu$m
continuum and the 8--11.5$\,\mu$m silicate feature domains are
unaffected by the PAH features. Thus in the following we disregard the
presence of PAH features in our silicate emission sources and
concentrate on the 10$\,\mu$m silicate feature only.

According to the standard picture of a two-layer passive disk
\citep{cg}, the mid-IR continuum emission comes from warm dust grains
both in the optically thick disk interior and in the superheated
surface layer. The energy source is the central star, which
illuminates the disk surface. In case of an actively accreting disk,
there is an extra heating by the viscous energy released in the disk
midplane, in the accretion columns, and in accretion hot spots on the
stellar surface as well. The 10$\,\mu$m silicate feature is thought to
arise from the superheated silicate grains in the disk atmosphere. The
radial extent of the disk atmosphere contributing to the mid-IR flux
both in the continuum and in the 10$\,\mu$m feature is within 10\,AU
for a low-mass star. The contribution of the disk interior is confined
to within 1\,AU. For intermediate mass stars, these dimensions are
scaled up, an inner hole may be present, and the inner rim of the
disk, directly illuminated by the star, may be puffed-up
\citep{dullemond2001}. The rim may cast a shadow on parts of the disk,
strongly affecting the emitting area of the 10$\,\mu$m silicate
feature. \citet{vinkovic2006} and \citet{vinkovic2007} argue that many
low-luminosity YSOs (both T\,Tauri and Herbig stars) have compact
($\approx$10\,AU), optically thin, dusty halos around their inner disk
regions. The dust temperature in the halo is set by the stellar
heating. According to their modeling, the emission of the halo is
dominant in the 2--5$\,\mu$m regime, but it can also have important
contribution at longer wavelengths. In certain cases
(e.g.~HD\,163296), there are claims that the halo alone suffices to
explain the observed 10$\,\mu$m silicate feature \citep{vinkovic2006}.

The distributions of low- and intermediate mass silicate emission
stars in Fig.~\ref{graph} is somewhat different from each other. Most
of the Herbig stars show very little variability in the 6$-$8$\,\mu$m
continuum, while the 10$\,\mu$m feature can vary as much as 0.6\,mag
(Fig.~\ref{graph} upper right panel). On the other hand, T\,Tauri
stars may vary also in the continuum (Fig.~\ref{graph} lower left
panel). The number of T\,Tauri stars exhibiting wavelength independent
flux change (diagonal strip) and showing larger changes in the
silicate feature than in the continuum is comparable. The different
variability patterns of low- and intermediate mass stars may indicate
a difference also in their circumstellar structure. Another possible
factor which might affect the variability statistics is the age of the
objects: most low-mass stars in our sample are located in young
star-forming regions, while the intermediate mass sample includes also
isolated, somewhat older, and thus less variable, stars.

First we discuss possible physical reasons for the wavelength
independent flux changes. We will basically consider the same
scenarios as for the variability of PAH dominated objects. Extinction
of the mid-IR emitting region by a passing dust clump in the outer
disk or envelope can be excluded because the accompanying optical flux
changes would be unreasonably high (up to 20\,mag). Wavelength
independent flux changes can be the result of varying illumination by
the central star, possibly caused by time variable accretion onto the
stellar surface. This scenario predicts similar amplitude changes also
at shorter wavelengths. In a few cases, it was possible to test this
by comparing two ISOPHOT-S spectra over the whole 2.5--11.6$\,\mu$m
wavelength range. Indeed, we found approximately wavelength
independent behavior, for DR\,Tau, Glass\,I, HD\,104237, and
EX\,Lup. Although simultaneous optical photometry does not exist,
based on V-band light curves in the ASAS database
\citep{pojmanski1997}, the typical amplitude of the optical
variability of these stars, which is often considered as a proxy for
accretion variations, is similar to the observed ${\le}\,$0.6\,mag
mid-IR variability.

For a surprisingly large number of our sources (including both
T\,Tauri and Herbig stars), the 6$-$8$\,\mu$m continuum is less
variable than the 10$\,\mu$m silicate feature. Similar phenomena were
observed and reported already in the literature \citep{hutchinson1994,
  sitko2008, muzerolle2009, bary2009}. As we concluded in
Section~\ref{sec:statistics}, the variation in the silicate emission
is in all cases a change in the strength of the feature, and not a
change in the spectral shape.

For Herbig stars, an obvious explanation for the observed flux changes
is that the inner rim casts a shadow on parts of the disk. Using a
very simple model, \citet{dullemond2001} investigated the effect of
increasing rim height and found that it can cause increasing continuum
and decreasing 10$\,\mu$m feature, i.e.~an anti-correlation between
shorter and longer mid-IR wavelengths. \citet{juhasz2007} performed a
similar study using a radiative transfer code, and their model
predicted a similar anti-correlation between 3$-$5$\,\mu$m and
$>$8$\,\mu$m fluxes (although their target, SV\,Cep, displayed a
constant 10$\,\mu$m flux, making it necessary to introduce an inner
envelope in the system). \citet{sitko2008} proposed a slightly
different model to explain mid-IR variability of the Herbig stars
HD\,31648 and HD\,163296, suggesting structural changes in the gas
disk close to or inside the dust sublimation zone. This gas disk may
influence how much illumination reaches the inner rim from the central
star. This model also predicts anti-correlation. Physical reasons for
changes in the disk include thermal or magnetorotational instabilities
and planetary perturbations.

In certain cases, inner dust rims may also exist for T\,Tauri stars
\citep{muzerolle2003}, and may play a role in the mid-IR
variability. Such a picture was discussed for LRLL\,31, a T\,Tauri
star where \citet{muzerolle2009} reported anti-correlation between
shorter and longer mid-IR wavelengths, with a constant pivot point at
8.5$\,\mu$m. They proposed that, similarly to Herbig stars, the height
of the inner rim increases due to rising accretion luminosity, casting
a shadow on the outer disk. Another scenario for LRLL\,31 is that the
shadowing is due to dynamical perturbations by a stellar or planetary
companion. These perturbations may induce warps or spiral density
waves in the inner disk, causing variable shadowing
\citep{flaherty2011}. CR\,Cha (anti-correlation between
$\lambda\,{<}\,$5$\,\mu$m and $\lambda\,{>}\,$8$\,\mu$m), as well as
WW\,Cha and XX\,Cha (anti-correlation between
$\lambda\,{=}\,$6$-$8$\,\mu$m and $\lambda\,{>}\,$8$\,\mu$m) are
promising candidates where one of these phenomena might take
place. Note that spectral variability of CR\,Cha was also investigated
by \citet{espaillat2011} who found considerable change in the slope of
the Spitzer/IRS spectra around 6$\,\mu$m.

The high variability amplitudes of the silicate feature of T\,Tauri
stars might also be explained in the framework of a recent model of
turbulent disk accretion. \citet{turner2010} and \citet{hirose2011}
performed magnetohydrodynamical calculations of a stratified shearing
box of the disk, and found that the magnetic activity intermittently
lifts clouds of small grains into the disk atmosphere. The photosphere
height changes by up to one-third over timescales of a few orbits,
resulting also in changes of the mid-infrared surface brightness (see
Fig.~18 in \citealt{turner2010}). They also suggest that the changing
shadows cast by the dust clouds on the outer disk are a cause of the
daily to monthly mid-infrared variability found in many young
stars. Since the dust clouds are optically thin, they may contribute
primarily to the strength of the 10$\,\mu$m silicate feature, while
they affect less the continuum emission. In the model, significant
variations in the 10$\,\mu$m emission occur on the orbital timescale
\citep{turner2010}. In a typical T\,Tauri star, the disk area
responsible for the bulk of the silicate emission is within the inner
1--2\,AU, where the orbital period is less than a few years. Since our
results on the variability also suggest timescales shorter than a few
years (Section~\ref{sec:timescales}), the Turner et al.~model may be
consistent with our findings concerning the variability timescale of
the silicate feature. In order to check whether this mechanism could
also account for the amplitudes of the observed variations, we
computed the predicted fluctuations in the following
way. \citet{hirose2011} found that the starlight-absorbing surface
moves up and down by a factor of two. The maximum possible 10$\,\mu$m
variability amplitude will occur if unit starlight optical depth is
reached where the temperature is near the 300\,K needed to place the
peak reprocessed emission at 10$\,\mu$m. Under these favorable
conditions, the 10$\,\mu$m flux could change by a factor two, or
0.75\,mag, similar to the mid-IR flux changes we observe in our
silicate emission sample. We note that the disk photosphere height
changes almost simultaneously at all azimuthal locations in the disk
annulus in question, because the differential rotation rapidly shears
out any non-axisymmetric structure, implying significant changes in
the emission of the system as a whole. Since the Turner et al.~model
predicts almost independent variability in the mid-infrared continuum
and in the silicate feature, with higher amplitudes of the latter one,
we suggest it as a promising explanation for our observations.

There might be other effects which change the silicate feature and the
adjacent continuum. One possibility is that the system is binary or
multiple with multiple disks. The silicate features in each component
may be different, and one of them may be variable, resulting in
changes in the combined spectrum. Another option is what is happening
for DG\,Tau, a single T\,Tauri-type star: there is both emission and
absorption along the line-of-sight, and one of them is changing
\citep{bary2009}. There is also a possibility that the 10$\,\mu$m
silicate feature and the continuum changes independently from each
other. This might occur if the continuum emission is dominated by
viscous heating and the emission feature is dominated by illumination
of the disk surface. The accretion in the disk midplane and the
accretion onto the stellar surface changes independently, or at least
with significant time difference. Finally, the example of CoKu\,Tau/4
demonstrates that even for non-accreting sources, wavelength-dependent
mid-IR variability can occur. This system consists of a close
eccentric binary surrounded by a circumbinary disk. As the distance of
the orbiting stars from the inner edge of the dust disk continuously
changes in time, the irradiation and consequently the mid-IR emission
of the inner disk wall also varies \citep{nagel2010}. Unfortunately,
the present dataset does not allow us to decide between these
scenarios. Future multi-epoch simultaneous optical/near-IR/mid-IR
observations, and detailed modeling of each source would be the key to
conclusive results.

\subsection{Objects with silicate/ice absorption}

Stars with silicate and occasionally with ice absorption bands are all
low-mass stars in our sample. The increasing SED and the absorption
features indicate that these objects are deeply embedded in a cold,
dense, extended envelope. The mid-IR emission is probably coming from
a circumstellar disk as well as from the inner warm part of the
envelope. The fact that they are mostly distributed along the diagonal
stripe in Fig.~\ref{graph} suggests scenarios that cause
wavelength-independent flux changes. The short wavelength
($\lambda\,{<}\,$5$\,\mu$m) flux changes, where available, support
this idea.

In accordance with this result, the concept of shadowing as discussed
for disks is not applicable for envelopes. This is not surprising,
since the envelope covers a large solid angle, thus the shadowing
material within the inner edge of the envelope should cover a large
solid angle. In principle, shadowing effects might occur in the disk,
but these changes would be tempered by the backwarming of the outer
disk by the surrounding envelope \citep{natta1993}. Similarly to PAH
dominated and silicate emission objects in the diagonal stripe, for
silicate absorption objects, the most obvious explanation for the
varying central luminosity is variable accretion rate. A good example
for this is the well-studied triple system T\,Tau, where the Sa
component is responsible for the 10$\,\mu$m silicate absorption and
for the mid-infrared variability. \citet{vanboekel2010} show that the
short-term (daily-weekly) mid-infrared variability of T\,Tau\,Sa can
be attributed to variable accretion, while the long-term (yearly) flux
changes can be either due to variable accretion, variable foreground
extinction, or the combination of these two mechanisms.

\section{Summary and conclusions}
\label{sec:summary}

In this paper we present low-resolution mid-IR spectra of 68 low- and
intermediate mass young stars obtained with ISOPHOT-S between 1996 and
1998 and with Spitzer/IRS between 2004 and 2007. Utilizing
self-developed software packages, we interactively re-processed the
spectra for improved removal of instrumental artifacts. By comparing
multi-epoch spectra of each object, we analyzed the statistics and the
trends of mid-IR spectral variability and its implications on the
geometry of the circumstellar material. Our study has three novel
aspects: (1) the spectroscopic observations enable the study of
spectral variations in a relatively extended mid-IR wavelength range;
(2) instead of focusing on a certain star forming region, individual
well-known YSOs were observed; (3) and the timescale of our
ISO/Spitzer comparison is longer than previous mid-IR variability
studies, covering a full decade. Our main achievements are the
following:
\begin{itemize}
\item The mid-infrared spectral atlas constructed for this project
  presents an improved, final data reduction of ISOPHOT-S spectra. The
  atlas contains all low- and intermediate mass stars ever observed
  with the ISOPHOT-S instrument, thus, it constitutes the legacy of
  the ISO.
\item Our spectra show that mid-IR variability among low- and
  intermediate mass YSOs is ubiquitous. We calculated synthetic
  photometry in the 6--8$\,\mu$m and the 8--11.5$\,\mu$m wavelength
  range, and found that 79\% of the sources vary more than 0.1\,mag,
  while 43\% is variable above the 0.3\,mag level. A comparison of the
  variability characteristics on annual and decadal timescales
  revealed no significant differences, constraining the physical
  timescale of the variability to less than a few years.
\item For intermediate mass stars with spectra dominated by PAH
  emission, we found relatively low incidence of variability. Flux
  changes in this group are mostly wavelength-independent, and can be
  interpreted in terms of non-steady irradiation of the disk due to
  fluctuating accretion. We propose a simple flared disk geometry to
  model these sources.
\item Intermediate mass stars exhibiting silicate emission at
  10$\,\mu$m often show higher variability amplitude in the silicate
  feature than in the adjacent continuum. Shadowing the disk by
  vertical variations of a puffed-up inner disk rim is invoked to
  explain the wavelength-dependence of the flux changes. The deduced
  geometry is a modestly flared or flat disk.
\item T\,Tauri stars are the most frequent variables in our
  sample. For those which exhibit more pronounced variability in the
  10$\,\mu$m silicate feature than in the continuum we propose to
  consider a new model: strong magnetohydrodynamical turbulence may
  intermittently lift clouds of small grains into the disk atmosphere
  resulting in extra silicate emission
  \citep{turner2010}. Nevertheless, based on our spectra alone, we
  cannot exclude that other scenarios may also be applicable.
\item Sources exhibiting silicate or ice absorption are objects
  embedded in a dense envelope. They typically show wavelength
  independent flux changes, probably due to varying accretion rate.
\end{itemize}
Our results suggest that mid-IR variability is widespread among
YSOs. Interpreting the amplitude, wavelength-dependence, and
timescales of these flux changes is a new and promising possibility to
explore the structure of the circumstellar disks and their dynamical
processes.

\acknowledgments

We thank the anonymous referee for providing a thorough and very
helpful report. The ISOPHOT data presented in this paper were reduced
using the ISOPHOT Interactive Analysis package PIA, which is a joint
development by the ESA Astrophysics Division and the ISOPHOT
Consortium, lead by the Max-Planck-Institut f\"ur Astronomie
(MPIA). This work is partly based on observations made with the
Spitzer Space Telescope, which is operated by the Jet Propulsion
Laboratory, California Institute of Technology under a contract with
NASA. This research has made use of the SIMBAD database, operated at
CDS, Strasbourg, France; NASA's Astrophysics Data System; and the
NASA/IPAC Infrared Science Archive, which is operated by the Jet
Propulsion Laboratory, California Institute of Technology, under
contract with NASA. \'A.K. acknowledges support from the Netherlands
Organization for Scientific Research (NWO). This work was partly
supported by the grant OTKA-101393 of the Hungarian Scientific
Research Fund.

{\it Facilities:} \facility{ISO (ISOPHOT-S)}, \facility{Spitzer (IRS)}.

\appendix

\section{Interactive post-processing of ISO observations}

Though the ISO Legacy Archive contains reliable results for most
ISOPHOT-S observations, several instrumental artifacts were discovered
after the closure of the Archive in 2001. We have developed an
IDL-based processing package aiming at correcting these
artifacts. Possible correction algorithms were tested and optimized on
a set of 43 normal star measurements, and the same data set was used
to estimate typical measurement uncertainties via comparison with
photospheric models. The processing scheme consists of the following
steps:

\paragraph{Deglitching.} 
As a preparation, we checked whether PIA had successfully removed all
cosmic glitches from the data. Although the PIA built-in deglitching
algorithms are quite efficient, we often experience that glitches are
still present in observations of very faint sources, producing
artificial spikes in the spectra. The reason is that at low signal
level the recovery from a cosmic hit takes a relatively long time, and
the disturbance following the hit may have a characteristic timescale
comparable to the full measurement time (for an example see
Fig.\,\ref{fig_glitch}). In such a situation, the statistical
algorithms of PIA -- which assume that most of the data points are not
affected by glitches -- may not work properly. In order to identify
the remaining glitches, the signal evolution of each pixel was
visually inspected at the intermediate Signal Raw Data (SRD)
processing level, and -- if necessary -- the affected data points were
manually discarded using PIA.

\paragraph{Detector temperature.} It was suggested during the ISO
mission that when the detector temperature was outside the nominal
2.8--3.1\,K range, the transient behavior might have differed from the
normal one, leading to systematic photometric errors
\citep{acosta2003}. We checked all observations presented here, and
found that one measurement (HD\,142666, TDT: 10402847) had non-nominal
detector temperature. In the case of this measurement we computed for
every pixel the dispersion of the calibrated flux values derived from
the individual observing ramps by the dynamic calibration
procedure. We checked whether the dispersion was higher in this
affected measurement than in other data sets of similarly bright stars
measured in the nominal temperature range. We indeed found a
2--3$\sigma$ increase in the noise, however, it is only 3--4\% and
1--2\% of the total measured flux in the ISOPHOT-SS and ISOPHOT-SL
wavelength range, respectively. Thus, we conclude that non-nominal
detector temperature is not a major source of uncertainty in the
ISOPHOT-S observation of HD\,142666, and we will neglect this effect
in the processing.

\paragraph{Orbital phase.} It was suspected that observations
obtained either at the very beginning (orbital phase $<0.25$) or at
the very end of the orbit (orbital phase $>0.80$) might have suffered
from reduced accuracy due to a higher dark current and to a higher
cosmic glitch rate, respectively. In our sample, 9 measurements were
taken during early, and 18 measurement during late orbital phase. In
order to check if there is an increase of uncertainty related to the
dark current subtraction at early orbital phases, we tested the
distribution of dedicated dark current calibration measurements around
a fitted curve that represents the orbital variation of the dark
current per pixel, and found no higher scatter at early orbital phases
\citep[see also Fig.~1 in][]{acosta2003}. This means that the
subtraction of the orbital dependent dark signal does not introduce a
higher noise component at early phases than later in the orbit. Close
to the end of the science window of the orbit, the rate of cosmic
glitches could somewhat rise, possibly making the data noisier than at
earlier orbital phases. However, this noise is probably random in
nature, and there is no reason to assume correlated behavior of groups
of pixels. Since our analysis is based on synthetic photometry derived
from the spectra, our results are unaffected by the increased noise at
late orbital phase.

\paragraph{Signal memory from preceding observation.} We found
the appearance of artificial spectral features in some spectra due to
memory effects from a preceding observation of a bright source
(Fig.~\ref{fig_memo_a}). In order to identify affected spectra, we
studied the short (32\,s) dark measurements performed before each
ISOPHOT-S observation. After reducing the complete sample of these
measurements from the mission, we determined their average signal
levels and typical measurement uncertainties per pixel (dashed and
dotted lines in Fig.~\ref{fig_memo_b}). If the measured short dark
signal of a particular observation exceeded the average level by more
than 1$\sigma$ for a continuous section of the spectrum (i.e.~for a
group of neighboring pixels, for an example, see
Fig.~\ref{fig_memo_b}), then the observation was flagged for memory
effect.

From tests on cases where external information on the spectral shape
(theoretical models, TIMMI2\footnote{Thermal Infrared Multimode
  Instrument} observations) was available, we concluded that
subtracting the excess of the short dark signal from the astronomical
observation per pixel can efficiently correct for the memory
effect. In order to determine the accuracy of this correction, we
analyzed the only standard star observation that was strongly affected
by signal memory effect. We found that for pixels where the correction
was below 50\% of the measured signal, the uncertainty of the
corrected signal remained below 8--10\%, i.e less than the photometric
uncertainty assigned to the measurements in the final step of our
processing scheme. In the 50--100\% memory excess range, the precision
of the corrected fluxes was 20\%, and even pixels suffering from very
high memory contamination, in the range of 100--300\%, could be
corrected to have a standard deviation of 33\%. In our sample, we
performed memory correction on 19 observations of 17 different sources
(marked with M in Column (7) of Table~\ref{Table1}). In almost all
cases the correction was in the 5--20\% range, and the associated
uncertainly was probably lower than the general photometric
accuracy. The two more strongly affected cases (marked with M! in
Table~\ref{Table1}) are Ced\,111\,IRS\,5 (TDT 62501412), and CK\,2
(TDT 10803228). Out of these, the SL part of the spectra of
Ced\,111\,IRS\,5 could be used for variability studies, while CK\,2
was discarded from our analysis.

\paragraph{Background subtraction.} For 61 observations 
out of the 94 ISOPHOT-S spectra presented here, no corresponding sky
background measurement has been performed. In the wavelength range of
ISOPHOT-S the dominant background component is the zodiacal light
(apart from localized regions like HII regions, reflection nebulae, or
the galactic plane). The zodiacal flux was clearly measurable for
ISOPHOT-S in the 5.8--11.7\,$\mu$m regime \citep{leinert2002}, and its
contribution is not negligible for fainter stars
(Fig.\,\ref{fig_bgd}). We created a model which was able to predict
the spectrum of the zodiacal background towards a given direction and
on a given date. The algorithm used 4.9 and 12\,$\mu$m photometric
points from COBE/DIRBE photometry as input, fitted a Planck curve to
them, and finally scaled with a wavelength- and exposure
time-dependent factor optimized for the 29 high quality ISOPHOT-S
spectra of the zodiacal light presented in \citet{leinert2002}. Those
61 measurements where background was predicted in this way are marked
with B in Column (7) of Table~\ref{Table1}.

We also performed an error analysis of the correction by computing a
predicted background spectrum for those 18 staring observations where
a background measurement was available, and compare the two spectra
pixel-by-pixel. The results demonstrated that the average error bar
for the ISOPHOT-SS part is between 0.04 and 0.07\,Jy, while for the
ISOPHOT-SL domain is between 0.03 and 0.13\,Jy, monotonically
increasing towards longer wavelengths. For most of our targets, this
additional uncertainty is well below the overall 10\% precision of the
ISOPHOT-S calibration. For the five faintest sources (among those
where predicted background was subtracted), however, the uncertainty
of the predicted background exceeds 10\% of the measured flux. These
cases are BARN\,5\,IRS\,1, LDN\,1551\,IRS\,5, Ced\,111\,IRS\,5,
[SVS76]\,Ser\,4, and CK\,2. For the first three objects, only the
ISOPHOT-SS part is affected, while for the latter two, the whole
ISOPHOT-S spectrum is uncertain. In Column 7 of Table~\ref{Table1} we
mark six measurements of these five objects with an exclamation mark
(B!). We conclude that variability can still be discussed for the
first three objects using the ISOPHOT-SL part (and Spitzer when
available), while for the two latter cases, variability study is not
possible.

\paragraph{Special processing of spectral maps.}
There are a few measurements which were performed in mapping mode
(usually 1D scans). The target is usually situated on the central map
position, and the neighboring raster steps can be used to obtain a
background spectrum interpolated for the source's position. The
mapping mode has several advantages over the typical ON--OFF staring
observations, because it provides a well-measured background, and the
subtraction of the interpolated background automatically cancels any
memory effect, thus the algorithm described above is not used here.
It is a problem, however, that the dynamic response calibration method
\citep{acosta2003} was developed for staring observations, and maps are
standardly calibrated by means of a static spectral response
function. This method, however, cannot cope with the different
transient timescales of the different detector pixels, and may produce
obvious spectral artifacts. In order to find an alternative solution,
we tested a simple algorithm which considers the time sequence of the
map as a long staring observation, applies Dynamic response
calibration on it, and creates a map from the calibrated flux
values. The results were in good agreement with the ISOPHOT photometry
and lacked any obvious spectral artifacts. Raster observations of
LDN\,1489\,IRS and OO\,Ser were processed using this algorithm.

\paragraph{Off-center position of the source in the beam.}
The footprints of the ISOPHOT-S pixels were peaked rather than
flat-top, therefore observing a compact source outside the optical
axis of the pixel could change the measured signal. Since the
calibration was set up for the pixel center, off-center observations
have to be corrected in order to derive correct flux values. In
addition, because the footprint profiles varied with wavelength (there
were especially large jumps at 3.7 and 8.8\,$\mu$m) off-center
position led to the appearance of spectral artifacts (an example is
shown in Fig.\,\ref{fig_off}). In order to correct the flux values and
the spectral artifacts, we determined the offset of the source from
the center by comparing the accurate position of the object with ISO's
coordinates. The 2-dimensional footprint map (we utilized the ones
derived by \citealt{leinert2002}) was then sampled at the offset
position, and the measured flux was scaled with the ratio between
footprint values at the center and at the offset location. This
procedure was repeated for each detector pixel independently.

In order to estimate the uncertainty related to the correction, we
derived error bars for each point of the footprint map by comparing
those two original footprint measurements of the calibration star HR
7924 which were averaged to create the final footprint map. We found
that in the majority of the measurements the estimated error was less
than 5\% of the total flux in the whole ISOPHOT-S wavelength
range. Offset correction was applied to 66 ISOPHOT-S spectra of 50
different sources (marked with O in Column (7) of
Table~\ref{Table1}). There are only two seriously affected ($>10$\%)
data sets: the short wavelength part of one of the two spectra of
WW\,Vul (TDT: 17600465) and the spectrum of MWC\,863 (TDT:
28900460). Both measurements are flagged with an exclamation mark (O!)
in Table~\ref{Table1}, and are discarded from the variability
analysis.

\paragraph{Empirical photometric correction.}
In order to check for any remaining systematic effects in the final
photometry, we queried the ISO Archive, collected all normal star
observations, and reduced them in the way described
above. Fig.~\ref{fig_std} shows a typical result, where the measured
spectrum is compared with the expected one. The latter spectrum was
taken from the ISO calibration data base which includes a compilation
of predicted spectra provided by M.~Cohen or P.~Hammersley
(http://iso.esac.esa.int/users/expl\_lib/ISO/wwwcal/). For those
objects not included in the data base we took one of the available
models of a star of identical spectral type, and scaled the values to
the K-band magnitude of our object. In our sample of 43 observations
of 28 normal stars, no excess due to hot circumstellar dust is
expected (the sample also includes the calibrator stars).

In Fig.~\ref{fig_photcheck} we show the measured-to-predicted ratios
of 3 representative pixels for the whole normal star sample. For each
pixel, the ratios above a brightness threshold (usually 5\,Jy) were
averaged and their standard deviation was computed. For the faint
stars (below 1\,Jy), the ratio values became too noisy, and we
computed the average and the standard deviation of differences between
the measured and predicted fluxes. The averages derived this way
represent the typical systematic errors. We decided to correct for
these systematic offsets by dividing by the average flux ratio (for
bright sources) or by subtracting the average flux difference (for
faint sources) from all observations. This {\it empirical photometric
  correction} was the last step of our post-processing sequence.

\paragraph{Error budget.}
The formal uncertainties provided by the PIA reduction do not contain
several important sources of uncertainties, e.g.~the ones related to
the corrections in the post-processing. The average uncertainties in
Fig.~\ref{fig_unc}, on the other hand, do not represent the quality
differences among observations (e.g.~the effect of a particularly
energetic cosmic glitch). As final photometric uncertainties, we
decided to adopt the maximum of the two types of error bars. In
practice, for most pixels/measurements the average uncertainty values
are the higher. Their typical values are 5--10\% for bright sources,
30--100\,mJy for faint sources.

\section{Notes on individual sources}

In the following, we collected literature information on some of the
sources, in order to help the reader with the understanding and the
interpretation of the spectra plotted in Fig.~\ref{map_var}. We mostly
focused on three points: information on (1) variability in the
optical-IR wavelength regime, (2) multiplicity, and (3) extended
emission at optical-IR wavelengths. Single sources with no extended
emission, and whose observations are unaffected by technical artifacts
are not discussed here. Note that here we also included sources where
variability could not be investigated.

\paragraph{IRAS\,03260+3111} (also known as NGC\,1333\,SVS\,3) is a
young Herbig binary illuminating a small reflection nebula
\citep{strom1976}. The binary is composed of a B5 and an F2 star, with
a separation of 3$\farcs$62 \citep{straizys2002}. The secondary is 32
times fainter than the primary in the K band and 24 times in the L
band. \citep{haisch2004}. The system appears extended with a
fan-shaped nebulosity at 10$\,\mu$m \citep{haisch2006}. There is no
evidence of emission from silicate grains in the mid-IR spectrum of
IRAS 03260+3111. On the other hand, PAHs emission is observed not only
at the position of the Herbig star, but also at various locations in
the NGC 1333 nebula. The different PAH features change with changing
distance from the central source due to the changing relative
populations of ionized and neutral PAHs \citep{bregman1993, roche1994,
  joblin1996, sloan1999}. The MIR spectra published in these papers
all have different flux levels, and they are also different from our
ISOPHOT-S spectrum. These differences are due to the different beam
sizes of the instruments and the exact pointing.

\paragraph{BARN 5 IRS 1} is a low-mass embedded young stellar
object in the Barnard\,5 dark cloud \citep{beichman1984}. It is
surrounded by a very faint reflection nebula detectable at
near-infrared wavelengths \citep{moore1992}. We found no evidence in
the literature for extended mid-infrared emission. JHKL photometric
monitoring revealed that the source has faded significantly between
1983 and 1993 \citep{moore1992, moore1994}. In the case of the
ISOPHOT-S spectrum, predicted background was subtracted, making the
short wavelength part more uncertain. However, the SED of the source
rises steeply towards longer wavelengths, and the long-wavelength part
of the ISOPHOT-S spectrum can be safely compared with the Spitzer/IRS
spectrum.

\paragraph{LDN\,1489\,IRS} is a single embedded YSO
\citep{connelley2008}. The source looks extended in the 2MASS J, H,
and K images, as well as in the 3.6$\,\mu$m Spitzer/IRAC
image. Although it is point-like in the IRAC images at
$\geq$5.8$\,\mu$m, we cannot exclude the possibility of extended
emission, which may be the reason for the slight difference between
the ISOPHOT-S and the Spitzer/IRS flux levels. Thus, this source is
not included in the discussion about variability.

\paragraph{T\,Tau} is a triple system consisting of T\,Tau\,N,
T\,Tau\,Sa, and T\,Tau\,Sb. The largest separation between the
components is 0$\farcs$7 \citep{kohler2008}, therefore neither ISO nor
Spitzer resolves the system: the observed flux represents the sum of
all components. T\,Tau\,Sa is a Herbig star, while T\,Tau\,Sb and
T\,Tau\,N are low-mass objects \citep{duchene2006}. \citet{skemer2008}
presented photometry at 8.7, 10.55, and 11.86$\,\mu$m for each of the
three components, while \citet{ratzka2009} show 8--13$\,\mu$m spectra
for T\,Tau\,S and T\,Tau\,N. Based on these results, it seems that
only T\,Tau\,Sa displays silicate absorption, while T\,Tau\,N and
T\,Tau\,Sb have more-or-less featureless 10$\,\mu$m spectra (T\,Tau\,N
may display slight emission). \citet{beck2004} resolve T\,Tau\,N and
T\,Tau\,S in the 2--4$\,\mu$m range, and claim that T\,Tau\,S shows
deep, broad H$_2$O ice absorption at 3.1$\,\mu$m, while T\,Tau\,N is
featureless in this wavelength range. Photometric monitoring between 2
and 12$\,\mu$m by \citet{beck2004} and by \citet{vanboekel2010}
indicate that T\,Tau\,N is constant within the uncertainties, while
T\,Tau\,S shows significant variability.

\paragraph{DG\,Tau} is a single classical T\,Tauri star
\citep{leinert1991, connelley2008}. \citet{wooden2000} report on
multi-epoch mid-IR spectra where they found that the 10$\,\mu$m
silicate feature showed significant variability on time scales of
months to years, while the underlying continuum remained relatively
constant. They claim that prior to 1996, DG\,Tau had a featureless
10$\,\mu$m spectrum, which then turned into silicate emission, then
went into absorption at the end of 1998. \citet{bary2009} also
observed that the silicate feature displayed variability on monthly
and yearly timescales. All our ISOPHOT-S and Spitzer spectra are
featureless, but they indicate that the continuum is variable.

\paragraph{Haro\,6-10} (also known as GV\,Tau) is a T\,Tauri binary
with a separation of 1$\farcs$2, P.A. 355$^{\circ}$
\citep{leinert1989}. Both the ISO and the Spitzer spectra contain the
full flux from both components (the Spitzer slit is oriented along the
binary). At wavelengths shorter than 4$\,\mu$m, the southern component
dominates, while most of the emission above 4$\,\mu$m comes from the
extremely red northern component
\citep{leinert1989}. \citet{leinert2001} reported J, H, K, L$'$, and
M-band variability, as well as the variation of the 3.1$\,\mu$m
ice-band absorption. Their resolved observations prove that both
components vary significantly on timescales of a few months. Our ISO
and Spitzer spectra confirm that the system is variable.

\paragraph{LDN\,1551\,IRS\,5} is a deeply embedded low-mass
triple system consisting of a binary with projected separation of
0$\farcs$3, and a third component at a distance of 0$\farcs$09 of the
northern component \citep{lim2006}, each star harboring a small
circumstellar disk. The whole system is surrounded by a circumbinary
disk and a flattened envelope as well
\citep{osorio2003}. \citet{liu1996} observed the source at 12$\,\mu$m,
and by comparing the observed flux with previous 12$\,\mu$m and N-band
photometry, reported significant mid-IR variability. Although
LDN\,1551\,IRS\,5 displays some extended emission at all Spitzer/IRAC
images, the larger aperture ISO spectrum has a lower flux than the
Spitzer spectrum, indicating real variability. We note that in the
case of the ISOPHOT-S spectrum, predicted background was
subtracted. The flux of LDN\,1551\,IRS\,5 steeply rises towards longer
wavelengths, but below 5$\,\mu$m, the source is so faint that the
uncertainty of the background subtraction dominates the error budget,
and the overall error is above 10\%.

\paragraph{HL\,Tau} is a single classical T\,Tauri star. At optical
wavelengths, the object is a compact reflection nebula with
complicated morphology and diameter of $\approx$3$''$; no point source
is visible \citep{stapelfeldt1995}. In the JHK images of
\citet{murakawa2008}, the central source can be seen, but it is still
surrounded by extended nebulosity (about 3$''$ in
extension). According to our knowledge, there is no information on the
literature on whether the source is extended at mid-IR
wavelengths. The fact that the larger aperture ISO spectrum has a
lower flux level than the Spitzer spectrum indicate real
variability. The 8--13$\,\mu$m UKIRT/CGS3 spectrum obtained in 1993 by
\citet{bowey2001} is consistent with the ISOPHOT-S brightness level.

\paragraph{UZ\,Tau} is a hierarchical quadruple T\,Tauri
system. UZ\,Tau\,E is a spectroscopic binary with an orbital period of
19.1\,days. At a distance of 3\farcs8 can be found UZ\,Tau\,W, which
itself is a 0\farcs34 binary \citep[][and references
  therein]{prato2002}. Due to periodic accretion from the circumbinary
disk, UZ\,Tau\,E shows optical variability of about 0.4--1.0\,mag in
the V, R, and I bands, while UZ\,Tau\,W is constant within 0.02\,mag
\citep{jensen2007, kospal2011}. UZ\,Tau\,E is also variable at
millimeter wavelengths, possibly related to non-thermal emission from
magnetospheric reconnection events above the binary components'
photospheres \citep{kospal2011}. The Spitzer/IRS slit was centered on
UZ\,Tau\,E, although the W component also has some contribution to the
observed flux. The ISOPHOT-S measurements were centered between the E
and W components, and essentially measured the sum of the two
components' fluxes. For this reason, the Spitzer/IRS and ISOPHOT-S
spectra cannot be compared, but variability can be studied by
comparing the two ISOPHOT-S spectra, which indicate some flux changes
in the silicate feature around 10$\,\mu$m.

\paragraph{VY\,Tau} is a subarcsecond binary with a separation of
0$\farcs$66, P.A. of 317$^{\circ}$, K-band flux ratio of 3.8, and
L-band flux ratio of 6.1 \citep{leinert1993,
  mccabe2006}. \citet{richichi1994} note that the K-band brightness
ratio of the two stars is time-variable. VY\,Tau is an EXor-type
object, exhibiting sporadic 1-4\,mag optical outbursts, but has been
in quiescence at V$\approx$13.8\,mag since about 1972
\citep{herbig1977, herbig1990, herbst1994}. The source shows
photospheric emission in the JHK bands \citep{shiba1993, sipos2011},
but displays clear excess emission from about 3$\,\mu$m and a clear
10$\,\mu$m silicate emission feature \citep{furlan2006, sargent2009,
  watson2009, sipos2011}. We found no indication in the literature for
extended IR emission. Both the ISO and the Spitzer spectra contain the
full flux from the binary components, thus they can be compared.

\paragraph{DR\,Tau} is a highly accreting single classical
T\,Tauri star with a long history of optical and near-IR photometric
and spectroscopic variability, which point to magnetospheric accretion
with fluctuating accretion rate \citep[e.g.][]{gotz1980, isobe1988,
  hessman1997, smith1999, alencar2001}. \citet{kenyon1994} present
light curves from the B to the L band, and explain the observed flux
variations with the presence of a hot spot (the base of the accretion
flow) on the stellar surface.  DR\,Tau is sometimes classified as an
EXor, a young star exhibiting short accretion outbursts
\citep{lorenzetti2009}. Our observations indicate that DR\,Tau is also
variable in the whole 2.5--15$\,\mu$m wavelength range.

\paragraph{HD\,34700} is a quadruple T\,Tauri system. The primary
itself is a double-line spectroscopic binary (HD\,34700\,Aa and
HD\,34700\,Ab), consisting of two equal-mass G0-type stars
\citep{arellano2003, torres2004}. Two faint stars at a distance of
5$\farcs$2 (HD\,34700\,B, spectral type: M1-2, classical T\,Tauri) and
9$\farcs$2 (HD\,34700\,C, spectral type: M3-4, weak-line T\,Tauri) are
also associated with HD\,34700\,A \citep{sterzik2005}. The IRS slit
only includes the spectroscopic binary HD\,34700\,A, while the
ISOPHOT-S aperture includes all four components. Although HD\,34700\,B
seems to be actively accreting, due to their faintness and distance
from component A, components B and C have negligible contribution to
the ISOPHOT-S flux. Thus, both instruments essentially measure
HD\,34700\,A.

\paragraph{RR\,Tau} is a Herbig Ae star. \citet{grinin2002} observed a
deep optical and near-IR minimum in the light curve of RR Tau in
2000--2001. They found that the UBVRIJH fluxes showed synchronous
behavior, and the color changes indicated that the dimming was caused
by a circumstellar dust cloud crossing the line of sight. However, the
fading in the optical and JH bands were accompanied with a flux
increase in the K and L band, suggesting that flux changes at longer
wavelength have a different physical mechanism than variable
extinction.

\paragraph{Reipurth\,50\,N\,IRS\,1.} Reipurth\,50 was first reported as a
nebulous object by \citet{reipurth1985}. \citet{reipurth1986} reported
the sudden appearance of a highly variable, conical nebula
Reipurth\,50\,N, possibly due to an FU\,Ori-like
eruption. \citet{scarrott1988} identified Reipurth\,50\,N\,IRS\,1
(a.k.a.~HBC\,494) as the illuminating source of the Reipurth\,50\,N
reflection nebula. Near-IR imaging, imaging polarimetry, and CVF
spectroscopy by \citet{casali1991} revealed that there is only one
single source in the Reipurth\,50\,N cloud, IRS\,1, which is a point
source at 3.6$\,\mu$m.

\paragraph{SX\,Cha} is a T\,Tauri-type binary with a separation of
2$\farcs$2, and P.A. of 288$^{\circ}$ \citep{reipurth1993}. The east
and west components have spectral types of M0.5 and M3.5, respectively
\citep{muzerolle2005b}. The 2MASS images show that while the west
component is brighter in the J band, the east component is brighter in
the K band. The Spitzer slit is centered on the east component, and
the west component is 1$\farcs$2-1$\farcs$6 away from the slit
centerline. This means that 55-70\% of the flux of the west component
is also included in the Spitzer measurement. The ISOPHOT-S data,
however, contains the total flux of both components. Thus, due to beam
confusion, the spectra of the two instruments cannot be directly
compared.

\paragraph{HD\,95881} is a Herbig Ae star. Based on spectro-astrometric
observations, \citet{baines2006} consider it a possible subarcsecond
binary. We found no indication in the literature for extended
emission at any wavelength, thus the ISOPHOT-S and Spitzer spectra can
be safely compared. The two spectra agree within the measurement
uncertainties until about 10$\,\mu$m. Between 10 and 11.5$\,\mu$m, the
ISOPHOT-S spectrum is higher than the Spitzer at a 1$\sigma$
level. The TIMMI2 spectrum published by \citet{vanboekel2005} is even
higher than the ISOPHOT-S, indicating possible variability.

\paragraph{CT\,Cha} is a single T\,Tauri-type star
\citep{guenther2007}. Recently, \citet{schmidt2008} found a candidate
sub-stellar companion (brown dwarf or planet) by direct imaging, at a
projected separation of 2$\farcs$67. Since the companion is very faint
(K-band flux ratio is 313), it does not affect our spectra, and the
ISOPHOT-S and Spitzer/IRS data can be safely compared.

\paragraph{VW\,Cha} is a triple system. The separation of the A and B
components is 0$\farcs$66, P.A. 177$^{\circ}$, and their K-band flux
ratio is 25.6. The separation of the B and C components is
0$\farcs$10, P.A. 233$^{\circ}$, and their K-band flux ratio is 1.34
\citep{brandeker2001}. There is also an even fainter, fourth component
at a distance of 16.78$''$ from VW\,Cha\,A \citep{correia2006}. Both
our ISO and Spitzer spectra contain the flux from the three brightest
components (A, B, and C).

\paragraph{CU\,Cha} (also known as HD\,97048) is a single Herbig star
with spectral type of B9.5/A0
\citep{ghez1997,bailey1998,doucet2007}. Based on ISOCAM-CVF data,
\citet{siebenmorgen2000} found that the emission of CU\,Cha in the PAH
bands is extended on scales of about 5-10$''$. According to
\citet{doucet2007}, this emission comes from transiently heated PAH
molecules in an extended envelope around the star-disk
system. \citet{lagage2006} claimed that CU\,Cha is extended in the
8.6$\,\mu$m PAH band, and the brightness profile is well reproduced by
emission from the surface layer of a vertically optically thick,
inclined, flared disk, extending at least up to 370\,AU (or
2$\farcs$1). \citet{doucet2007} found that the object is also extended
in the 11.3$\,\mu$m PAH band, while \citet{vanboekel2004} could
resolve the disk also at continuum wavelengths. While the differences
between the ISOPHOT-S and Spitzer/IRS spectra might be due to the
different beam sizes, the fact that the three ISOPHOT-S spectra differ
by 20-40\% points to real variability. Long-term optical monitoring
indicates that the amplitude of optical variability of CU\,Cha is $<$
0.1-0.3 mag \citep{manfroid1991, herbst1999},

\paragraph{Glass\,I} is a T\,Tauri-type binary with a separation of
2.67$''$, P.A. of 285$^{\circ}$, and K-band flux ratio of 5.9
\citep{haisch2004}. It is also known as Ced\,111\,IRS\,4 or
Cha\,I\,T33. The primary is brighter in the optical, but it is a
``naked'' T\,Tauri; most of the IR emission comes from the deeply
embedded Class I secondary \citep{feigelson1989,
  prusti1991}. \citet{gurtler1999} already reported strong mid-IR
variability based on two ISO spectra. The Spitzer slit in 2006 was
positioned halfway between the two stars, but the slit and the P.A. of
the binary is misaligned, leading to some flux loss. We do not attempt
to correct for this effect. The slit in 2008 was positioned exactly on
the secondary component, which dominates the IR emission. Our
variability study for this source is based on the ISOPHOT-S spectra,
which contain the total flux of both components.

\paragraph{Ced\,111\,IRS\,5} is a single T\,Tauri-type star
\citep{haisch2004}. It is extended on an arcminute scale, even at the
8$\,\mu$m IRAC image. It is also called Chamaeleon IR nebula. This
explains the difference between the ISO and Spitzer spectrum. However,
variability can be studied by comparing the long-wavelength parts of
the two ISOPHOT-S spectra, which have nearly identical pointings and
are not strongly affected by memory effects or uncertainties related
to the subtraction of the predicted background.

\paragraph{HD\,97300} is a B9-type binary Herbig star
\citep{siebenmorgen1998}. In the HR diagram, it is located close to
the ZAMS, and it has no significant IR excess below 24$\,\mu$m
\citep{kospal2012}. The separation of the binary components is
0$\farcs$8, and their K-band flux ratio is 17 \citep{ghez1997,
  lafreniere2008}. \citet{siebenmorgen1998} noticed that at certain IR
wavelengths, two emission peaks can be seen, one coinciding with HD
97300, one being at a distance of about 3$''$. They also detected an
extended, ring-like structure, whose spectrum is dominated by PAH
emission. The size of the ring is about 50$''{\times}$33$''$. It is
not centered on the star, and is probably made of interstellar matter,
whose density is enhanced by the interaction of the stellar wind from
HD\,97300 with the environment \citep[see also][]{kospal2012}. The PAH
molecules in the ring are transiently heated by HD\,97300. This
extended emission explains the factor of 3-4 difference we detect
between the ISOPHOT-S and IRS spectra (the ISOCAM-CVF spectrum
published by \citealt{siebenmorgen2000}, representing the total flux
of the system, is another factor of 2 higher than the ISOPHOT-S
spectrum). The good agreement between the two ISOPHOT-S spectra
indicates that the emission of the central source, and consequently
that of the PAH molecules heated by it, is constant in time. Indeed,
long-term optical and near-IR photometric monitoring indicates that
the brightness of HD 97300 is constant within 0.1-0.2 mag
\citep{manfroid1991, davies1990}.

\paragraph{Ced\,112\,IRS\,4} (also known as FM\,Cha) is a single
T\,Tauri-type star with no indication of extended near-IR or mid-IR
emission in its immediate vicinity \citep{haisch2004,
  haisch2006}. \citet{alexander2003} presented an ISOCAM-CVF spectrum
of the source and noticed that no ice absorption features are present,
while the 10$\,\mu$m region indicate the combination of silicate
emission and absorption. Since our ISOPHOT-S and Spitzer/IRS spectra
are consistent with this finding, we decided not to assign any type to
this source. All the ISOCAM-CVF, ISOPHOT-S, and Spitzer spectra
indicate significant flux changes and changes in the spectral slope.

\paragraph{WX\,Cha} is a T\,Tauri-type binary with a separation of
0$\farcs$79 and P.A. of 55$^{\circ}$ \citep{ghez1997}. Both the ISO
and Spitzer spectra contain all the flux from both components, thus,
they can be safely compared.

\paragraph{CV\,Cha} is a single T\,Tauri-type star \citep{guenther2007,
  melo2003}, although it shows variable radial velocity
\citep{reipurth2002}. The Spitzer/IRS slit contained only CV\,Cha,
while the ISOPHOT-S beam included also CW\,Cha (at a separation of
11$''$). Resolved IRAC photometry from \citet{luhman2008} indicates that
the flux ratio of CW\,Cha to CV\,Cha is 0.12-0.16 between 3.6 and
8$\,\mu$m. This, and the fact that CW\,Cha falls at the very edge of
the ISOPHOT-S beam ensures that CW\,Cha has negligible contribution to
the ISOPHOT-S spectrum. Thus, we conclude that both the ISOPHOT-S and
the Spitzer/IRS spectra represent the flux from CV\,Cha only, thus they
can be compared.

\paragraph{HD\,98800} (also known as TV\,Crt) is a quadruple T\,Tauri
system. It consists of a binary with a projected separation of
0$\farcs$8 (or 38\,AU), P.A. 0$^{\circ}$, and each component is a
spectroscopic binary with a separation of about 1\,AU
\citep{prato2001, boden2005}. Both the ISO aperture and the IRS slit
contain all four objects, so they can be compared. The IR excess is
attributed to HD\,98800\,B \citep{prato2001}. HD\,98800\,B displays no
excess emission below about 5.5$\,\mu$m, indicating an inner
cleared-out region \citep{furlan2007}.

\paragraph{HD\,100453} is an isolated binary Herbig star. The primary
has a spectral type of A9\,Ve. The companion, HD\,100453\,B, is
situated at a separation of 1$\farcs$06 (or 120\,AU), and is 110 times
fainter in the K$_{\rm s}$ band than the primary
\citep{habart2006}. The companion is a M4.0--M4.5V star with no
significant excess in the K or L band \citep{collins2009}, thus it has
no contribution to the plotted ISOPHOT-S or Spitzer/IRS
spectra. \citet{meeus2002} explain the lack of the 10$\,\mu$m silicate
feature with the depletion of small, hot silicate grains in the disk
of HD\,100453 (a factor of 100-500 depletion compared to AB\,Aur is
needed in their model to be able to consistent with the observed
ISO-SWS spectrum). Using higher signal-to-noise ratio TIMMI2 data,
\citet{vanboekel2005} confirmed the lack of the 10$\,\mu$m silicate
emission, and attributed it to grain growth. The comparison of our
ISOPHOT-S and IRS spectra reveal some differences in the 8--12$\,\mu$m
wavelength range, while the flux at $\lambda < 8\,\mu$m is
constant. It is possible that this variability is due to a weak,
variable silicate emission feature (Fig.~\ref{hd100453}). Confirmation
of this idea would require a detailed modeling of the IRS spectrum,
which has the highest signal-to-noise ratio among the 10$\,\mu$m
spectra published so far for HD\,100453. The star shows no strong
accretion activity and is photometrically stable \citep{collins2009,
  garcialopez2006, guimaraes2006}.

\paragraph{HD\,104237} is a spectroscopic binary consisting of a
Herbig A4V primary and a K3-type companion \citep{bohm2004}. Based on
X-ray observations, \citet{testa2008} identified four more low-mass
stars in the vicinity, HD\,104237\,B-E, with separations between
1$\farcs$365 and 14$\farcs$88, while \citet{grady2004} found five
low-mass sources in their optical, near-IR and mid-IR images. In
addition to the Herbig Ae star itself, HD\,104237-6 (corresponding to
the X-ray source HD\,104237\,E) and HD\,104237-2 have IR excesses, but
the Herbig star HD\,104237 dominates the mid-IR integrated light of
the region \citep{grady2004}. While HD\,104237-6 falls outside of both
the ISOPHOT-S beam and the Spitzer/IRS slit, HD\,104237-2 is included
in both. The Spitzer slit is centered between the Herbig star and
HD\,104237-2. Offset correction was applied using the position of the
Herbig star. Although the observed Spitzer/IRS spectrum may contain
75-100\% of the flux of HD\,104237-2, its contribution to the much
brighter primary is negligible, thus we conclude that the ISOPHOT-S
and Spitzer/IRS spectra can be compared. Nevertheless, the ISOPHOT-S
spectra alone already shows significant variability (see also
Fig.~\ref{hd104237}).

\paragraph{DK\,Cha} is a deeply embedded active Herbig Ae star. It
displays brightness variations of more than 1\,mag in the K band
\citep{hughes1991}. The source is redder when fainter, and the
amplitude of the brightness fluctuations decreases with increasing
wavelength. They explain the observed flux changes with a combination
of variable circumstellar extinction and intrinsic
variability. \citet{prusti1994} found no evidence for extended mid-IR
emission (with multi-diaphragm observations between 3.8 and
18.6$\,\mu$m, meaning that the emitting area is less than 5.4$''$). We
note that there is a significant discrepancy between the ISOPHOT-S
(this work) and the ISO-SWS \citep{acke2004} spectra of the source,
pointing to possible variability. The Spitzer data points are close to
saturation and are not very reliable.

\paragraph{HD\,135344\,B} is an F4\,Ve-type Herbig star
\citep{vanboekel2005}. \citet{doucet2006} resolved the source at
20.5$\,\mu$m, and modeled the emission with a disk with an outer
radius of 200 AU (or 1$\farcs$4, which is within both the ISOPHOT-S
and the Spitzer/IRS aperture). \citet{grady2009} observed significant
changes between 1--10$\,\mu$m, both in flux and in the shape of the
SED. At wavelengths shortward of 4.75$\,\mu$m, at least 13\%
variations are present, while at 10$\,\mu$m, they found 60\%
variability.

\paragraph{HD\,139614} is an Herbig Ae star
\citep{dunkin1997}. \citet{yudin1999} detected polarimetric
variability on a time-scale of days, which they explain in terms of a
model of a pole-on disk with dust clouds or comet-like bodies moving
in a circumstellar envelope around the star. We found no evidence in
the literature that the source might be extended at mid-IR
wavelengths, thus we conclude that the ISOPHOT-S and the Spitzer
spectra can be safely compared. The two spectra agree within the
measurement uncertainties, moreover, the TIMMI2 spectrum of
\citet{vanboekel2005} is also identical with them. Thus, HD\,139614 is
constant at mid-IR wavelengths (nor does it show any significant
variability in the optical, see \citealt{meeus1998}).

\paragraph{HD\,141569} is a B9.5/A0 Herbig star \citep{jaschek1992,
  dunkin1997}. Coronagraphic observations of the scattered light by
\citet{augereau1999}, \citet{weinberger1999}, and \citet{mouillet2001}
revealed that its disk is asymmetric, and consists of two rings with a
gap between them. \citet{weinberger2000} claims that the disk is
composed of secondary debris material, although it still contains some
gas \citep{zuckerman1995}. HD\,141569 has two M-type companions:
HD\,141569\,B (M2V) and HD\,141569\,C (M4V) at projected distances of
7$\farcs$6 (A-B) and 1$\farcs$4 (B-C); K-band flux ratios are: 5.3
(A/B) and 11.9 (A/C) \citep{pirzkal1997, weinberger2000}. The
Spitzer/IRS slit included only the primary (A), while the ISOPHOT-S
aperture included all three components. Due to their faintness and
distance from component A, components B and C have negligible
contribution to the ISOPHOT-S flux, thus the ISOPHOT-S and Spitzer/IRS
fluxes can be compared.

\paragraph{HD\,142666} is a Herbig Ae star
\citep{vanboekel2005}. Mid-IR spectra of HD\,142666 obtained with
different instruments are published in \citet{sylvester1997,
  bouwman2001}, and \citet{vanboekel2005}. HD\,142666 displays large
photometric variations (\citealt{malfait1998} give a visual amplitude
of about 1.2\,mag, while \citealt{lecavelier2005} detected 0.3\,mag
variability with Stromgren photometry). The star appears redder when
fainter, which they interpret as variable dust extinction in a close
to edge-on disk with clumpy dust distribution. Since beam effects do
not play a role here, we can safely compare the ISOPHOT-S and Spitzer
spectra.

\paragraph{HD\,142527} is an isolated Fe-type Herbig star
\citep[and references therein]{vanboekel2005}. Based on
spectro-astrometric observations, \citet{baines2006} consider it a
possible subarcsecond binary. The circumstellar environment of the
star was resolved at various wavelengths (see e.g.~near-IR images in
\citealt{fukagawa2006}, mid-IR images in \citealt{fujiwara2006}, or
millimeter images in \citealt{verhoeff2011}). Based on these
observations, the star is surrounded by an inner disk, an outer disk,
and a halo around the inner disk regions. Spatially resolved
18$\,\mu$m images in \citet{fujiwara2006} and in \citet{verhoeff2011}
indicate that the emission is coming from within about 1$''$ of the
star, which is fully included in both the ISOPHOT-S beam and the
Spitzer/IRS slit. Thus, the spectra can be compared, and they indicate
significant variability. An additional 8--13$\,\mu$m spectrum from
\citet{vanboekel2005} also confirms this (Fig.~\ref{hd142527}).

\paragraph{EX\,Lup} is the prototype of EXors, a class of young
stars showing sporadic optical outburst separated by extended
quiescent periods \citep{herbig1977}. The outbursts are powered by
enhanced accretion from the circumstellar disk onto the star
\citep{herbig2001}. EX\,Lup itself brightened by
${\Delta}V\,{\approx}\,$5\,mag in 1955-56 and again in 2008, and
produced several smaller, 1--3\,mag brightenings in-between. EX\,Lup
was extensively studied during its outburst in 2008, and the
observations indicate that it brightened in the whole 0.4--100$\,\mu$m
wavelength range \citep{juhasz2012}. By comparing pre-outburst and
outburst Spitzer spectra, \citet{abraham2009} observed a significant
change in the shape of the 10$\,\mu$m silicate feature: while before
the outburst, the feature indicated mostly amorphous silicates, during
outburst, several narrow peaks and shoulders appeared, indicating the
presence of crystalline silicates. \citet{abraham2009} explain the
appearance of the crystalline silicates by annealing in the surface
layer of the inner disk by heat from the outburst. In
Fig.~\ref{map_var}, we plot ISOPHOT-S spectra from 1997 and Spitzer
spectra from 2004 and 2005. These are all quiescent spectra, when the
visual magnitude of EX\,Lup was between 12.5\,mag and 14\,mag.

\paragraph{HD\,144432} is a pre-main sequence binary consisting of
HD\,144432\,A, an A9/F0-type Herbig star primary and HD\,144432\,B, a
K0-5-type T\,Tauri secondary, with a separation of 1.4$''$,
P.A. 0$^{\circ}$ \citep{perez2004, stelzer2009, fukagawa2010}. The
secondary is a factor of 9-14 fainter than the primary in the K and H
bands \citep{perez2004, fukagawa2010}. The secondary displays
photospheric emission until K band, and the IR excess is attributed
exclusively to the primary \citep{perez2004}. Both the ISOPHOT-S
aperture and the Spitzer/IRS slit include both components and the
observed spectra represent the total flux of the two objects (the
Spitzer slit is oriented along the binary, north-south). Keck
segment-tilting observations by \citet{monnier2009} resolved the
source at 10.7$\,\mu$m, which is $\approx$40\,mas. There is a
$\approx$10\% difference between the ISOPHOT-S and the Spitzer/IRS
spectra plotted in Fig.~\ref{map_var}. Several other mid-IR spectra
are available on HD\,144432 in the literature, see
e.g.~\citet{sylvester1996, leinert2004, vanboekel2005}. The first of
these is consistent with the Spitzer/IRS flux level, the second one is
consistent with the ISOPHOT-S flux level, and the third one is lower
than the others by another $\approx$10\%, supporting our finding that
the source is variable in the mid-IR regime.

\paragraph{HR\,5999} is a binary star consisting of a Herbig Ae
primary, and a T\,Tauri secondary called Rossiter\,3930
\citep{rossiter1955}. The separation is 1$\farcs$4, the P.A. is
111$^{\circ}$, and the K-band flux ratio is between 20-35
\citep{stecklum1995, ghez1997, leinert1997}. The primary itself may be
a spectroscopic binary \citep{tjinadjie1989}, although observations by
\citet{corporon1999} do not confirm this. HR\,5999 is a known optical
variable, whose ``bursts'' are attributed to unsteady accretion
\citep{Perez1992}. Multi-epoch BVRIJHKLMNQ band photometry by
\citet{hutchinson1994} indicated that the object is indeed variable at
optical wavelengths but not much above 4.8$\,\mu$m. An ISOCAM CVF
spectrum published in \citet{acke2004} is consistent with our
ISOPHOT-S spectrum until 8.5$\,\mu$m, but is lower by about 20\% at
longer wavelengths, indicating a possibly variable silicate
feature. \citet{siebenmorgen2000} noted that HR\,5999 is a point
source at these wavelengths. Using VLTI/MIDI, \citet{preibisch2006}
resolved the mid-IR emission from HR\,5999 and they give sizes between
5 and 25\,mas (1-5\,AU), depending on the fitted model.

\paragraph{WL\,16} is a single B8-A7 Herbig star \citep{simon1995,
  ratzka2005, luhman1999}. Images at different mid-IR wavelengths
between 7.9 and 24.5$\,\mu$m (both in the continuum and in the PAH
bands) published by \citet{ressler2003} revealed that the central star
is surrounded by a bright disk with a size of 7$''\times$3$\farcs$5
(corresponding to a disk diameter of ~900 AU), viewed at an
inclination of 62.2\degr. \citet{geers2007} also resolve both the
8--13$\,\mu$m continuum and the 8.6, 11.2, and 12.7$\,\mu$m PAH
features, but give smaller radial extents (50-60\,AU, or
0$\farcs$4-0$\farcs$5) than \citet{ressler2003}. In their study, the
3.3$\,\mu$m feature is undetected, indicating the dominance of ionized
PAHs. Our ISOPHOT-S spectrum of WL\,16 in Fig.~1 agrees well with the
ISO-SWS spectrum, and the slight differences between these and the
10$\,\mu$m spectrum published in \citet{hanner1992} are probably due
to the different beam sizes of the instruments.

\paragraph{MWC\,863} is a pre-main sequence binary, consisting of a
Herbig Ae primary, HD\,150193\,A, and an F9-type T\,Tauri secondary,
HD\,150193\,B \citep{carmona2007}. Their separation is 1$\farcs$1, the
P.A. is 227$^{\circ}$ \citep{reipurth1993}, flux ratios are 6-8 and
4-17 in the K and L$^{\prime}$ bands, respectively
\citep{zinnecker1991, koresko2002}. The secondary component has no IR
excess \citep[][and references therein]{fukagawa2003}. Using
coronagraphic images, \citet{fukagawa2003} resolved the disk around
HD\,150193\,A in the H band, and gave a radius of 1$\farcs$3
(190\,AU). During Keck segment-tilting observations by
\citet{monnier2009} the source remained unresolved at 10.7$\,\mu$m,
which gives an upper limit of FWHM=35\,mas for the emitting region. We
note that while the ISO-SWS spectrum published in \citet{acke2004} is
identical with out ISOPHOT-S spectrum, the UKIRT/CGS3 spectrum taken
by \citet{sylvester2000} is significantly different both in shape an
in flux level, pointing to possible changes in the composition and
size distribution of the silicate dust grains.

\paragraph{AK\,Sco} is a double-line spectroscopic binary consisting
of approximately equal mass F5-type stars with a projected separation
of 0.14\,AU ($\approx$1\,mas); the two stars are surrounded by a
circumbinary disk \citep{alencar2003}. Both the ISO and Spitzer
spectra include both components. Its optical light curve shows a
``roughly constant maximum light level being interrupted at irregular
intervals by minima about a magnitude deep'' \citep{andersen1989}. Our
ISOPHOT-S and Spitzer/IRS spectra agree within the measurement
uncertainties, and a TIMMI2 spectrum by \citet{przygodda2003} also has
the same flux level.

\paragraph{MWC\,865} (also known as V921\,Sco and
CD$-$42$\degr$11721) is a B0IV-type star embedded in a small, thick
dark cloud (\citealt{boersma2009} and references therein). It
illuminates a reflection nebula \citep{vandenbergh1975}. Its distance
is quite uncertain (values between 400 and 2500 pc appear in the
literature, e.g.~\citealt{boersma2009, lopes1992}), and there is also
some debate about its pre-main sequence nature
\citep{borgesfernandes2007}, although it is generally considered to be
a Herbig Be star. \citet{habart2003} found four IR companions within
13$''$ of MWC\,865, while \citet{wang2007} found about 10 YSO
candidates in the vicinity (within about 70$''$) of MWC\,865 using
Spitzer/IRAC images. \citet{boersma2009} shows Spitzer/IRAC images
between 3.6 and 8.0$\,\mu$m and TIMMI2 images at 11.9$\,\mu$m of
MWC\,865 and its surroundings. These images reveal two bright peaks,
one coinciding with the star, the other one being a close-by patch
4$\farcs$3 to the north. There is an arc stretching towards the
south-east as well, at a distance of 11$''$ from the star. TIMMI2
long-slit spectra indicate that the central star has a featureless
spectrum, while the close-by patch and the arc display strong PAH
emission. \citet{boersma2009} combine the spectra of these components
and compare it with an ISO-SWS spectrum. They find that the ISO-SWS
spectrum shows a similar shape, but has higher flux levels due to the
larger aperture of ISO-SWS and the extent of the PAH emission. Our
ISOPHOT-S spectrum of MWC\,865 also has a very similar shape to the
ISO-SWS spectrum, but even higher flux levels. We conclude that these
differences are also due to the difference in the apertures and their
exact position with respect to the PAH-emitting patches. We note that
significant offset correction was applied to the ISOPHOT-S spectrum of
MWC\,865, introducing extra uncertainty in the absolute flux levels.

\paragraph{51\,Oph} is a single Herbig star
\citep{baines2006}, which is photometrically constant
\citep{lecavelier2005}. The source appeared point-like in the
18$\,\mu$m images of \citet{jayawardhana2001}, giving an upper limit
of 0$\farcs$5 (65\,AU) for the radius of the emitting
region. \citet{leinert2004} resolved the source at 12.5$\,\mu$m with
VLTI/MIDI and gave a half-light radius of 7\,mas (0.5\,AU). Our
conclusion is that the ISOPHOT-S and the Spitzer/IRS spectra can be
compared and the observed variability is real. We note that
ground-based mid-IR spectra published in \citet{sylvester1996} and
\citet{leinert2004} are consistent with the Spitzer/IRS flux level.

\paragraph{HD\,163296} is a well-known single
\citep{corporon1999,pirzkal1997} Herbig Ae star. Using optical and
near-IR coronagraphic images, \citet{fukagawa2010} and
\citet{grady2000} could detect the disk in scattered light out to a
radius of 3$\farcs$6 (440\,AU). \citet{wisniewski2008} found that the
optical scattered light is variable in time due to variable
self-shadowing. During Keck segment-tilting observations by
\citet{monnier2009} the source remained unresolved at 10.7$\,\mu$m,
which gives an upper limit of FWHM=35\,mas for the emitting
region. \citet{doucet2006} resolved the disk thermal emission at
20.5$\,\mu$m and derived an outer disk radius of 1$\farcs$6
(200\,AU). The source was resolved by \citet{leinert2004} using
VLTI/MIDI data, and the half-light radius at 12.5$\,\mu$m is 7\,mas
(0.8\,AU). The source appeared point-like in the 18$\,\mu$m images of
\citet{jayawardhana2001}, giving an upper limit of 0$\farcs$5 (60\,AU)
for the radius of the emitting region. \citet{sitko2008} conducted a
long-term 3$-$13$\,\mu$m spectroscopic monitoring of HD\,163296. Most
of their data are consistent with no variability larger than 10\%,
except at one epoch when the 1$-$5$\,\mu$m flux significantly
increased, accompanied by a slight increase in the flux of the
10$\,\mu$m silicate feature. Although no Spitzer/IRS low resolution
spectra are available for this source, our ISOPHOT-S spectrum is
consistent with an ISO-SWS spectrum (taken on the same day), and also
consistent with the TIMMI2 spectrum published in
\citet{vanboekel2005}, obtained in March 2003, and the MIDI spectrum
published in \citet{leinert2004}, obtained in June 2003, while it is
10-40\% higher than that obtained by \citet{kessler2005} with the
Keck/LWS in August 1999 and in June 2000.

\paragraph{HD\,169142} is a Herbig Ae star \citep{sylvester1996}. High
spatial resolution mid-IR observations at 3.3, 10.8, and 18.2$\,\mu$m
show that the emission comes from a disk with an outer radius of 150
AU (1$''$) at most \citep{habart2006, jayawardhana2001}. SED modeling
gave disk radii in the range of 100-300 AU (0$\farcs$7-2$''$)
\citep{dominik2003, dent2006}. HD\,169142 has a possible companion,
2MASS J18242929$-$2946559, located at a distance of 9$\farcs$3
\citep{grady2007}. The 2MASS source is a binary weak-line T\,Tauri
with 130\,mas separation, and is 11 times fainter than HD\,169142 in
the K band. The Spitzer/IRS slit contains only HD\,169142, while the
ISOPHOT-S beam includes both components. However, the contribution of
the 2MASS source at mid-IR wavelengths is probably negligible, thus
the Spitzer and the ISO spectra can be compared. The spectra plotted
in Fig.~1 agree within the measurement uncertainties, thus we conclude
that this source is constant in the 5-11.5$\,\mu$m regime. The flux
levels of the UKIRT/CGS3 spectrum from \citet{sylvester1996} and the
TIMMI2 spectrum from \citet{vanboekel2005} also agree with our spectra
until $\approx$9$\,\mu$m, while they differ by up to 50\% above
9$\,\mu$m (Fig.~\ref{hd169142}). The source is photometrically stable
at optical wavelengths \citep{guimaraes2006}.

\paragraph{VV\,Ser} is a single \citep{pirzkal1997, leinert1997} Herbig
star. It displays UXor-type optical variability \citep{herbst1999,
  rostopchina2001}, i.e. 1-3\,mag deep minima in the V band due to
variable obscuration by dust clumps in a nearly edge-on circumstellar
disk. \citet{habart2003} found no other source emitting at 10$\,\mu$m
in the 1$'\,{\times}\,$1$'$ vicinity of VV\,Ser. \citet{li1994} found
no extended emission around the source in JHK filters, while
\citet{leinert2001b} found faint extended emission with a FWHM of
0.6$''$ in JHK with speckle interferometry. \citet{eisner2004}
resolved the source at 2.2$\,\mu$m with the PTI, and determined a size
of 1.5-4.5\,mas (depending on the model used) for the emitting
region. \citet{pontoppidan2007} discovered nebulous mid-IR emission
extending over 4$'$ centered on VV\,Ser. The nebulosity is due to
transiently heated dust grains. Although we cannot exclude the
possibility that this extended emission has different contributions to
the ISOPHOT-S and the Spitzer/IRS spectra, the fact that the
Spitzer/IRS flux level is higher points to real variability.

\paragraph{OO\,Ser} is a deeply embedded YSO that produced an outburst
in 1995 and became 4.6\,mag brighter in the K band \citep{hodapp1996},
reaching its maximum brightness in 1995 October
\citep{hodapp1999}. The eruption caused brightening at a wide
wavelength range from 2.2$\,\mu$m to 100$\,\mu$m
\citep{kospal2007}. After peak brightness the object started a gradual
fading. The ISOPHOT-S spectra plotted in Fig.~\ref{map_var} cover a
period of 20 months, and the first one was obtained 4 months after the
maximum brightness occurred. The gradual fading, which is discussed in
detail in \citet{kospal2007}, is well visible in the figure. The
source is surrounded by a small elongated nebula, approximately 15$''$
in diameter, best seen in the K band, but also discernible at 3.6, 4.5
and 5.8$\,\mu$m \citep{hodapp1996, kospal2007}.

\paragraph{CK\,1} is an embedded protostellar binary with a separation
of 1$\farcs$5, and P.A.~of 10$^{\circ}$ \citep{eiroa1987}. The
southern component is brighter than the northern one, and their flux
ratio decreases from 4.2 at 0.9$\,\mu$m to 3.3 at 3.5$\,\mu$m to
1.1-2.8 between 8--12.9$\,\mu$m \citep{eiroa1987a, eiroa1987,
  ciardi2005}. Both the ISO and the Spitzer spectra represent the sum
of the spectra of CK\,1 north and CK\,1 south (the Spitzer slit is
oriented along the separation of the binary), thus, they can be
compared. \citet{ciardi2005} present resolved 8--13$\,\mu$m spectra of
the two components, and the sum of the two spectra as well. This sum
agrees very well with our ISOPHOT-S and Spitzer/IRS data.

\paragraph{[SVS76]\,Ser\,4} is a small, dense cluster of deeply
embedded low- and intermediate mass YSOs
\citep{pontoppidan2004}. Judging from the JHK photometry of
\citet{eiroa1989} and the Spitzer 3.6--24.0$\,\mu$m fluxes from
\citet{harvey2007}, most of the flux in the Spitzer slit comes from
SVS4/9 and SVS4/10, while in addition to these two sources, SVS4/4,
SVS4/8, SVS4/11 and SVS4/5 also contribute to the ISO flux. SVS4/5 is
especially bright above 5$\,\mu$m (very red source). Source confusion
explains the differences between the ISO and Spitzer spectra. We note
that in the case of the ISOPHOT-S spectrum, predicted background was
subtracted, making the absolute level of the measured flux probably
more uncertain than 10\%.

\paragraph{CK\,2} is an embedded YSO in the Serpens star-forming
region. We found no indication of extended mid-IR emission in the
literature. The source is close to the sensitivity limit of ISOPHOT-S,
resulting in a very low signal-to-noise, especially above
6$\,\mu$m. The ISOPHOT-S spectrum also suffers from memory effect, and
a predicted background is subtracted, which, considering how faint the
source is, introduces additional uncertainties. Thus, we do not
analyze the variability of CK\,2. Nevertheless, the ISOCAM CVF
spectrum of \citet{alexander2003} is significantly lower than our
Spitzer/IRS spectrum, indicating possible variability.

\paragraph{S\,CrA} is a PMS binary with a separation of
1\farcs3--1\farcs5, P.A of 147$^{\circ}$--160$^{\circ}$, K-band flux
ratio of 1.9--2.7, and L-band flux ratio of 2.5 \citep{baier1985,
  zinnecker1991, chelli1995, prato2003}. \citet{carmona2007} obtained
spatially resolved optical spectra for the two components and
determined a spectral type of G5\,Ve for the primary (S\,CrA\,NW) and
K5Ve for the secondary (S\,CrA\,SE). S\,CrA is a well-known optical
variable, with V-band variations of up to $\approx$2\,mag
\citep[e.g.][]{kardopolov1981}. \citep{graham1992} interpreted the
photometric and spectroscopic changes of the object as variable
accretion onto the star. Both the ISOPHOT-S aperture and the
Spitzer/IRS slit included the whole flux of both components, thus,
they can be compared.

\paragraph{R\,CrA, T\,CrA, and HH\,100\,IRS} are part of {\it
  Coronet}, a small cluster of YSOs. The two brightest sources, R\,CrA
and T\,CrA, are Herbig stars, while the remaining dozen (HH\,100\,IRS
among others) are deeply embedded low-mass protostars \citep[][and
  references therein]{forbrich2007}. R\,CrA illuminates the reflection
nebula NGC\,6729. Both R\,CrA and the nebula are highly variable at
optical wavelengths \citep{graham1987}. HH\,100\,IRS is also highly
variable, with amplitudes up to 2.5\,mag in the K band
\citep{axon1982, reipurth1983}. Both R\,CrA and T\,CrA are possible
spectro-astrometric binaries \citep{takami2003}. Using speckle
interferometry, \citet{dewarf1993} found that R\,CrA is unresolved at
2.2 and 3.8$\,\mu$m. \citet{prusti1994} found no evidence for extended
mid-IR emission larger than $\approx$5$''$ around R\,CrA and T\,CrA.
Interestingly R\,CrA displays the 4.27$\,\mu$m feature of solid CO$_2$
ice in absorption and the 10$\,\mu$m silicate feature in emission at
the same time \citep{nummelin2001, acke2004}. No Spitzer spectra are
available for R\,CrA and T\,CrA, but the comparison of our ISOPHOT-S
spectra in Fig.~\ref{map_var} and the ISO-SWS spectra published in
\citet{acke2004} indicate very similar flux levels. For HH\,100\,IRS,
variability can be studied by comparing the two ISOPHOT-S spectra.

\paragraph{VV\,CrA} is a T\,Tauri-type binary with a separation of
2$\farcs$10, P.A. of 44$^{\circ}$, and K-band flux ratio of 8.6. The
optical secondary becomes brighter around the J band, and dominates
the flux at longer wavelengths \citep{chelli1995}. \citet{ratzka2008}
report that the secondary faded significantly and ``it is now fainter
than the primary even in the N-band''. They also show 10$\,\mu$m MIDI
spectra for each component: VV\,CrA\,NE shows deep silicate
absorption, while VV\,CrA\,SW is featureless. Our ISO spectrum
contains flux from both components, while the Spitzer/IRS spectrum in
centered on the NE component, but contains significant contribution
from the SW as well. For this reason, we do not analyze variability
for this source.

\paragraph{WW\,Vul} is an isolated Herbig Ae star, which shows
UXor-like fadings at optical wavelengths, although its disk probably
has a larger inclination than for other UXors \citep{pugach2004,
  kozlova2006}. WW\,Vul was observed twice with ISOPHOT-S and once
with Spitzer. Both ISOPHOT-S spectra are offset-corrected, but the
large correction applied for the earlier one (TDT: 17600465) makes
this more uncertain than the 10\% assigned as a final uncertainty of
the ISOPHOT-S spectra. For this reason, this spectrum in not used in
our variability analysis, and we base our conclusions on the
comparison of the other ISOPHOT-S spectrum (TDT: 51300108) and the
Spitzer measurement.

\paragraph{BD\,+40\,4124 and LkHa\,224} are members of a
cluster of young stars \citep[][and references
  therein]{henning1998}. The brightest members of the cluster are the
B2-type BD\,+40\,4124 and the B5-type LkHa\,224, but several fainter
FGKM stars can be found within a few arcminutes; some of these are
included in the ISOPHOT-S aperture. No Spitzer spectrum is obtained
for these sources, however an ISO-SWS is available. Comparison of our
ISOPHOT-S spectrum with the ISO-SWS reveals no significant differences.
BD\,+40\,4124 displays small-scale ($<$0.4 mag) quasi-periodic
variability at optical wavelengths \citep{herbst1999, strom1972}.

\paragraph{PV\,Cep} is an embedded YSO surrounded by the cometary
reflection nebula RNO\,125 \citep{the1994}. Both the star and the
nebula are strongly variable
\citep[e.g.][]{scarrott1991,lorenzetti2011}. \citet{kun2011} analyzed
the brightness changes of PV\,Cep in the whole optical--infrared
regime, and found significant flux variations below 10$\,\mu$m. They
conclude that the reason for the variability is a combination of
changing accretion rate and changing circumstellar extinction. We
found no indication in the literature that the object is extended at
mid-IR wavelengths. The fact that the Spitzer spectrum has a higher
flux than the ISOPHOT-S points to real variability.

\paragraph{SV\,Cep} is spectro-astrometric binary Herbig star
\citep{wheelwright2010}. Its optical light curve is characteristic of
UXor-type variables with irregular, quasi-periodic variations due to
inhomogeneities in a nearly edge-on circumstellar disk
\citep[e.g.][]{rostopchina2000}. \citet{juhasz2007} investigated the
long-term 0.55--100$\,\mu$m variability of SV\,Cep and found a
correlation between the optical and far-IR light curves. In the
mid-IR regime, they found lower variability amplitudes (a peak-to-peak
variability of $\approx$40\% at 10$\,\mu$m) and a hint for
anti-correlation with the optical changes. They interpret the
variability using a self-shadowed disk and a puffed-up inner rim with
variable inner rim height. Their model also contained an optically thin
envelope to account for the constant 25$\,\mu$m flux. Our ISOPHOT-S
spectrum plotted in Fig.~\ref{map_var} and the ISO-SWS spectrum
obtained 9 months previously \citep{acke2004} are consistent with each
other.

\bibliographystyle{apj}
\bibliography{paper}{}

\begin{thebibliography}{247}
\expandafter\ifx\csname natexlab\endcsname\relax\def\natexlab#1{#1}\fi

\bibitem[{{{\'A}brah{\'a}m} {et~al.}(2004){{\'A}brah{\'a}m}, {K{\'o}sp{\'a}l},
  {Csizmadia}, {Kun}, {Mo{\'o}r}, \& {Prusti}}]{abraham2004}
{{\'A}brah{\'a}m}, P., {K{\'o}sp{\'a}l}, {\'A}., {Csizmadia}, S., {Kun}, M.,
  {Mo{\'o}r}, A., \& {Prusti}, T. 2004, \aap, 428, 89

\bibitem[{{{\'A}brah{\'a}m} {et~al.}(2009){{\'A}brah{\'a}m}, {Juh{\'a}sz},
  {Dullemond}, {K{\'o}sp{\'a}l}, {van Boekel}, {Bouwman}, {Henning},
  {Mo{\'o}r}, {Mosoni}, {Sicilia-Aguilar}, \& {Sipos}}]{abraham2009}
{{\'A}brah{\'a}m}, P., {et~al.} 2009, \nat, 459, 224

\bibitem[{{Acke} \& {van den Ancker}(2004)}]{acke2004}
{Acke}, B., \& {van den Ancker}, M.~E. 2004, \aap, 426, 151

\bibitem[{{Acosta-Pulido} \& {{\'A}brah{\'a}m}(2003)}]{acosta2003}
{Acosta-Pulido}, J.~A., \& {{\'A}brah{\'a}m}, P. 2003, in ESA Special
  Publication, Vol. 481, The Calibration Legacy of the ISO Mission, ed.
  {L.~Metcalfe, A.~Salama, S.~B.~Peschke, \& M.~F.~Kessler}, 95

\bibitem[{{Alencar} {et~al.}(2001){Alencar}, {Johns-Krull}, \&
  {Basri}}]{alencar2001}
{Alencar}, S.~H.~P., {Johns-Krull}, C.~M., \& {Basri}, G. 2001, \aj, 122, 3335

\bibitem[{{Alencar} {et~al.}(2003){Alencar}, {Melo}, {Dullemond}, {Andersen},
  {Batalha}, {Vaz}, \& {Mathieu}}]{alencar2003}
{Alencar}, S.~H.~P., {Melo}, C.~H.~F., {Dullemond}, C.~P., {Andersen}, J.,
  {Batalha}, C., {Vaz}, L.~P.~R., \& {Mathieu}, R.~D. 2003, \aap, 409, 1037

\bibitem[{{Alexander} {et~al.}(2003){Alexander}, {Casali}, {Andr{\'e}},
  {Persi}, \& {Eiroa}}]{alexander2003}
{Alexander}, R.~D., {Casali}, M.~M., {Andr{\'e}}, P., {Persi}, P., \& {Eiroa},
  C. 2003, \aap, 401, 613

\bibitem[{{Andersen} {et~al.}(1989){Andersen}, {Lindgren}, {Hazen}, \&
  {Mayor}}]{andersen1989}
{Andersen}, J., {Lindgren}, H., {Hazen}, M.~L., \& {Mayor}, M. 1989, \aap, 219,
  142

\bibitem[{{Arellano Ferro} \& {Giridhar}(2003)}]{arellano2003}
{Arellano Ferro}, A., \& {Giridhar}, S. 2003, \aap, 408, L29

\bibitem[{{Augereau} {et~al.}(1999){Augereau}, {Lagrange}, {Mouillet}, \&
  {M{\'e}nard}}]{augereau1999}
{Augereau}, J.~C., {Lagrange}, A.~M., {Mouillet}, D., \& {M{\'e}nard}, F. 1999,
  \aap, 350, L51

\bibitem[{{Axon} {et~al.}(1982){Axon}, {Allen}, {Bailey}, {Hough}, {Ward}, \&
  {Jameson}}]{axon1982}
{Axon}, D.~J., {Allen}, D.~A., {Bailey}, J., {Hough}, J.~H., {Ward}, M.~J., \&
  {Jameson}, R.~F. 1982, \mnras, 200, 239

\bibitem[{{Baier} {et~al.}(1985){Baier}, {Keller}, {Weigelt}, {Bastian}, \&
  {Mundt}}]{baier1985}
{Baier}, G., {Keller}, E., {Weigelt}, G., {Bastian}, U., \& {Mundt}, R. 1985,
  \aap, 153, 278

\bibitem[{{Bailey}(1998)}]{bailey1998}
{Bailey}, J. 1998, \mnras, 301, 161

\bibitem[{{Baines} {et~al.}(2006){Baines}, {Oudmaijer}, {Porter}, \&
  {Pozzo}}]{baines2006}
{Baines}, D., {Oudmaijer}, R.~D., {Porter}, J.~M., \& {Pozzo}, M. 2006, \mnras,
  367, 737

\bibitem[{{Barsony} {et~al.}(2005){Barsony}, {Ressler}, \&
  {Marsh}}]{barsony2005}
{Barsony}, M., {Ressler}, M.~E., \& {Marsh}, K.~A. 2005, \apj, 630, 381

\bibitem[{{Bary} {et~al.}(2009){Bary}, {Leisenring}, \& {Skrutskie}}]{bary2009}
{Bary}, J.~S., {Leisenring}, J.~M., \& {Skrutskie}, M.~F. 2009, \apjl, 706,
  L168

\bibitem[{{Beck} {et~al.}(2004){Beck}, {Schaefer}, {Simon}, {Prato}, {Stoesz},
  \& {Howell}}]{beck2004}
{Beck}, T.~L., {Schaefer}, G.~H., {Simon}, M., {Prato}, L., {Stoesz}, J.~A., \&
  {Howell}, R.~R. 2004, \apj, 614, 235

\bibitem[{{Beichman} {et~al.}(1984){Beichman}, {Jennings}, {Emerson}, {Baud},
  {Harris}, {Rowan-Robinson}, {Aumann}, {Gautier}, {Gillett}, {Habing},
  {Marsden}, {Neugebauer}, \& {Young}}]{beichman1984}
{Beichman}, C.~A., {et~al.} 1984, \apjl, 278, L45

\bibitem[{{Boden} {et~al.}(2005){Boden}, {Sargent}, {Akeson}, {Carpenter},
  {Torres}, {Latham}, {Soderblom}, {Nelan}, {Franz}, \&
  {Wasserman}}]{boden2005}
{Boden}, A.~F., {et~al.} 2005, \apj, 635, 442

\bibitem[{{Boersma} {et~al.}(2009){Boersma}, {Peeters},
  {Mart{\'{\i}}n-Hern{\'a}ndez}, {van der Wolk}, {Verhoeff}, {Tielens},
  {Waters}, \& {Pel}}]{boersma2009}
{Boersma}, C., {Peeters}, E., {Mart{\'{\i}}n-Hern{\'a}ndez}, N.~L., {van der
  Wolk}, G., {Verhoeff}, A.~P., {Tielens}, A.~G.~G.~M., {Waters}, L.~B.~F.~M.,
  \& {Pel}, J.~W. 2009, \aap, 502, 175

\bibitem[{{B{\"o}hm} {et~al.}(2004){B{\"o}hm}, {Catala}, {Balona}, \&
  {Carter}}]{bohm2004}
{B{\"o}hm}, T., {Catala}, C., {Balona}, L., \& {Carter}, B. 2004, \aap, 427,
  907

\bibitem[{{Bontemps} {et~al.}(2001){Bontemps}, {Andr{\'e}}, {Kaas}, {Nordh},
  {Olofsson}, {Huldtgren}, {Abergel}, {Blommaert}, {Boulanger}, {Burgdorf},
  {Cesarsky}, {Cesarsky}, {Copet}, {Davies}, {Falgarone}, {Lagache},
  {Montmerle}, {P{\'e}rault}, {Persi}, {Prusti}, {Puget}, \&
  {Sibille}}]{bontemps2001}
{Bontemps}, S., {et~al.} 2001, \aap, 372, 173

\bibitem[{{Borges Fernandes} {et~al.}(2007){Borges Fernandes}, {Kraus}, {Lorenz
  Martins}, \& {de Ara{\'u}jo}}]{borgesfernandes2007}
{Borges Fernandes}, M., {Kraus}, M., {Lorenz Martins}, S., \& {de Ara{\'u}jo},
  F.~X. 2007, \mnras, 377, 1343

\bibitem[{{Bouwman} {et~al.}(2001){Bouwman}, {Meeus}, {de Koter}, {Hony},
  {Dominik}, \& {Waters}}]{bouwman2001}
{Bouwman}, J., {Meeus}, G., {de Koter}, A., {Hony}, S., {Dominik}, C., \&
  {Waters}, L.~B.~F.~M. 2001, \aap, 375, 950

\bibitem[{{Bowey} \& {Adamson}(2001)}]{bowey2001}
{Bowey}, J.~E., \& {Adamson}, A.~J. 2001, \mnras, 320, 131

\bibitem[{{Brandeker} {et~al.}(2001){Brandeker}, {Liseau}, {Artymowicz}, \&
  {Jayawardhana}}]{brandeker2001}
{Brandeker}, A., {Liseau}, R., {Artymowicz}, P., \& {Jayawardhana}, R. 2001,
  \apjl, 561, L199

\bibitem[{{Bregman} {et~al.}(1993){Bregman}, {Rank}, {Sandford}, \&
  {Temi}}]{bregman1993}
{Bregman}, J., {Rank}, D., {Sandford}, S.~A., \& {Temi}, P. 1993, \apj, 410,
  668

\bibitem[{{Brinch} {et~al.}(2007){Brinch}, {Crapsi}, {Hogerheijde}, \&
  {J{\o}rgensen}}]{brinch2007}
{Brinch}, C., {Crapsi}, A., {Hogerheijde}, M.~R., \& {J{\o}rgensen}, J.~K.
  2007, \aap, 461, 1037

\bibitem[{{Carmona} {et~al.}(2007){Carmona}, {van den Ancker}, \&
  {Henning}}]{carmona2007}
{Carmona}, A., {van den Ancker}, M.~E., \& {Henning}, T. 2007, \aap, 464, 687

\bibitem[{{Casali}(1991)}]{casali1991}
{Casali}, M.~M. 1991, \mnras, 248, 229

\bibitem[{{Chelli} {et~al.}(1995){Chelli}, {Cruz-Gonzalez}, \&
  {Reipurth}}]{chelli1995}
{Chelli}, A., {Cruz-Gonzalez}, I., \& {Reipurth}, B. 1995, \aaps, 114, 135

\bibitem[{{Chiang} \& {Goldreich}(1997)}]{cg}
{Chiang}, E.~I., \& {Goldreich}, P. 1997, \apj, 490, 368

\bibitem[{{Ciardi} {et~al.}(2005){Ciardi}, {Telesco}, {Packham}, {G{\'o}mez
  Martin}, {Radomski}, {De Buizer}, {Phillips}, \& {Harker}}]{ciardi2005}
{Ciardi}, D.~R., {Telesco}, C.~M., {Packham}, C., {G{\'o}mez Martin}, C.,
  {Radomski}, J.~T., {De Buizer}, J.~M., {Phillips}, C.~J., \& {Harker}, D.~E.
  2005, \apj, 629, 897

\bibitem[{{Cohen}(2003)}]{cohen2003}
{Cohen}, M. 2003, in ESA Special Publication, Vol. 481, The Calibration Legacy
  of the ISO Mission, ed. {L.~Metcalfe, A.~Salama, S.~B.~Peschke, \&
  M.~F.~Kessler}, 135

\bibitem[{{Collins} {et~al.}(2009){Collins}, {Grady}, {Hamaguchi},
  {Wisniewski}, {Brittain}, {Sitko}, {Carpenter}, {Williams}, {Mathews},
  {Williger}, {van Boekel}, {Carmona}, {Henning}, {van den Ancker}, {Meeus},
  {Chen}, {Petre}, \& {Woodgate}}]{collins2009}
{Collins}, K.~A., {et~al.} 2009, \apj, 697, 557

\bibitem[{{Connelley} {et~al.}(2008){Connelley}, {Reipurth}, \&
  {Tokunaga}}]{connelley2008}
{Connelley}, M.~S., {Reipurth}, B., \& {Tokunaga}, A.~T. 2008, \aj, 135, 2496

\bibitem[{{Corporon} \& {Lagrange}(1999)}]{corporon1999}
{Corporon}, P., \& {Lagrange}, A. 1999, \aaps, 136, 429

\bibitem[{{Correia} {et~al.}(2006){Correia}, {Zinnecker}, {Ratzka}, \&
  {Sterzik}}]{correia2006}
{Correia}, S., {Zinnecker}, H., {Ratzka}, T., \& {Sterzik}, M.~F. 2006, \aap,
  459, 909

\bibitem[{{Cutri} {et~al.}(2003){Cutri}, {Skrutskie}, {van Dyk}, {Beichman},
  {Carpenter}, {Chester}, {Cambresy}, {Evans}, {Fowler}, {Gizis}, {Howard},
  {Huchra}, {Jarrett}, {Kopan}, {Kirkpatrick}, {Light}, {Marsh}, {McCallon},
  {Schneider}, {Stiening}, {Sykes}, {Weinberg}, {Wheaton}, {Wheelock}, \&
  {Zacarias}}]{cutri2003}
{Cutri}, R.~M., {et~al.} 2003, VizieR Online Data Catalog, 2246, 0

\bibitem[{{Davies} {et~al.}(1990){Davies}, {Evans}, {Bode}, \&
  {Whittet}}]{davies1990}
{Davies}, J.~K., {Evans}, A., {Bode}, M.~F., \& {Whittet}, D.~C.~B. 1990,
  \mnras, 247, 517

\bibitem[{{Decin} {et~al.}(2004){Decin}, {Morris}, {Appleton}, {Charmandaris},
  {Armus}, \& {Houck}}]{decin2004}
{Decin}, L., {Morris}, P.~W., {Appleton}, P.~N., {Charmandaris}, V., {Armus},
  L., \& {Houck}, J.~R. 2004, \apjs, 154, 408

\bibitem[{{Dent} {et~al.}(2006){Dent}, {Torrelles}, {Osorio}, {Calvet}, \&
  {Anglada}}]{dent2006}
{Dent}, W.~R.~F., {Torrelles}, J.~M., {Osorio}, M., {Calvet}, N., \& {Anglada},
  G. 2006, \mnras, 365, 1283

\bibitem[{{D{\'e}sert} {et~al.}(1990){D{\'e}sert}, {Boulanger}, \&
  {Puget}}]{desert1990}
{D{\'e}sert}, F., {Boulanger}, F., \& {Puget}, J.~L. 1990, \aap, 237, 215

\bibitem[{{Dewarf} \& {Dyck}(1993)}]{dewarf1993}
{Dewarf}, L.~E., \& {Dyck}, H.~M. 1993, \aj, 105, 2211

\bibitem[{{Dominik} {et~al.}(2003){Dominik}, {Dullemond}, {Waters}, \&
  {Walch}}]{dominik2003}
{Dominik}, C., {Dullemond}, C.~P., {Waters}, L.~B.~F.~M., \& {Walch}, S. 2003,
  \aap, 398, 607

\bibitem[{{Doucet} {et~al.}(2007){Doucet}, {Habart}, {Pantin}, {Dullemond},
  {Lagage}, {Pinte}, {Duch{\^e}ne}, \& {M{\'e}nard}}]{doucet2007}
{Doucet}, C., {Habart}, E., {Pantin}, E., {Dullemond}, C., {Lagage}, P.~O.,
  {Pinte}, C., {Duch{\^e}ne}, G., \& {M{\'e}nard}, F. 2007, \aap, 470, 625

\bibitem[{{Doucet} {et~al.}(2006){Doucet}, {Pantin}, {Lagage}, \&
  {Dullemond}}]{doucet2006}
{Doucet}, C., {Pantin}, E., {Lagage}, P.~O., \& {Dullemond}, C.~P. 2006, \aap,
  460, 117

\bibitem[{{Duch{\^e}ne} {et~al.}(2006){Duch{\^e}ne}, {Beust}, {Adjali},
  {Konopacky}, \& {Ghez}}]{duchene2006}
{Duch{\^e}ne}, G., {Beust}, H., {Adjali}, F., {Konopacky}, Q.~M., \& {Ghez},
  A.~M. 2006, \aap, 457, L9

\bibitem[{{Dullemond} {et~al.}(2001){Dullemond}, {Dominik}, \&
  {Natta}}]{dullemond2001}
{Dullemond}, C.~P., {Dominik}, C., \& {Natta}, A. 2001, \apj, 560, 957

\bibitem[{{Dullemond} {et~al.}(2007){Dullemond}, {Henning}, {Visser}, {Geers},
  {van Dishoeck}, \& {Pontoppidan}}]{dullemond2007}
{Dullemond}, C.~P., {Henning}, T., {Visser}, R., {Geers}, V.~C., {van
  Dishoeck}, E.~F., \& {Pontoppidan}, K.~M. 2007, \aap, 473, 457

\bibitem[{{Dunkin} {et~al.}(1997){Dunkin}, {Barlow}, \& {Ryan}}]{dunkin1997}
{Dunkin}, S.~K., {Barlow}, M.~J., \& {Ryan}, S.~G. 1997, \mnras, 290, 165

\bibitem[{{Eaton} \& {Herbst}(1994)}]{uxors}
{Eaton}, N.~L., \& {Herbst}, W. 1994, in Bulletin of the American Astronomical
  Society, Vol.~26, American Astronomical Society Meeting Abstracts \#184, 933

\bibitem[{{Eiroa} \& {Casali}(1989)}]{eiroa1989}
{Eiroa}, C., \& {Casali}, M.~M. 1989, \aap, 223, L17

\bibitem[{{Eiroa} \& {Leinert}(1987)}]{eiroa1987}
{Eiroa}, C., \& {Leinert}, C. 1987, \aap, 188, 46

\bibitem[{{Eiroa} {et~al.}(1987){Eiroa}, {Lenzen}, {Leinert}, \&
  {Hodapp}}]{eiroa1987a}
{Eiroa}, C., {Lenzen}, R., {Leinert}, C., \& {Hodapp}, K. 1987, \aap, 179, 171

\bibitem[{{Eisner} {et~al.}(2004){Eisner}, {Lane}, {Hillenbrand}, {Akeson}, \&
  {Sargent}}]{eisner2004}
{Eisner}, J.~A., {Lane}, B.~F., {Hillenbrand}, L.~A., {Akeson}, R.~L., \&
  {Sargent}, A.~I. 2004, \apj, 613, 1049

\bibitem[{{Espaillat} {et~al.}(2011){Espaillat}, {Furlan}, {D'Alessio},
  {Sargent}, {Nagel}, {Calvet}, {Watson}, \& {Muzerolle}}]{espaillat2011}
{Espaillat}, C., {Furlan}, E., {D'Alessio}, P., {Sargent}, B., {Nagel}, E.,
  {Calvet}, N., {Watson}, D.~M., \& {Muzerolle}, J. 2011, \apj, 728, 49

\bibitem[{{Feigelson} \& {Kriss}(1989)}]{feigelson1989}
{Feigelson}, E.~D., \& {Kriss}, G.~A. 1989, \apj, 338, 262

\bibitem[{{Flaherty} {et~al.}(2011){Flaherty}, {Muzerolle}, {Rieke},
  {Gutermuth}, {Balog}, {Herbst}, {Megeath}, \& {Kun}}]{flaherty2011}
{Flaherty}, K.~M., {Muzerolle}, J., {Rieke}, G., {Gutermuth}, R., {Balog}, Z.,
  {Herbst}, W., {Megeath}, S.~T., \& {Kun}, M. 2011, \apj, 732, 83

\bibitem[{{Forbrich} {et~al.}(2007){Forbrich}, {Preibisch}, {Menten},
  {Neuh{\"a}user}, {Walter}, {Tamura}, {Matsunaga}, {Kusakabe}, {Nakajima},
  {Brandeker}, {Fornasier}, {Posselt}, {Tachihara}, \& {Broeg}}]{forbrich2007}
{Forbrich}, J., {et~al.} 2007, \aap, 464, 1003

\bibitem[{{Fujiwara} {et~al.}(2006){Fujiwara}, {Honda}, {Kataza}, {Yamashita},
  {Onaka}, {Fukagawa}, {Okamoto}, {Miyata}, {Sako}, {Fujiyoshi}, \&
  {Sakon}}]{fujiwara2006}
{Fujiwara}, H., {et~al.} 2006, \apjl, 644, L133

\bibitem[{{Fukagawa} {et~al.}(2003){Fukagawa}, {Tamura}, {Itoh}, {Hayashi}, \&
  {Oasa}}]{fukagawa2003}
{Fukagawa}, M., {Tamura}, M., {Itoh}, Y., {Hayashi}, S.~S., \& {Oasa}, Y. 2003,
  \apjl, 590, L49

\bibitem[{{Fukagawa} {et~al.}(2006){Fukagawa}, {Tamura}, {Itoh}, {Kudo},
  {Imaeda}, {Oasa}, {Hayashi}, \& {Hayashi}}]{fukagawa2006}
{Fukagawa}, M., {Tamura}, M., {Itoh}, Y., {Kudo}, T., {Imaeda}, Y., {Oasa}, Y.,
  {Hayashi}, S.~S., \& {Hayashi}, M. 2006, \apjl, 636, L153

\bibitem[{{Fukagawa} {et~al.}(2010){Fukagawa}, {Tamura}, {Itoh}, {Oasa},
  {Kudo}, {Hayashi}, {Kato}, {Ootsubo}, {Itoh}, {Shibai}, \&
  {Hayashi}}]{fukagawa2010}
{Fukagawa}, M., {et~al.} 2010, \pasj, 62, 347

\bibitem[{{Furlan} {et~al.}(2006){Furlan}, {Hartmann}, {Calvet}, {D'Alessio},
  {Franco-Hern{\'a}ndez}, {Forrest}, {Watson}, {Uchida}, {Sargent}, {Green},
  {Keller}, \& {Herter}}]{furlan2006}
{Furlan}, E., {et~al.} 2006, \apjs, 165, 568

\bibitem[{{Furlan} {et~al.}(2007){Furlan}, {Sargent}, {Calvet}, {Forrest},
  {D'Alessio}, {Hartmann}, {Watson}, {Green}, {Najita}, \& {Chen}}]{furlan2007}
---. 2007, \apj, 664, 1176

\bibitem[{{Gabriel} {et~al.}(1997){Gabriel}, {Acosta-Pulido}, {Heinrichsen},
  {Morris}, \& {Tai}}]{pia}
{Gabriel}, C., {Acosta-Pulido}, J., {Heinrichsen}, I., {Morris}, H., \& {Tai},
  W. 1997, in Astronomical Society of the Pacific Conference Series, Vol. 125,
  Astronomical Data Analysis Software and Systems VI, ed. {G.~Hunt \&
  H.~Payne}, 108--111

\bibitem[{{Garcia Lopez} {et~al.}(2006){Garcia Lopez}, {Natta}, {Testi}, \&
  {Habart}}]{garcialopez2006}
{Garcia Lopez}, R., {Natta}, A., {Testi}, L., \& {Habart}, E. 2006, \aap, 459,
  837

\bibitem[{{Geers} {et~al.}(2007){Geers}, {van Dishoeck}, {Visser},
  {Pontoppidan}, {Augereau}, {Habart}, \& {Lagrange}}]{geers2007}
{Geers}, V.~C., {van Dishoeck}, E.~F., {Visser}, R., {Pontoppidan}, K.~M.,
  {Augereau}, J., {Habart}, E., \& {Lagrange}, A.~M. 2007, \aap, 476, 279

\bibitem[{{Geers} {et~al.}(2006){Geers}, {Augereau}, {Pontoppidan},
  {Dullemond}, {Visser}, {Kessler-Silacci}, {Evans}, {van Dishoeck}, {Blake},
  {Boogert}, {Brown}, {Lahuis}, \& {Mer{\'{\i}}n}}]{geers2006}
{Geers}, V.~C., {et~al.} 2006, \aap, 459, 545

\bibitem[{{Ghez} {et~al.}(1997){Ghez}, {McCarthy}, {Patience}, \&
  {Beck}}]{ghez1997}
{Ghez}, A.~M., {McCarthy}, D.~W., {Patience}, J.~L., \& {Beck}, T.~L. 1997,
  \apj, 481, 378

\bibitem[{{G{\"o}tz}(1980)}]{gotz1980}
{G{\"o}tz}, W. 1980, Information Bulletin on Variable Stars, 1747, 1

\bibitem[{{Grady} {et~al.}(2000){Grady}, {Devine}, {Woodgate}, {Kimble},
  {Bruhweiler}, {Boggess}, {Linsky}, {Plait}, {Clampin}, \&
  {Kalas}}]{grady2000}
{Grady}, C.~A., {et~al.} 2000, \apj, 544, 895

\bibitem[{{Grady} {et~al.}(2004){Grady}, {Woodgate}, {Torres}, {Henning},
  {Apai}, {Rodmann}, {Wang}, {Stecklum}, {Linz}, {Williger}, {Brown},
  {Wilkinson}, {Harper}, {Herczeg}, {Danks}, {Vieira}, {Malumuth}, {Collins},
  \& {Hill}}]{grady2004}
---. 2004, \apj, 608, 809

\bibitem[{{Grady} {et~al.}(2007){Grady}, {Schneider}, {Hamaguchi}, {Sitko},
  {Carpenter}, {Hines}, {Collins}, {Williger}, {Woodgate}, {Henning},
  {M{\'e}nard}, {Wilner}, {Petre}, {Palunas}, {Quirrenbach}, {Nuth},
  {Silverstone}, \& {Kim}}]{grady2007}
---. 2007, \apj, 665, 1391

\bibitem[{{Grady} {et~al.}(2009){Grady}, {Schneider}, {Sitko}, {Williger},
  {Hamaguchi}, {Brittain}, {Ablordeppey}, {Apai}, {Beerman}, {Carpenter},
  {Collins}, {Fukagawa}, {Hammel}, {Henning}, {Hines}, {Kimes}, {Lynch},
  {M{\'e}nard}, {Pearson}, {Russell}, {Silverstone}, {Smith}, {Troutman},
  {Wilner}, {Woodgate}, \& {Clampin}}]{grady2009}
---. 2009, \apj, 699, 1822

\bibitem[{{Graham}(1992)}]{graham1992}
{Graham}, J.~A. 1992, \pasp, 104, 479

\bibitem[{{Graham} \& {Phillips}(1987)}]{graham1987}
{Graham}, J.~A., \& {Phillips}, A.~C. 1987, \pasp, 99, 91

\bibitem[{{Grinin} {et~al.}(2002){Grinin}, {Shakhovskoi}, {Shenavrin},
  {Rostopchina}, \& {Tambovtseva}}]{grinin2002}
{Grinin}, V.~P., {Shakhovskoi}, D.~N., {Shenavrin}, V.~I., {Rostopchina},
  A.~N., \& {Tambovtseva}, L.~V. 2002, Astronomy Reports, 46, 646

\bibitem[{{Guenther} {et~al.}(2007){Guenther}, {Esposito}, {Mundt}, {Covino},
  {Alcal{\'a}}, {Cusano}, \& {Stecklum}}]{guenther2007}
{Guenther}, E.~W., {Esposito}, M., {Mundt}, R., {Covino}, E., {Alcal{\'a}},
  J.~M., {Cusano}, F., \& {Stecklum}, B. 2007, \aap, 467, 1147

\bibitem[{{Guimar{\~a}es} {et~al.}(2006){Guimar{\~a}es}, {Alencar}, {Corradi},
  \& {Vieira}}]{guimaraes2006}
{Guimar{\~a}es}, M.~M., {Alencar}, S.~H.~P., {Corradi}, W.~J.~B., \& {Vieira},
  S.~L.~A. 2006, \aap, 457, 581

\bibitem[{{G{\"u}rtler} {et~al.}(1999){G{\"u}rtler}, {Schreyer}, {Henning},
  {Lemke}, \& {Pfau}}]{gurtler1999}
{G{\"u}rtler}, J., {Schreyer}, K., {Henning}, T., {Lemke}, D., \& {Pfau}, W.
  1999, \aap, 346, 205

\bibitem[{{Habart} {et~al.}(2004){Habart}, {Natta}, \&
  {Kr{\"u}gel}}]{habart2004}
{Habart}, E., {Natta}, A., \& {Kr{\"u}gel}, E. 2004, \aap, 427, 179

\bibitem[{{Habart} {et~al.}(2006){Habart}, {Natta}, {Testi}, \&
  {Carbillet}}]{habart2006}
{Habart}, E., {Natta}, A., {Testi}, L., \& {Carbillet}, M. 2006, \aap, 449,
  1067

\bibitem[{{Habart} {et~al.}(2003){Habart}, {Testi}, {Natta}, \&
  {Vanzi}}]{habart2003}
{Habart}, E., {Testi}, L., {Natta}, A., \& {Vanzi}, L. 2003, \aap, 400, 575

\bibitem[{{Haisch} {et~al.}(2006){Haisch}, {Barsony}, {Ressler}, \&
  {Greene}}]{haisch2006}
{Haisch}, Jr., K.~E., {Barsony}, M., {Ressler}, M.~E., \& {Greene}, T.~P. 2006,
  \aj, 132, 2675

\bibitem[{{Haisch} {et~al.}(2004){Haisch}, {Greene}, {Barsony}, \&
  {Stahler}}]{haisch2004}
{Haisch}, Jr., K.~E., {Greene}, T.~P., {Barsony}, M., \& {Stahler}, S.~W. 2004,
  \aj, 127, 1747

\bibitem[{{Hammersley} \& {Jourdain de Muizon}(2003)}]{hammersley2003}
{Hammersley}, P.~L., \& {Jourdain de Muizon}, M. 2003, in ESA Special
  Publication, Vol. 481, The Calibration Legacy of the ISO Mission, ed.
  {L.~Metcalfe, A.~Salama, S.~B.~Peschke, \& M.~F.~Kessler}, 129

\bibitem[{{Hanner} {et~al.}(1992){Hanner}, {Tokunaga}, \&
  {Geballe}}]{hanner1992}
{Hanner}, M.~S., {Tokunaga}, A.~T., \& {Geballe}, T.~R. 1992, \apjl, 395, L111

\bibitem[{{Harvey} {et~al.}(2007){Harvey}, {Mer{\'{\i}}n}, {Huard}, {Rebull},
  {Chapman}, {Evans}, \& {Myers}}]{harvey2007}
{Harvey}, P., {Mer{\'{\i}}n}, B., {Huard}, T.~L., {Rebull}, L.~M., {Chapman},
  N., {Evans}, II, N.~J., \& {Myers}, P.~C. 2007, \apj, 663, 1149

\bibitem[{{Henning}(2010)}]{henning2010}
{Henning}, T. 2010, \araa, 48, 21

\bibitem[{{Henning} {et~al.}(1998){Henning}, {Burkert}, {Launhardt}, {Leinert},
  \& {Stecklum}}]{henning1998}
{Henning}, T., {Burkert}, A., {Launhardt}, R., {Leinert}, C., \& {Stecklum}, B.
  1998, \aap, 336, 565

\bibitem[{{Herbig}(1977)}]{herbig1977}
{Herbig}, G.~H. 1977, \apj, 217, 693

\bibitem[{{Herbig}(1990)}]{herbig1990}
---. 1990, \apj, 360, 639

\bibitem[{{Herbig} {et~al.}(2001){Herbig}, {Aspin}, {Gilmore}, {Imhoff}, \&
  {Jones}}]{herbig2001}
{Herbig}, G.~H., {Aspin}, C., {Gilmore}, A.~C., {Imhoff}, C.~L., \& {Jones},
  A.~F. 2001, \pasp, 113, 1547

\bibitem[{{Herbst} {et~al.}(1994){Herbst}, {Herbst}, {Grossman}, \&
  {Weinstein}}]{herbst1994}
{Herbst}, W., {Herbst}, D.~K., {Grossman}, E.~J., \& {Weinstein}, D. 1994, \aj,
  108, 1906

\bibitem[{{Herbst} \& {Shevchenko}(1999)}]{herbst1999}
{Herbst}, W., \& {Shevchenko}, V.~S. 1999, \aj, 118, 1043

\bibitem[{{Hessman} \& {Guenther}(1997)}]{hessman1997}
{Hessman}, F.~V., \& {Guenther}, E.~W. 1997, \aap, 321, 497

\bibitem[{{Hirose} \& {Turner}(2011)}]{hirose2011}
{Hirose}, S., \& {Turner}, N.~J. 2011, \apjl, 732, L30

\bibitem[{{Hodapp} {et~al.}(1996){Hodapp}, {Hora}, {Rayner}, {Pickles}, \&
  {Ladd}}]{hodapp1996}
{Hodapp}, K., {Hora}, J.~L., {Rayner}, J.~T., {Pickles}, A.~J., \& {Ladd},
  E.~F. 1996, \apj, 468, 861

\bibitem[{{Hodapp}(1999)}]{hodapp1999}
{Hodapp}, K.~W. 1999, \aj, 118, 1338

\bibitem[{{Houck} {et~al.}(2004){Houck}, {Roellig}, {Van Cleve}, {Forrest},
  {Herter}, {Lawrence}, {Matthews}, {Reitsema}, {Soifer}, {Watson}, {Weedman},
  {Huisjen}, {Troeltzsch}, {Barry}, {Bernard-Salas}, {Blacken}, {Brandl},
  {Charmandaris}, {Devost}, {Gull}, {Hall}, {Henderson}, {Higdon}, {Pirger},
  {Schoenwald}, {Sloan}, {Uchida}, {Appleton}, {Armus}, {Burgdorf},
  {Fajardo-Acosta}, {Grillmair}, {Ingalls}, {Morris}, \& {Teplitz}}]{houck2004}
{Houck}, J.~R., {et~al.} 2004, in Presented at the Society of Photo-Optical
  Instrumentation Engineers (SPIE) Conference, Vol. 5487, Society of
  Photo-Optical Instrumentation Engineers (SPIE) Conference Series, ed.
  {J.~C.~Mather}, 62--76

\bibitem[{{Hughes} {et~al.}(1991){Hughes}, {Hartigan}, {Graham}, {Emerson}, \&
  {Marang}}]{hughes1991}
{Hughes}, J.~D., {Hartigan}, P., {Graham}, J.~A., {Emerson}, J.~P., \&
  {Marang}, F. 1991, \aj, 101, 1013

\bibitem[{{Hutchinson} {et~al.}(1994){Hutchinson}, {Albinson}, {Barrett},
  {Davies}, {Evans}, {Goldsmith}, \& {Maddison}}]{hutchinson1994}
{Hutchinson}, M.~G., {Albinson}, J.~S., {Barrett}, P., {Davies}, J.~K.,
  {Evans}, A., {Goldsmith}, M.~J., \& {Maddison}, R.~C. 1994, \aap, 285, 883

\bibitem[{{Isobe} {et~al.}(1988){Isobe}, {Norimoto}, \& {Kitamura}}]{isobe1988}
{Isobe}, S., {Norimoto}, Y., \& {Kitamura}, T. 1988, \pasj, 40, 89

\bibitem[{{Jaschek} \& {Jaschek}(1992)}]{jaschek1992}
{Jaschek}, C., \& {Jaschek}, M. 1992, \aaps, 95, 535

\bibitem[{{Jayawardhana} {et~al.}(2001){Jayawardhana}, {Fisher}, {Telesco},
  {Pi{\~n}a}, {Barrado y Navascu{\'e}s}, {Hartmann}, \&
  {Fazio}}]{jayawardhana2001}
{Jayawardhana}, R., {Fisher}, R.~S., {Telesco}, C.~M., {Pi{\~n}a}, R.~K.,
  {Barrado y Navascu{\'e}s}, D., {Hartmann}, L.~W., \& {Fazio}, G.~G. 2001,
  \aj, 122, 2047

\bibitem[{{Jensen} {et~al.}(2007){Jensen}, {Dhital}, {Stassun}, {Patience},
  {Herbst}, {Walter}, {Simon}, \& {Basri}}]{jensen2007}
{Jensen}, E.~L.~N., {Dhital}, S., {Stassun}, K.~G., {Patience}, J., {Herbst},
  W., {Walter}, F.~M., {Simon}, M., \& {Basri}, G. 2007, \aj, 134, 241

\bibitem[{{Joblin} {et~al.}(1996){Joblin}, {Tielens}, {Geballe}, \&
  {Wooden}}]{joblin1996}
{Joblin}, C., {Tielens}, A.~G.~G.~M., {Geballe}, T.~R., \& {Wooden}, D.~H.
  1996, \apjl, 460, L119

\bibitem[{{Juh{\'a}sz} {et~al.}(2007){Juh{\'a}sz}, {Prusti}, {{\'A}brah{\'a}m},
  \& {Dullemond}}]{juhasz2007}
{Juh{\'a}sz}, A., {Prusti}, T., {{\'A}brah{\'a}m}, P., \& {Dullemond}, C.~P.
  2007, \mnras, 374, 1242

\bibitem[{{Juh{\'a}sz} {et~al.}(2010){Juh{\'a}sz}, {Bouwman}, {Henning},
  {Acke}, {van den Ancker}, {Meeus}, {Dominik}, {Min}, {Tielens}, \&
  {Waters}}]{juhasz2010herbig}
{Juh{\'a}sz}, A., {et~al.} 2010, \apj, 721, 431

\bibitem[{{Juh{\'a}sz} {et~al.}(2012){Juh{\'a}sz}, {Dullemond}, {van Boekel},
  {Bouwman}, {{\'A}brah{\'a}m}, {Acosta-Pulido}, {Henning}, {K{\'o}sp{\'a}l},
  {Sicilia-Aguilar}, {Jones}, {Mo{\'o}r}, {Mosoni}, {Reg{\'a}ly}, {Szokoly}, \&
  {Sipos}}]{juhasz2012}
---. 2012, \apj, 744, 118

\bibitem[{{Kardopolov} \& {Filipev}(1981)}]{kardopolov1981}
{Kardopolov}, V.~I., \& {Filipev}, G.~K. 1981, \sovast, 25, 457

\bibitem[{{Keller} {et~al.}(2008){Keller}, {Sloan}, {Forrest}, {Ayala},
  {D'Alessio}, {Shah}, {Calvet}, {Najita}, {Li}, {Hartmann}, {Sargent},
  {Watson}, \& {Chen}}]{keller2008}
{Keller}, L.~D., {et~al.} 2008, \apj, 684, 411

\bibitem[{{Kenyon} {et~al.}(1994){Kenyon}, {Hartmann}, {Hewett}, {Carrasco},
  {Cruz-Gonzalez}, {Recillas}, {Salas}, {Serrano}, {Strom}, {Strom}, \&
  {Newton}}]{kenyon1994}
{Kenyon}, S.~J., {et~al.} 1994, \aj, 107, 2153

\bibitem[{{Kessler-Silacci} {et~al.}(2005){Kessler-Silacci}, {Hillenbrand},
  {Blake}, \& {Meyer}}]{kessler2005}
{Kessler-Silacci}, J.~E., {Hillenbrand}, L.~A., {Blake}, G.~A., \& {Meyer},
  M.~R. 2005, \apj, 622, 404

\bibitem[{{K{\"o}hler} {et~al.}(2008){K{\"o}hler}, {Ratzka}, {Herbst}, \&
  {Kasper}}]{kohler2008}
{K{\"o}hler}, R., {Ratzka}, T., {Herbst}, T.~M., \& {Kasper}, M. 2008, \aap,
  482, 929

\bibitem[{{Koresko}(2002)}]{koresko2002}
{Koresko}, C.~D. 2002, \aj, 124, 1082

\bibitem[{{K{\'o}sp{\'a}l} {et~al.}(2007){K{\'o}sp{\'a}l}, {{\'A}brah{\'a}m},
  {Prusti}, {Acosta-Pulido}, {Hony}, {Mo{\'o}r}, \&
  {Siebenmorgen}}]{kospal2007}
{K{\'o}sp{\'a}l}, {\'A}., {{\'A}brah{\'a}m}, P., {Prusti}, T., {Acosta-Pulido},
  J., {Hony}, S., {Mo{\'o}r}, A., \& {Siebenmorgen}, R. 2007, \aap, 470, 211

\bibitem[{{K{\'o}sp{\'a}l} {et~al.}(2011){K{\'o}sp{\'a}l}, {Salter},
  {Hogerheijde}, {Mo{\'o}r}, \& {Blake}}]{kospal2011}
{K{\'o}sp{\'a}l}, {\'A}., {Salter}, D.~M., {Hogerheijde}, M.~R., {Mo{\'o}r},
  A., \& {Blake}, G.~A. 2011, \aap, 527, A96

\bibitem[{{K{\'o}sp{\'a}l} {et~al.}(2008){K{\'o}sp{\'a}l}, {{\'A}brah{\'a}m},
  {Apai}, {Ardila}, {Grady}, {Henning}, {Juh{\'a}sz}, {Miller}, \&
  {Mo{\'o}r}}]{par21}
{K{\'o}sp{\'a}l}, {\'A}., {et~al.} 2008, \mnras, 383, 1015

\bibitem[{{K{\'o}sp{\'a}l} {et~al.}(2012){K{\'o}sp{\'a}l}, {Prusti}, {Cox},
  {Pilbratt}, {Alves de Oliveira}, {Winston}, {Mer{\'\i}n}, {Ribas}, {Royer},
  {Vavrek}, \& {Waelkens}}]{kospal2012}
{K{\'o}sp{\'a}l}, {\'A.}., {et~al.} 2012, \aap

\bibitem[{{Kozlova} {et~al.}(2006){Kozlova}, {Shakhovskoi}, {Rostopchina}, \&
  {Alekseev}}]{kozlova2006}
{Kozlova}, O.~V., {Shakhovskoi}, D.~N., {Rostopchina}, A.~N., \& {Alekseev},
  I.~Y. 2006, Astrophysics, 49, 151

\bibitem[{{Kun} {et~al.}(2011){Kun}, {Szegedi-Elek}, {Mo{\'o}r},
  {K{\'o}sp{\'a}l}, {{\'A}brah{\'a}m}, {Apai}, {Kiss}, {Klagyivik}, {Magakian},
  {Mez{\H o}}, {Movsessian}, {P{\'a}l}, {R{\'a}cz}, \& {Rogers}}]{kun2011}
{Kun}, M., {et~al.} 2011, \mnras, 413, 2689

\bibitem[{{Lafreni{\`e}re} {et~al.}(2008){Lafreni{\`e}re}, {Jayawardhana},
  {Brandeker}, {Ahmic}, \& {van Kerkwijk}}]{lafreniere2008}
{Lafreni{\`e}re}, D., {Jayawardhana}, R., {Brandeker}, A., {Ahmic}, M., \& {van
  Kerkwijk}, M.~H. 2008, \apj, 683, 844

\bibitem[{{Lagage} {et~al.}(2006){Lagage}, {Doucet}, {Pantin}, {Habart},
  {Duch{\^e}ne}, {M{\'e}nard}, {Pinte}, {Charnoz}, \& {Pel}}]{lagage2006}
{Lagage}, P., {et~al.} 2006, Science, 314, 621

\bibitem[{{Laureijs} {et~al.}(2003){Laureijs}, {Klaas}, {Richards}, {Schulz},
  \& {Abraham}}]{isophothandbook}
{Laureijs}, R.~J., {Klaas}, U., {Richards}, P.~J., {Schulz}, B., \& {Abraham},
  P. 2003, {The ISO Handbook, Volume IV - PHT - The Imaging Photo-Polarimeter},
  ed. {Kessler, M.~F., Mueller, T.~G., Leech, K., Arviset, C., Garcia-Lario,
  P., Metcalfe, L., Pollock, A., Prusti, T., \& Salama, A.}

\bibitem[{{Lecavelier Des Etangs} {et~al.}(2005){Lecavelier Des Etangs},
  {Nitschelm}, {Olsen}, {Vidal-Madjar}, \& {Ferlet}}]{lecavelier2005}
{Lecavelier Des Etangs}, A., {Nitschelm}, C., {Olsen}, E.~H., {Vidal-Madjar},
  A., \& {Ferlet}, R. 2005, \aap, 439, 571

\bibitem[{{Leinert} {et~al.}(2002){Leinert}, {{\'A}brah{\'a}m},
  {Acosta-Pulido}, {Lemke}, \& {Siebenmorgen}}]{leinert2002}
{Leinert}, C., {{\'A}brah{\'a}m}, P., {Acosta-Pulido}, J., {Lemke}, D., \&
  {Siebenmorgen}, R. 2002, \aap, 393, 1073

\bibitem[{{Leinert} {et~al.}(2001{\natexlab{a}}){Leinert}, {Beck}, {Ligori},
  {Simon}, {Woitas}, \& {Howell}}]{leinert2001}
{Leinert}, C., {Beck}, T.~L., {Ligori}, S., {Simon}, M., {Woitas}, J., \&
  {Howell}, R.~R. 2001{\natexlab{a}}, \aap, 369, 215

\bibitem[{{Leinert} \& {Haas}(1989)}]{leinert1989}
{Leinert}, C., \& {Haas}, M. 1989, \apjl, 342, L39

\bibitem[{{Leinert} {et~al.}(2001{\natexlab{b}}){Leinert}, {Haas},
  {{\'A}brah{\'a}m}, \& {Richichi}}]{leinert2001b}
{Leinert}, C., {Haas}, M., {{\'A}brah{\'a}m}, P., \& {Richichi}, A.
  2001{\natexlab{b}}, \aap, 375, 927

\bibitem[{{Leinert} {et~al.}(1991){Leinert}, {Haas}, {Mundt}, {Richichi}, \&
  {Zinnecker}}]{leinert1991}
{Leinert}, C., {Haas}, M., {Mundt}, R., {Richichi}, A., \& {Zinnecker}, H.
  1991, \aap, 250, 407

\bibitem[{{Leinert} {et~al.}(1997){Leinert}, {Richichi}, \&
  {Haas}}]{leinert1997}
{Leinert}, C., {Richichi}, A., \& {Haas}, M. 1997, \aap, 318, 472

\bibitem[{{Leinert} {et~al.}(1993){Leinert}, {Zinnecker}, {Weitzel},
  {Christou}, {Ridgway}, {Jameson}, {Haas}, \& {Lenzen}}]{leinert1993}
{Leinert}, C., {Zinnecker}, H., {Weitzel}, N., {Christou}, J., {Ridgway},
  S.~T., {Jameson}, R., {Haas}, M., \& {Lenzen}, R. 1993, \aap, 278, 129

\bibitem[{{Leinert} {et~al.}(2004){Leinert}, {van Boekel}, {Waters},
  {Chesneau}, {Malbet}, {K{\"o}hler}, {Jaffe}, {Ratzka}, {Dutrey}, {Preibisch},
  {Graser}, {Bakker}, {Chagnon}, {Cotton}, {Dominik}, {Dullemond},
  {Glazenborg-Kluttig}, {Glindemann}, {Henning}, {Hofmann}, {de Jong},
  {Lenzen}, {Ligori}, {Lopez}, {Meisner}, {Morel}, {Paresce}, {Pel},
  {Percheron}, {Perrin}, {Przygodda}, {Richichi}, {Sch{\"o}ller}, {Schuller},
  {Stecklum}, {van den Ancker}, {von der L{\"u}he}, \& {Weigelt}}]{leinert2004}
{Leinert}, C., {et~al.} 2004, \aap, 423, 537

\bibitem[{{Li} {et~al.}(1994){Li}, {Evans}, {Harvey}, \& {Colom{\'e}}}]{li1994}
{Li}, W., {Evans}, II, N.~J., {Harvey}, P.~M., \& {Colom{\'e}}, C. 1994, \apj,
  433, 199

\bibitem[{{Lim} \& {Takakuwa}(2006)}]{lim2006}
{Lim}, J., \& {Takakuwa}, S. 2006, \apj, 653, 425

\bibitem[{{Liu} {et~al.}(1996){Liu}, {Graham}, {Ghez}, {Meixner}, {Skinner},
  {Keto}, {Ball}, {Arens}, \& {Jernigan}}]{liu1996}
{Liu}, M.~C., {et~al.} 1996, \apj, 461, 334

\bibitem[{{Lopes} {et~al.}(1992){Lopes}, {Damineli Neto}, \& {de Freitas
  Pacheco}}]{lopes1992}
{Lopes}, D.~F., {Damineli Neto}, A., \& {de Freitas Pacheco}, J.~A. 1992, \aap,
  261, 482

\bibitem[{{Lorenzetti} {et~al.}(2009){Lorenzetti}, {Larionov}, {Giannini},
  {Arkharov}, {Antoniucci}, {Nisini}, \& {Di Paola}}]{lorenzetti2009}
{Lorenzetti}, D., {Larionov}, V.~M., {Giannini}, T., {Arkharov}, A.~A.,
  {Antoniucci}, S., {Nisini}, B., \& {Di Paola}, A. 2009, \apj, 693, 1056

\bibitem[{{Lorenzetti} {et~al.}(2011){Lorenzetti}, {Giannini}, {Larionov},
  {Arkharov}, {Antoniucci}, {Di Paola}, {Konstantinova}, {Kopatskaya}, {Causi},
  \& {Nisini}}]{lorenzetti2011}
{Lorenzetti}, D., {et~al.} 2011, \apj, 732, 69

\bibitem[{{Luhman} {et~al.}(2010){Luhman}, {Allen}, {Espaillat}, {Hartmann}, \&
  {Calvet}}]{luhman2010}
{Luhman}, K.~L., {Allen}, P.~R., {Espaillat}, C., {Hartmann}, L., \& {Calvet},
  N. 2010, \apjs, 186, 111

\bibitem[{{Luhman} \& {Rieke}(1999)}]{luhman1999}
{Luhman}, K.~L., \& {Rieke}, G.~H. 1999, \apj, 525, 440

\bibitem[{{Luhman} {et~al.}(2008){Luhman}, {Allen}, {Allen}, {Gutermuth},
  {Hartmann}, {Mamajek}, {Megeath}, {Myers}, \& {Fazio}}]{luhman2008}
{Luhman}, K.~L., {et~al.} 2008, \apj, 675, 1375

\bibitem[{{Magakian} \& {Movsesian}(2001)}]{magakian2001}
{Magakian}, T.~Y., \& {Movsesian}, T.~A. 2001, Astrophysics, 44, 419

\bibitem[{{Malfait} {et~al.}(1998){Malfait}, {Bogaert}, \&
  {Waelkens}}]{malfait1998}
{Malfait}, K., {Bogaert}, E., \& {Waelkens}, C. 1998, \aap, 331, 211

\bibitem[{{Manfroid} {et~al.}(1991){Manfroid}, {Sterken}, {Bruch}, {Burger},
  {de Groot}, {Duerbeck}, {Duemmler}, {Figer}, {Hageman}, {Hensberge},
  {Jorissen}, {Madejsky}, {Mandel}, {Ott}, {Reitermann}, {Schulte-Ladbeck},
  {Stahl}, {Steenman}, {Vander Linden}, \& {Zickgraf}}]{manfroid1991}
{Manfroid}, J., {et~al.} 1991, \aaps, 87, 481

\bibitem[{{McCabe} {et~al.}(2006){McCabe}, {Ghez}, {Prato}, {Duch{\^e}ne},
  {Fisher}, \& {Telesco}}]{mccabe2006}
{McCabe}, C., {Ghez}, A.~M., {Prato}, L., {Duch{\^e}ne}, G., {Fisher}, R.~S.,
  \& {Telesco}, C. 2006, \apj, 636, 932

\bibitem[{{Meeus} {et~al.}(2002){Meeus}, {Bouwman}, {Dominik}, {Waters}, \& {de
  Koter}}]{meeus2002}
{Meeus}, G., {Bouwman}, J., {Dominik}, C., {Waters}, L.~B.~F.~M., \& {de
  Koter}, A. 2002, \aap, 392, 1039

\bibitem[{{Meeus} {et~al.}(1998){Meeus}, {Waelkens}, \& {Malfait}}]{meeus1998}
{Meeus}, G., {Waelkens}, C., \& {Malfait}, K. 1998, \aap, 329, 131

\bibitem[{{Meeus} {et~al.}(2001){Meeus}, {Waters}, {Bouwman}, {van den Ancker},
  {Waelkens}, \& {Malfait}}]{meeus2001}
{Meeus}, G., {Waters}, L.~B.~F.~M., {Bouwman}, J., {van den Ancker}, M.~E.,
  {Waelkens}, C., \& {Malfait}, K. 2001, \aap, 365, 476

\bibitem[{{Melo}(2003)}]{melo2003}
{Melo}, C.~H.~F. 2003, \aap, 410, 269

\bibitem[{{Monnier} {et~al.}(2009){Monnier}, {Tuthill}, {Ireland}, {Cohen},
  {Tannirkulam}, \& {Perrin}}]{monnier2009}
{Monnier}, J.~D., {Tuthill}, P.~G., {Ireland}, M., {Cohen}, R., {Tannirkulam},
  A., \& {Perrin}, M.~D. 2009, \apj, 700, 491

\bibitem[{{Moore} \& {Emerson}(1992)}]{moore1992}
{Moore}, T.~J.~T., \& {Emerson}, J.~P. 1992, \mnras, 259, 381

\bibitem[{{Moore} \& {Emerson}(1994)}]{moore1994}
---. 1994, \mnras, 271, 243

\bibitem[{{Morales-Calder{\'o}n} {et~al.}(2009){Morales-Calder{\'o}n},
  {Stauffer}, {Rebull}, {Whitney}, {Barrado y Navascu{\'e}s}, {Ardila}, {Song},
  {Brooke}, {Hartmann}, \& {Calvet}}]{morales2009}
{Morales-Calder{\'o}n}, M., {et~al.} 2009, \apj, 702, 1507

\bibitem[{{Morales-Calder{\'o}n} {et~al.}(2011){Morales-Calder{\'o}n},
  {Stauffer}, {Hillenbrand}, {Gutermuth}, {Song}, {Rebull}, {Plavchan},
  {Carpenter}, {Whitney}, {Covey}, {Alves de Oliveira}, {Winston},
  {McCaughrean}, {Bouvier}, {Guieu}, {Vrba}, {Holtzman}, {Marchis}, {Hora},
  {Wasserman}, {Terebey}, {Megeath}, {Guinan}, {Forbrich}, {Hu{\'e}lamo},
  {Riviere-Marichalar}, {Barrado}, {Stapelfeldt}, {Hern{\'a}ndez}, {Allen},
  {Ardila}, {Bayo}, {Favata}, {James}, {Werner}, \& {Wood}}]{morales2011}
---. 2011, \apj, 733, 50

\bibitem[{{Mouillet} {et~al.}(2001){Mouillet}, {Lagrange}, {Augereau}, \&
  {M{\'e}nard}}]{mouillet2001}
{Mouillet}, D., {Lagrange}, A.~M., {Augereau}, J.~C., \& {M{\'e}nard}, F. 2001,
  \aap, 372, L61

\bibitem[{{Murakawa} {et~al.}(2008){Murakawa}, {Oya}, {Pyo}, \&
  {Ishii}}]{murakawa2008}
{Murakawa}, K., {Oya}, S., {Pyo}, T., \& {Ishii}, M. 2008, \aap, 492, 731

\bibitem[{{Muzerolle} {et~al.}(2003){Muzerolle}, {Calvet}, {Hartmann}, \&
  {D'Alessio}}]{muzerolle2003}
{Muzerolle}, J., {Calvet}, N., {Hartmann}, L., \& {D'Alessio}, P. 2003, \apjl,
  597, L149

\bibitem[{{Muzerolle} {et~al.}(2005{\natexlab{a}}){Muzerolle}, {Luhman},
  {Brice{\~n}o}, {Hartmann}, \& {Calvet}}]{muzerolle2005b}
{Muzerolle}, J., {Luhman}, K.~L., {Brice{\~n}o}, C., {Hartmann}, L., \&
  {Calvet}, N. 2005{\natexlab{a}}, \apj, 625, 906

\bibitem[{{Muzerolle} {et~al.}(2005{\natexlab{b}}){Muzerolle}, {Megeath},
  {Flaherty}, {Gordon}, {Rieke}, {Young}, \& {Lada}}]{muzerolle2005}
{Muzerolle}, J., {Megeath}, S.~T., {Flaherty}, K.~M., {Gordon}, K.~D., {Rieke},
  G.~H., {Young}, E.~T., \& {Lada}, C.~J. 2005{\natexlab{b}}, \apjl, 620, L107

\bibitem[{{Muzerolle} {et~al.}(2009){Muzerolle}, {Flaherty}, {Balog}, {Furlan},
  {Smith}, {Allen}, {Calvet}, {D'Alessio}, {Megeath}, {Muench}, {Rieke}, \&
  {Sherry}}]{muzerolle2009}
{Muzerolle}, J., {et~al.} 2009, \apjl, 704, L15

\bibitem[{{Myers} {et~al.}(1987){Myers}, {Fuller}, {Mathieu}, {Beichman},
  {Benson}, {Schild}, \& {Emerson}}]{myers1987}
{Myers}, P.~C., {Fuller}, G.~A., {Mathieu}, R.~D., {Beichman}, C.~A., {Benson},
  P.~J., {Schild}, R.~E., \& {Emerson}, J.~P. 1987, \apj, 319, 340

\bibitem[{{Nagel} {et~al.}(2010){Nagel}, {D'Alessio}, {Calvet}, {Espaillat},
  {Sargent}, {Hern{\'a}ndez}, \& {Forrest}}]{nagel2010}
{Nagel}, E., {D'Alessio}, P., {Calvet}, N., {Espaillat}, C., {Sargent}, B.,
  {Hern{\'a}ndez}, J., \& {Forrest}, W.~J. 2010, \apj, 708, 38

\bibitem[{{Natta}(1993)}]{natta1993}
{Natta}, A. 1993, \apj, 412, 761

\bibitem[{{Nummelin} {et~al.}(2001){Nummelin}, {Whittet}, {Gibb}, {Gerakines},
  \& {Chiar}}]{nummelin2001}
{Nummelin}, A., {Whittet}, D.~C.~B., {Gibb}, E.~L., {Gerakines}, P.~A., \&
  {Chiar}, J.~E. 2001, \apj, 558, 185

\bibitem[{{Osorio} {et~al.}(2003){Osorio}, {D'Alessio}, {Muzerolle}, {Calvet},
  \& {Hartmann}}]{osorio2003}
{Osorio}, M., {D'Alessio}, P., {Muzerolle}, J., {Calvet}, N., \& {Hartmann}, L.
  2003, \apj, 586, 1148

\bibitem[{{P{\'e}rez} {et~al.}(2004){P{\'e}rez}, {van den Ancker}, {de Winter},
  \& {Bopp}}]{perez2004}
{P{\'e}rez}, M.~R., {van den Ancker}, M.~E., {de Winter}, D., \& {Bopp}, B.~W.
  2004, \aap, 416, 647

\bibitem[{{Perez} {et~al.}(1992){Perez}, {Webb}, \& {Th{\'e}}}]{Perez1992}
{Perez}, M.~R., {Webb}, J.~R., \& {Th{\'e}}, P.~S. 1992, \aap, 257, 209

\bibitem[{{Pirzkal} {et~al.}(1997){Pirzkal}, {Spillar}, \&
  {Dyck}}]{pirzkal1997}
{Pirzkal}, N., {Spillar}, E.~J., \& {Dyck}, H.~M. 1997, \apj, 481, 392

\bibitem[{{Pojmanski}(1997)}]{pojmanski1997}
{Pojmanski}, G. 1997, \actaa, 47, 467

\bibitem[{{Pontoppidan} {et~al.}(2007){Pontoppidan}, {Dullemond}, {Blake},
  {Evans}, {Geers}, {Harvey}, \& {Spiesman}}]{pontoppidan2007}
{Pontoppidan}, K.~M., {Dullemond}, C.~P., {Blake}, G.~A., {Evans}, II, N.~J.,
  {Geers}, V.~C., {Harvey}, P.~M., \& {Spiesman}, W. 2007, \apj, 656, 991

\bibitem[{{Pontoppidan} {et~al.}(2004){Pontoppidan}, {van Dishoeck}, \&
  {Dartois}}]{pontoppidan2004}
{Pontoppidan}, K.~M., {van Dishoeck}, E.~F., \& {Dartois}, E. 2004, \aap, 426,
  925

\bibitem[{{Prato} {et~al.}(2003){Prato}, {Greene}, \& {Simon}}]{prato2003}
{Prato}, L., {Greene}, T.~P., \& {Simon}, M. 2003, \apj, 584, 853

\bibitem[{{Prato} {et~al.}(2002){Prato}, {Simon}, {Mazeh}, {Zucker}, \&
  {McLean}}]{prato2002}
{Prato}, L., {Simon}, M., {Mazeh}, T., {Zucker}, S., \& {McLean}, I.~S. 2002,
  \apjl, 579, L99

\bibitem[{{Prato} {et~al.}(2001){Prato}, {Ghez}, {Pi{\~n}a}, {Telesco},
  {Fisher}, {Wizinowich}, {Lai}, {Acton}, \& {Stomski}}]{prato2001}
{Prato}, L., {et~al.} 2001, \apj, 549, 590

\bibitem[{{Preibisch} {et~al.}(2006){Preibisch}, {Kraus}, {Driebe}, {van
  Boekel}, \& {Weigelt}}]{preibisch2006}
{Preibisch}, T., {Kraus}, S., {Driebe}, T., {van Boekel}, R., \& {Weigelt}, G.
  2006, \aap, 458, 235

\bibitem[{{Prusti} {et~al.}(1991){Prusti}, {Clark}, {Whittet}, {Laureijs}, \&
  {Zhang}}]{prusti1991}
{Prusti}, T., {Clark}, F.~O., {Whittet}, D.~C.~B., {Laureijs}, R.~J., \&
  {Zhang}, C.~Y. 1991, \mnras, 251, 303

\bibitem[{{Prusti} {et~al.}(1994){Prusti}, {Natta}, \& {Palla}}]{prusti1994}
{Prusti}, T., {Natta}, A., \& {Palla}, F. 1994, \aap, 292, 593

\bibitem[{{Przygodda} {et~al.}(2003){Przygodda}, {van Boekel},
  {{\'A}brah{\'a}m}, {Melnikov}, {Waters}, \& {Leinert}}]{przygodda2003}
{Przygodda}, F., {van Boekel}, R., {{\'A}brah{\'a}m}, P., {Melnikov}, S.~Y.,
  {Waters}, L.~B.~F.~M., \& {Leinert}, C. 2003, \aap, 412, L43

\bibitem[{{Pugach}(2004)}]{pugach2004}
{Pugach}, A.~F. 2004, Astronomy Reports, 48, 470

\bibitem[{{Ratzka} {et~al.}(2005){Ratzka}, {K{\"o}hler}, \&
  {Leinert}}]{ratzka2005}
{Ratzka}, T., {K{\"o}hler}, R., \& {Leinert}, C. 2005, \aap, 437, 611

\bibitem[{{Ratzka} {et~al.}(2008){Ratzka}, {Leinert}, {Przygodda}, \&
  {Wolf}}]{ratzka2008}
{Ratzka}, T., {Leinert}, C., {Przygodda}, F., \& {Wolf}, S. 2008, in The Power
  of Optical/IR Interferometry: Recent Scientific Results and 2nd Generation,
  ed. {A.~Richichi, F.~Delplancke, F.~Paresce, \& A.~Chelli}, 519--522

\bibitem[{{Ratzka} {et~al.}(2009){Ratzka}, {Schegerer}, {Leinert},
  {{\'A}brah{\'a}m}, {Henning}, {Herbst}, {K{\"o}hler}, {Wolf}, \&
  {Zinnecker}}]{ratzka2009}
{Ratzka}, T., {et~al.} 2009, \aap, 502, 623

\bibitem[{{Rebull}(2011)}]{rebull2011}
{Rebull}, L.~M. 2011, ArXiv e-prints

\bibitem[{{Reipurth}(1985)}]{reipurth1985}
{Reipurth}, B. 1985, \aaps, 61, 319

\bibitem[{{Reipurth} \& {Bally}(1986)}]{reipurth1986}
{Reipurth}, B., \& {Bally}, J. 1986, \nat, 320, 336

\bibitem[{{Reipurth} {et~al.}(2002){Reipurth}, {Lindgren}, {Mayor},
  {Mermilliod}, \& {Cramer}}]{reipurth2002}
{Reipurth}, B., {Lindgren}, H., {Mayor}, M., {Mermilliod}, J., \& {Cramer}, N.
  2002, \aj, 124, 2813

\bibitem[{{Reipurth} \& {Wamsteker}(1983)}]{reipurth1983}
{Reipurth}, B., \& {Wamsteker}, W. 1983, \aap, 119, 14

\bibitem[{{Reipurth} \& {Zinnecker}(1993)}]{reipurth1993}
{Reipurth}, B., \& {Zinnecker}, H. 1993, \aap, 278, 81

\bibitem[{{Ressler} \& {Barsony}(2003)}]{ressler2003}
{Ressler}, M.~E., \& {Barsony}, M. 2003, \apj, 584, 832

\bibitem[{{Richichi} {et~al.}(1994){Richichi}, {Leinert}, {Jameson}, \&
  {Zinnecker}}]{richichi1994}
{Richichi}, A., {Leinert}, C., {Jameson}, R., \& {Zinnecker}, H. 1994, \aap,
  287, 145

\bibitem[{{Roche} {et~al.}(1994){Roche}, {Aitken}, \& {Smith}}]{roche1994}
{Roche}, P.~F., {Aitken}, D.~K., \& {Smith}, C.~H. 1994, \mnras, 269, 649

\bibitem[{{Rossiter}(1955)}]{rossiter1955}
{Rossiter}, R.~A. 1955, Publications of Michigan Observatory, 11, 1

\bibitem[{{Rostopchina} {et~al.}(2001){Rostopchina}, {Grinin}, \&
  {Shakhovskoi}}]{rostopchina2001}
{Rostopchina}, A.~N., {Grinin}, V.~P., \& {Shakhovskoi}, D.~N. 2001, Astronomy
  Reports, 45, 51

\bibitem[{{Rostopchina} {et~al.}(2000){Rostopchina}, {Grinin}, {Shakhovskoi},
  {Th{\'e}}, \& {Minikulov}}]{rostopchina2000}
{Rostopchina}, A.~N., {Grinin}, V.~P., {Shakhovskoi}, D.~N., {Th{\'e}}, P.~S.,
  \& {Minikulov}, N.~K. 2000, Astronomy Reports, 44, 365

\bibitem[{{Sandell} \& {Weintraub}(2001)}]{sandell2001}
{Sandell}, G., \& {Weintraub}, D.~A. 2001, \apjs, 134, 115

\bibitem[{{Sargent} {et~al.}(2009){Sargent}, {Forrest}, {Tayrien}, {McClure},
  {Watson}, {Sloan}, {Li}, {Manoj}, {Bohac}, {Furlan}, {Kim}, \&
  {Green}}]{sargent2009}
{Sargent}, B.~A., {et~al.} 2009, \apjs, 182, 477

\bibitem[{{Scarrott} {et~al.}(1991){Scarrott}, {Rolph}, \&
  {Tadhunter}}]{scarrott1991}
{Scarrott}, S.~M., {Rolph}, C.~D., \& {Tadhunter}, C.~N. 1991, \mnras, 249, 131

\bibitem[{{Scarrott} \& {Wolstencroft}(1988)}]{scarrott1988}
{Scarrott}, S.~M., \& {Wolstencroft}, R.~D. 1988, \mnras, 231, 1019

\bibitem[{{Schmidt} {et~al.}(2008){Schmidt}, {Neuh{\"a}user}, {Seifahrt},
  {Vogt}, {Bedalov}, {Helling}, {Witte}, \& {Hauschildt}}]{schmidt2008}
{Schmidt}, T.~O.~B., {Neuh{\"a}user}, R., {Seifahrt}, A., {Vogt}, N.,
  {Bedalov}, A., {Helling}, C., {Witte}, S., \& {Hauschildt}, P.~H. 2008, \aap,
  491, 311

\bibitem[{{Shiba} {et~al.}(1993){Shiba}, {Sato}, {Yamashita}, {Kobayashi}, \&
  {Takami}}]{shiba1993}
{Shiba}, H., {Sato}, S., {Yamashita}, T., {Kobayashi}, Y., \& {Takami}, H.
  1993, \apjs, 89, 299

\bibitem[{{Siebenmorgen} {et~al.}(1998){Siebenmorgen}, {Natta}, {Kruegel}, \&
  {Prusti}}]{siebenmorgen1998}
{Siebenmorgen}, R., {Natta}, A., {Kruegel}, E., \& {Prusti}, T. 1998, \aap,
  339, 134

\bibitem[{{Siebenmorgen} {et~al.}(2000){Siebenmorgen}, {Prusti}, {Natta}, \&
  {M{\"u}ller}}]{siebenmorgen2000}
{Siebenmorgen}, R., {Prusti}, T., {Natta}, A., \& {M{\"u}ller}, T.~G. 2000,
  \aap, 361, 258

\bibitem[{{Simon} {et~al.}(1995){Simon}, {Ghez}, {Leinert}, {Cassar}, {Chen},
  {Howell}, {Jameson}, {Matthews}, {Neugebauer}, \& {Richichi}}]{simon1995}
{Simon}, M., {et~al.} 1995, \apj, 443, 625

\bibitem[{{Sipos} {et~al.}(2011){Sipos}, {K{\'o}sp{\'a}l}, {Grossman}, \&
  {Weinstein}}]{sipos2011}
{Sipos}, N., {K{\'o}sp{\'a}l}, {\'A}., {Grossman}, E.~J., \& {Weinstein}, D.
  2011, \aap, in prep

\bibitem[{{Sitko} {et~al.}(2008){Sitko}, {Carpenter}, {Kimes}, {Wilde},
  {Lynch}, {Russell}, {Rudy}, {Mazuk}, {Venturini}, {Puetter}, {Grady},
  {Polomski}, {Wisnewski}, {Brafford}, {Hammel}, \& {Perry}}]{sitko2008}
{Sitko}, M.~L., {et~al.} 2008, \apj, 678, 1070

\bibitem[{{Skemer} {et~al.}(2008){Skemer}, {Close}, {Hinz}, {Hoffmann},
  {Kenworthy}, \& {Miller}}]{skemer2008}
{Skemer}, A.~J., {Close}, L.~M., {Hinz}, P.~M., {Hoffmann}, W.~F., {Kenworthy},
  M.~A., \& {Miller}, D.~L. 2008, \apj, 676, 1082

\bibitem[{{Sloan} {et~al.}(1999){Sloan}, {Hayward}, {Allamandola}, {Bregman},
  {Devito}, \& {Hudgins}}]{sloan1999}
{Sloan}, G.~C., {Hayward}, T.~L., {Allamandola}, L.~J., {Bregman}, J.~D.,
  {Devito}, B., \& {Hudgins}, D.~M. 1999, \apjl, 513, L65

\bibitem[{{Smith} {et~al.}(1999){Smith}, {Lewis}, {Bonnell}, {Bunclark}, \&
  {Emerson}}]{smith1999}
{Smith}, K.~W., {Lewis}, G.~F., {Bonnell}, I.~A., {Bunclark}, P.~S., \&
  {Emerson}, J.~P. 1999, \mnras, 304, 367

\bibitem[{{Stapelfeldt} {et~al.}(1995){Stapelfeldt}, {Burrows}, {Krist},
  {Trauger}, {Hester}, {Holtzman}, {Ballester}, {Casertano}, {Clarke}, {Crisp},
  {Evans}, {Gallagher}, {Griffiths}, {Hoessel}, {Mould}, {Scowen}, {Watson}, \&
  {Westphal}}]{stapelfeldt1995}
{Stapelfeldt}, K.~R., {et~al.} 1995, \apj, 449, 888

\bibitem[{{Stecklum} {et~al.}(1995){Stecklum}, {Eckart}, {Henning}, \&
  {Loewe}}]{stecklum1995}
{Stecklum}, B., {Eckart}, A., {Henning}, T., \& {Loewe}, M. 1995, \aap, 296,
  463

\bibitem[{{Stelzer} {et~al.}(2009){Stelzer}, {Robrade}, {Schmitt}, \&
  {Bouvier}}]{stelzer2009}
{Stelzer}, B., {Robrade}, J., {Schmitt}, J.~H.~M.~M., \& {Bouvier}, J. 2009,
  \aap, 493, 1109

\bibitem[{{Sterzik} {et~al.}(2005){Sterzik}, {Melo}, {Tokovinin}, \& {van der
  Bliek}}]{sterzik2005}
{Sterzik}, M.~F., {Melo}, C.~H.~F., {Tokovinin}, A.~A., \& {van der Bliek}, N.
  2005, \aap, 434, 671

\bibitem[{{Strai{\v z}ys} {et~al.}(2002){Strai{\v z}ys}, {Corbally},
  {Kazlauskas}, \& {{\v C}ernis}}]{straizys2002}
{Strai{\v z}ys}, V., {Corbally}, C.~J., {Kazlauskas}, A., \& {{\v C}ernis}, K.
  2002, Baltic Astronomy, 11, 261

\bibitem[{{Strom} {et~al.}(1972){Strom}, {Strom}, {Breger}, {Brooke}, {Yost},
  {Grasdalen}, \& {Carrasco}}]{strom1972}
{Strom}, K.~M., {Strom}, S.~E., {Breger}, M., {Brooke}, A.~L., {Yost}, J.,
  {Grasdalen}, G., \& {Carrasco}, L. 1972, \apjl, 173, L65

\bibitem[{{Strom} {et~al.}(1976){Strom}, {Vrba}, \& {Strom}}]{strom1976}
{Strom}, S.~E., {Vrba}, F.~J., \& {Strom}, K.~M. 1976, \aj, 81, 314

\bibitem[{{Sylvester} \& {Mannings}(2000)}]{sylvester2000}
{Sylvester}, R.~J., \& {Mannings}, V. 2000, \mnras, 313, 73

\bibitem[{{Sylvester} {et~al.}(1997){Sylvester}, {Skinner}, \&
  {Barlow}}]{sylvester1997}
{Sylvester}, R.~J., {Skinner}, C.~J., \& {Barlow}, M.~J. 1997, \mnras, 289, 831

\bibitem[{{Sylvester} {et~al.}(1996){Sylvester}, {Skinner}, {Barlow}, \&
  {Mannings}}]{sylvester1996}
{Sylvester}, R.~J., {Skinner}, C.~J., {Barlow}, M.~J., \& {Mannings}, V. 1996,
  \mnras, 279, 915

\bibitem[{{Takami} {et~al.}(2003){Takami}, {Bailey}, \&
  {Chrysostomou}}]{takami2003}
{Takami}, M., {Bailey}, J., \& {Chrysostomou}, A. 2003, \aap, 397, 675

\bibitem[{{Testa} {et~al.}(2008){Testa}, {Huenemoerder}, {Schulz}, \&
  {Ishibashi}}]{testa2008}
{Testa}, P., {Huenemoerder}, D.~P., {Schulz}, N.~S., \& {Ishibashi}, K. 2008,
  \apj, 687, 579

\bibitem[{{Th{\'e}} {et~al.}(1994){Th{\'e}}, {de Winter}, \& {Perez}}]{the1994}
{Th{\'e}}, P.~S., {de Winter}, D., \& {Perez}, M.~R. 1994, \aaps, 104, 315

\bibitem[{{Tielens}(2008)}]{tielens2008}
{Tielens}, A.~G.~G.~M. 2008, \araa, 46, 289

\bibitem[{{Tjin A Djie} {et~al.}(1989){Tjin A Djie}, {Th{\'e}}, {Andersen},
  {Nordstrom}, {Finkenzeller}, \& {Jankovics}}]{tjinadjie1989}
{Tjin A Djie}, H.~R.~E., {Th{\'e}}, P.~S., {Andersen}, J., {Nordstrom}, B.,
  {Finkenzeller}, U., \& {Jankovics}, I. 1989, \aaps, 78, 1

\bibitem[{{Torres}(2004)}]{torres2004}
{Torres}, G. 2004, \aj, 127, 1187

\bibitem[{{Turner} {et~al.}(2010){Turner}, {Carballido}, \&
  {Sano}}]{turner2010}
{Turner}, N.~J., {Carballido}, A., \& {Sano}, T. 2010, \apj, 708, 188

\bibitem[{{van Boekel} {et~al.}(2010){van Boekel}, {Juh{\'a}sz}, {Henning},
  {K{\"o}hler}, {Ratzka}, {Herbst}, {Bouwman}, \& {Kley}}]{vanboekel2010}
{van Boekel}, R., {Juh{\'a}sz}, A., {Henning}, T., {K{\"o}hler}, R., {Ratzka},
  T., {Herbst}, T., {Bouwman}, J., \& {Kley}, W. 2010, \aap, 517, A16

\bibitem[{{van Boekel} {et~al.}(2005){van Boekel}, {Min}, {Waters}, {de Koter},
  {Dominik}, {van den Ancker}, \& {Bouwman}}]{vanboekel2005}
{van Boekel}, R., {Min}, M., {Waters}, L.~B.~F.~M., {de Koter}, A., {Dominik},
  C., {van den Ancker}, M.~E., \& {Bouwman}, J. 2005, \aap, 437, 189

\bibitem[{{van Boekel} {et~al.}(2004){van Boekel}, {Waters}, {Dominik},
  {Dullemond}, {Tielens}, \& {de Koter}}]{vanboekel2004}
{van Boekel}, R., {Waters}, L.~B.~F.~M., {Dominik}, C., {Dullemond}, C.~P.,
  {Tielens}, A.~G.~G.~M., \& {de Koter}, A. 2004, \aap, 418, 177

\bibitem[{{van den Bergh} \& {Herbst}(1975)}]{vandenbergh1975}
{van den Bergh}, S., \& {Herbst}, W. 1975, \aj, 80, 208

\bibitem[{{van Dishoeck}(2004)}]{vandishoeck2004}
{van Dishoeck}, E.~F. 2004, \araa, 42, 119

\bibitem[{{Verhoeff} {et~al.}(2011){Verhoeff}, {Min}, {Pantin}, {Waters},
  {Tielens}, {Honda}, {Fujiwara}, {Bouwman}, {van Boekel}, {Dougherty}, {de
  Koter}, {Dominik}, \& {Mulders}}]{verhoeff2011}
{Verhoeff}, A.~P., {et~al.} 2011, \aap, 528, A91

\bibitem[{{Vinkovi{\'c}} {et~al.}(2006){Vinkovi{\'c}}, {Ivezi{\'c}},
  {Jurki{\'c}}, \& {Elitzur}}]{vinkovic2006}
{Vinkovi{\'c}}, D., {Ivezi{\'c}}, {\v Z}., {Jurki{\'c}}, T., \& {Elitzur}, M.
  2006, \apj, 636, 348

\bibitem[{{Vinkovi{\'c}} \& {Jurki{\'c}}(2007)}]{vinkovic2007}
{Vinkovi{\'c}}, D., \& {Jurki{\'c}}, T. 2007, \apj, 658, 462

\bibitem[{{Wang} \& {Looney}(2007)}]{wang2007}
{Wang}, S., \& {Looney}, L.~W. 2007, \apj, 659, 1360

\bibitem[{{Watson} {et~al.}(2009){Watson}, {Leisenring}, {Furlan}, {Bohac},
  {Sargent}, {Forrest}, {Calvet}, {Hartmann}, {Nordhaus}, {Green}, {Kim},
  {Sloan}, {Chen}, {Keller}, {d'Alessio}, {Najita}, {Uchida}, \&
  {Houck}}]{watson2009}
{Watson}, D.~M., {et~al.} 2009, \apjs, 180, 84

\bibitem[{{Weinberger} {et~al.}(1999){Weinberger}, {Becklin}, {Schneider},
  {Smith}, {Lowrance}, {Silverstone}, {Zuckerman}, \&
  {Terrile}}]{weinberger1999}
{Weinberger}, A.~J., {Becklin}, E.~E., {Schneider}, G., {Smith}, B.~A.,
  {Lowrance}, P.~J., {Silverstone}, M.~D., {Zuckerman}, B., \& {Terrile}, R.~J.
  1999, \apjl, 525, L53

\bibitem[{{Weinberger} {et~al.}(2000){Weinberger}, {Rich}, {Becklin},
  {Zuckerman}, \& {Matthews}}]{weinberger2000}
{Weinberger}, A.~J., {Rich}, R.~M., {Becklin}, E.~E., {Zuckerman}, B., \&
  {Matthews}, K. 2000, \apj, 544, 937

\bibitem[{{Wheelwright} {et~al.}(2010){Wheelwright}, {Oudmaijer}, \&
  {Goodwin}}]{wheelwright2010}
{Wheelwright}, H.~E., {Oudmaijer}, R.~D., \& {Goodwin}, S.~P. 2010, \mnras,
  401, 1199

\bibitem[{{Wisniewski} {et~al.}(2008){Wisniewski}, {Clampin}, {Grady},
  {Ardila}, {Ford}, {Golimowski}, {Illingworth}, \& {Krist}}]{wisniewski2008}
{Wisniewski}, J.~P., {Clampin}, M., {Grady}, C.~A., {Ardila}, D.~R., {Ford},
  H.~C., {Golimowski}, D.~A., {Illingworth}, G.~D., \& {Krist}, J.~E. 2008,
  \apj, 682, 548

\bibitem[{{Wooden} {et~al.}(2000){Wooden}, {Bell}, {Harker}, \&
  {Woodward}}]{wooden2000}
{Wooden}, D.~H., {Bell}, K.~R., {Harker}, D.~E., \& {Woodward}, C.~E. 2000, in
  Bulletin of the American Astronomical Society, Vol.~32, American Astronomical
  Society Meeting Abstracts, 1482

\bibitem[{{Yudin} {et~al.}(1999){Yudin}, {Clarke}, \& {Smith}}]{yudin1999}
{Yudin}, R.~V., {Clarke}, D., \& {Smith}, R.~A. 1999, \aap, 345, 547

\bibitem[{{Zinnecker} {et~al.}(1991){Zinnecker}, {Christou}, {Ridgway}, \&
  {Probst}}]{zinnecker1991}
{Zinnecker}, H., {Christou}, J.~C., {Ridgway}, S.~T., \& {Probst}, R.~G. 1991,
  in Astronomical Society of the Pacific Conference Series, Vol.~14,
  Astronomical Society of the Pacific Conference Series, ed. {R.~Elston},
  270--272

\bibitem[{{Zuckerman} {et~al.}(1995){Zuckerman}, {Forveille}, \&
  {Kastner}}]{zuckerman1995}
{Zuckerman}, B., {Forveille}, T., \& {Kastner}, J.~H. 1995, \nat, 373, 494

\end{thebibliography}

\clearpage

\begin{deluxetable}{l@{}c@{}c@{}c@{}cc@{}l@{}c@{}c@{}c@{}r@{}c@{}c@{}c@{}c@{}c@{}c@{}c@{}c}
\tablenum{1}
\tabletypesize{\tiny}
\tablewidth{0pt}
\tablecaption{Catalog of spectral observations of young stellar objects
  obtained with ISO and Spitzer\label{Table1}}
\rotate
\tablehead{
               &                & & \multicolumn{2}{c}{Coordinates} & & \multicolumn{3}{c}{ISO} & \colhead{} & \multicolumn{3}{c}{Spitzer} & & & & \\
\cline{4-5} \cline{7-9} \cline{11-13} & \\
\colhead{Name} & \colhead{Mass} & \colhead{Sp.type} & \colhead{RA(2000)} & \colhead{Dec(2000)} & &
\colhead{TDT} & \colhead{Corr.} & \colhead{Obs. date} & &
\colhead{AOR} & \colhead{Corr.} & \colhead{Obs. date} & &
\colhead{Si-O} & \colhead{Ice} & \colhead{PAH} & \colhead{Type} & \colhead{Var.}\\
\colhead{(1)} & \colhead{(2)} & \colhead{(3)} & \colhead{(4)} & \colhead{(5)} & & \colhead{(6)} & \colhead{(7)} & \colhead{(8)} & & \colhead{(9)}
& \colhead{(10)} & \colhead{(11)} & & \colhead{(12)} & \colhead{(13)} & \colhead{(14)} & \colhead{(15)} & \colhead{(16)}
}
\startdata
VX Cas            & I & A0              & 00 31 30.69 & $+$61 58 51.0 & & 58704023 & BO    & 25-Jul-1997 & &          &   &             & & em  & ... & em  & sil. em.  & ... \\
IRAS 03260+3111   & I & B5+F2           & 03 29 10.38 & $+$31 21 59.2 & & 65201755 & BO    & 29-Aug-1997 & &          &   &             & & ... & ... & em  & PAH dom.  & ... \\
BARN 5 IRS 1      & L & \dots           & 03 47 41.60 & $+$32 51 43.8 & & 63103703 & B!O   & 08-Aug-1997 & &  5635328 &   & 03-Feb-2004 & & abs & abs & ... & sil. abs. & yes \\
LDN 1489 IRS$^{\rm a}$ & L & \dots      & 04 04 43.07 & $+$26 18 56.4 & & 81401708 & O     & 07-Feb-1998 & &  3528960 &   & 07-Feb-2004 & & abs & abs & ... & sil. abs. & ... \\
T Tau             & L & G5              & 04 21 59.43 & $+$19 32 06.4 & & 67901255 & O     & 25-Sep-1997 & &  3530240 &   & 08-Feb-2004 & & abs & abs & ... & sil. abs. & yes \\
DG Tau            & L & G               & 04 27 04.70 & $+$26 06 16.3 & & 64501604 & B     & 22-Aug-1997 & &  3530496 & O & 02-Mar-2004 & & ... & ... & ... & ...       & yes \\
                  &   &                 &             &               & &          &       &             & & 15115520 &   & 22-Mar-2006 & &     &     &     &           &     \\
Haro 6-10         & L & K3              & 04 29 23.73 & $+$24 33 00.3 & & 65300605 & B     & 30-Aug-1997 & &  3531008 &   & 02-Mar-2004 & & abs & abs & ... & sil. abs. & yes \\
                  &   &                 &             &               & & 66901306 & O     & 15-Sep-1997 & & 25679872 &   & 14-Oct-2008 & &     &     &     &           &     \\
                  &   &                 &             &               & &          &       &             & & 31618304 & O & 04-Apr-2009 & &     &     &     &           &     \\
LDN 1551 IRS 5    & L & \dots           & 04 31 34.08 & $+$18 08 04.9 & & 64201706 & MB!O  & 19-Aug-1997 & &  3531776 &   & 04-Mar-2004 & & abs & abs & ... & sil. abs. & yes \\
HL Tau            & L & K9              & 04 31 38.44 & $+$18 13 57.7 & & 65602507 & MB    & 02-Sep-1997 & &  3531776 &   & 04-Mar-2004 & & abs & abs & ... & sil. abs. & yes \\
UZ Tau            & L & M2              & 04 32 43.04 & $+$25 52 31.1 & & 68401434 & O     & 30-Sep-1997 & &  3531264 &   & 28-Feb-2004 & & em  & ... & ... & sil. em.  & yes \\
                  &   &                 &             &               & & 83300749 & O     & 25-Feb-1998 & & 12288000 &   & 14-Feb-2005 & &     &     &     &           &     \\
VY Tau            & L & M0              & 04 39 17.41 & $+$22 47 53.4 & & 68101239 &       & 27-Sep-1997 & &  3547904 &   & 08-Feb-2004 & & em  & ... & ... & sil. em.  & yes \\
                  &   &                 &             &               & & 83300854 &       & 25-Feb-1998 & & 11565824 & O & 18-Feb-2005 & &     &     &     &           &     \\
                  &   &                 &             &               & & 86100859 &       & 25-Mar-1998 & &          &   &             & &     &     &     &           &     \\
DR Tau            & L & K4              & 04 47 06.21 & $+$16 58 42.8 & & 67901329 &       & 25-Sep-1997 & &  3533568 &   & 08-Feb-2004 & & em  & ... & ... & sil. em.  & yes \\ 
                  &   &                 &             &               & & 83300944 &       & 25-Feb-1998 & & 12287744 &   & 18-Feb-2005 & &     &     &     &           &     \\
UX Ori            & I & A3              & 05 04 29.99 & $-$03 47 14.3 & & 85801453 & B     & 22-Mar-1998 & &          &   &             & & em  & ... & ... & sil. em.  & ... \\
HD 34700          & L & G0              & 05 19 41.40 & $+$05 38 42.9 & & 63602294 & B     & 13-Aug-1997 & &  3584768 & O & 02-Oct-2004 & & ... & ... & em  & PAH dom.  & no  \\
BF Ori            & I & A5              & 05 37 13.27 & $-$06 35 00.5 & & 70101958 & B     & 17-Oct-1997 & &  5638144 &   & 03-Oct-2004 & & em  & ... & ... & sil. em.  & yes \\
                  &   &                 &             &               & &          &       &             & & 18835968 &   & 09-Mar-2007 & &     &     &     &           &     \\
RR Tau            & I & A2              & 05 39 30.52 & $+$26 22 27.0 & & 67000863 & BO    & 16-Sep-1997 & &  5638400 &   & 28-Sep-2004 & & ... & ... & em  & PAH dom.  & yes \\
                  &   &                 &             &               & & 86603163 & BO    & 30-Mar-1998 & &          &   &             & &     &     &     &           &     \\
Reipurth 50 N IRS 1 & L & \dots         & 05 40 27.45 & $-$07 27 30.1 & & 69703804 & BO    & 13-Oct-1997 & &          &   &             & & abs & abs & ... & sil. abs. & ... \\
SX Cha            & L & M0.5            & 10 55 59.73 & $-$77 24 39.9 & & 16600204 & O     & 01-May-1996 & & 12697345 &   & 26-May-2005 & & em  & ... & ... & sil. em.  & ... \\
CR Cha            & L & K4              & 10 59 06.99 & $-$77 01 40.4 & & 16600317 & O     & 01-May-1996 & & 12697345 &   & 26-May-2005 & & em  & ... & ... & sil. em.  & yes \\
                  &   &                 &             &               & & 62501703 & MBO   & 02-Aug-1997 & & 26143744 &   & 01-Jun-2008 & &     &     &     &           &     \\
                  &   &                 &             &               & &          &       &             & & 27186944 &   & 08-Jun-2008 & &     &     &     &           &     \\
HD 95881          & I & A1/A2           & 11 01 57.64 & $-$71 30 48.4 & & 10400919 & BO    & 29-Feb-1996 & & 11004928 &   & 15-Feb-2005 & & em  & ... & em  & sil. em.  & no  \\
CT Cha            & L & K7              & 11 04 09.09 & $-$76 27 19.4 & & 16600124 & O     & 01-May-1996 & & 12697345 &   & 26-May-2005 & & em  & ... & ... & sil. em.  & yes \\
VW Cha            & L & K8              & 11 08 01.49 & $-$77 42 28.9 & & 16700234 & O     & 02-May-1996 & & 12696832 &   & 12-Jul-2005 & & em  & ... & ... & sil. em.  & yes \\
CU Cha            & I & A0              & 11 08 03.30 & $-$77 39 17.4 & & 07900309 & MBO   & 04-Feb-1996 & & 12697088 &   & 07-Mar-2006 & & ... & ... & em  & PAH dom.  & yes \\
                  &   &                 &             &               & & 14101580 & BO    & 06-Apr-1996 & &          &   &             & &     &     &     &           &     \\
                  &   &                 &             &               & & 62501510 & MBO   & 02-Aug-1997 & &          &   &             & &     &     &     &           &     \\
Glass I           & L & K7              & 11 08 15.10 & $-$77 33 53.2 & & 07900410 & MO    & 04-Feb-1996 & & 12697088 &   & 07-Mar-2006 & & em  & ... & ... & sil. em.  & yes \\
                  &   &                 &             &               & & 16600138 & BO    & 01-May-1996 & & 25680640 &   & 16-Aug-2008 & &     &     &     &           &     \\
                  &   &                 &             &               & & 58401511 & BO    & 22-Jun-1997 & &          &   &             & &     &     &     &           &     \\
Ced 111 IRS 5$^{\rm b}$ & L & M5        & 11 08 39.43 & $-$77 43 52.1 & & 07900211 & B!O   & 04-Feb-1996 & & 12697345 & O!& 26-May-2005 & & ... & abs & ... & sil. abs. & no \\
                  &   &                 &             &               & & 62501412 & M!B!O & 02-Aug-1997 & &          &   &             & &     &     &     &           &     \\
VZ Cha            & L & K7              & 11 09 23.79 & $-$76 23 20.8 & & 27101145 & O     & 14-Aug-1996 & & 12696832 &   & 12-Jul-2005 & & em  & ... & ... & sil. em.  & yes \\
HD 97300          & I & B9              & 11 09 50.03 & $-$76 36 47.7 & & 07901912 & MBO   & 04-Feb-1996 & & 12697088 &   & 07-Mar-2006 & & ... & ... & em  & PAH dom.  & no  \\
                  &   &                 &             &               & & 62501316 & MBO   & 02-Aug-1997 & &          &   &             & &     &     &     &           &     \\
Ced 112 IRS 4     & L & K5              & 11 09 53.41 & $-$76 34 25.5 & & 62501217 & MBO   & 02-Aug-1997 & & 12697088 &   & 07-Mar-2006 & & ... & ... & ... & ...       & yes \\
WX Cha            & L & M1.25           & 11 09 58.74 & $-$77 37 08.9 & & 16600749 & O     & 01-May-1996 & & 12696576 &   & 10-Jul-2005 & & em  & ... & ... & sil. em.  & yes \\
WW Cha            & L & K5              & 11 10 00.11 & $-$76 34 57.9 & & 27101153 & O     & 14-Aug-1996 & & 12697088 &   & 07-Mar-2006 & & em  & ... & ... & sil. em.  & yes \\
                  &   &                 &             &               & &          &       &             & & 26142976 &   & 01-Jun-2008 & &     &     &     &           &     \\
                  &   &                 &             &               & &          &       &             & & 27185408 & O & 08-Jun-2008 & &     &     &     &           &     \\
XX Cha            & L & M2              & 11 11 39.66 & $-$76 20 15.3 & & 16600664 & O     & 01-May-1996 & & 12696576 &   & 10-Jul-2005 & & em  & ... & ... & sil. em.  & yes \\
CV Cha            & L & K0              & 11 12 27.72 & $-$76 44 22.3 & & 58401637 & O     & 22-Jun-1997 & & 12697088 &   & 07-Mar-2006 & & em  & ... & ... & sil. em.  & yes \\
                  &   &                 &             &               & & 60601421 & BO    & 14-Jul-1997 & &          &   &             & &     &     &     &           &     \\
HD 98800          & L & K4              & 11 22 05.30 & $-$24 46 39.4 & & 24001017 & B     & 14-Jul-1996 & &  3571968 &   & 25-Jun-2004 & & em  & ... & ... & sil. em.  & yes \\
                  &   &                 &             &               & &          &       &             & &  3571969 &   & 03-Jan-2005 & &     &     &     &           &     \\
HD 100453         & I & A9              & 11 33 05.59 & $-$54 19 28.5 & & 26000131 & B     & 02-Aug-1996 & &  3578880 &   & 06-Jun-2004 & & ... & ... & em  & PAH dom.  & no  \\
HD 100546         & I & B9              & 11 33 25.43 & $-$70 11 41.3 & & 10400537 & BO    & 29-Feb-1996 & &          &   &             & & em  & ... & em  & sil. em.  & ... \\
HD 104237         & I & A               & 12 00 05.12 & $-$78 11 34.7 & & 10400325 & BO    & 29-Feb-1996 & & 12677632 & O & 30-May-2005 & & em  & ... & ... & sil. em.  & yes \\
                  &   &                 &             &               & & 23300625 & BO    & 07-Jul-1996 & &          &   &             & &     &     &     &           &     \\
                  &   &                 &             &               & & 53300118 & MB    & 02-May-1997 & &          &   &             & &     &     &     &           &     \\
DK Cha            & I & F0              & 12 53 17.22 & $-$77 07 10.6 & & 07901717 & BO    & 04-Feb-1996 & & 12679168 &   & 26-May-2006 & & abs & abs & ... & sil. abs. & ... \\
HD 135344 B       & I & F8              & 15 15 48.45 & $-$37 09 16.0 & & 10401742 & B     & 29-Feb-1996 & &  3580672 &   & 08-Aug-2004 & & ... & ... & em  & PAH dom.  & yes \\
                  &   &                 &             &               & & 10401876 & B     & 29-Feb-1996 & &          &   &             & &     &     &     &           &     \\
HD 139614         & I & A7              & 15 40 46.39 & $-$42 29 53.6 & & 10402322 & BO    & 29-Feb-1996 & &  3580928 &   & 08-Aug-2004 & & em  & ... & em  & sil. em.  & yes \\
HD 141569         & I & A0              & 15 49 57.76 & $-$03 55 16.2 & & 62701662 & B     & 04-Aug-1997 & &  3560960 &   & 03-Mar-2004 & & ... & ... & em  & PAH dom.  & no  \\
HD 142666         & I & A8              & 15 56 40.02 & $-$22 01 40.0 & & 10402847 & MBO   & 29-Feb-1996 & &  3586816 &   & 08-Aug-2004 & & em  & ... & em  & sil. em.  & yes \\
HD 142527         & I & F6              & 15 56 41.89 & $-$42 19 23.3 & & 10402547 & B     & 29-Feb-1996 & & 11005696 &   & 18-Mar-2005 & & em  & ... & em  & sil. em.  & no  \\
EX Lup            & L & M0              & 16 03 05.49 & $-$40 18 25.4 & & 44700104 &       & 05-Feb-1997 & &  5645056 &   & 30-Aug-2004 & & em  & ... & ... & sil. em.  & yes \\
                  &   &                 &             &               & & 48801209 &       & 18-Mar-1997 & & 11570688 &   & 18-Mar-2005 & &     &     &     &           &     \\
                  &   &                 &             &               & & 64701714 & O     & 24-Aug-1997 & &          &   &             & &     &     &     &           &     \\
                  &   &                 &             &               & & 67403725 & O     & 19-Sep-1997 & &          &   &             & &     &     &     &           &     \\
HD 144432         & I & A9/F0           & 16 06 57.95 & $-$27 43 09.4 & & 10402662 & BO    & 29-Feb-1996 & &  3587072 &   & 08-Aug-2004 & & em  & ... & ... & sil. em.  & yes \\
HR 5999           & I & A7              & 16 08 34.28 & $-$39 06 18.2 & & 28901748 & MB    & 01-Sep-1996 & &          &   &             & & em  & ... & ... & sil. em.  & ... \\
WL 16             & I & \dots           & 16 27 02.34 & $-$24 37 27.2 & & 08100620 & BO    & 06-Feb-1996 & &          &   &             & & ... & ... & em  & PAH dom.  & ... \\
WL 6              & L & \dots           & 16 27 21.80 & $-$24 29 53.4 & & 08101921 & BO    & 06-Feb-1996 & & 12698880 &   & 16-Apr-2006 & & abs & abs & ... & sil. abs. & yes \\
MWC 863           & I & A1              & 16 40 17.92 & $-$23 53 45.2 & & 64102335 &       & 18-Aug-1997 & &          &   &             & & em  & ... & ... & sil. em.  & ... \\
AK Sco            & I & F5              & 16 54 44.85 & $-$36 53 18.5 & & 64402829 & O     & 21-Aug-1997 & & 12700160 &   & 16-Apr-2005 & & em  & ... & ... & sil. em.  & no  \\
MWC 865           & I & B               & 16 59 06.77 & $-$42 42 08.4 & & 28900460 & BO!   & 31-Aug-1996 & &          &   &             & & ... & ... & em  & PAH dom.  & ... \\
51 Oph            & I & A0              & 17 31 24.97 & $-$23 57 45.3 & & 10301104 & MBO   & 28-Feb-1996 & &  3582464 &   & 24-Mar-2004 & & em  & ... & ... & sil. em.  & yes \\
HD 163296         & I & A1              & 17 56 21.29 & $-$21 57 21.8 & & 32901192 & BO    & 10-Oct-1996 & &          &   &             & & em  & ... & ... & sil. em.  & ... \\
HD 169142         & I & B9              & 18 24 29.79 & $-$29 46 49.2 & & 13601437 & MBO   & 01-Apr-1996 & &  3587584 &   & 26-Mar-2004 & & ... & ... & em  & PAH dom.  & no  \\
VV Ser            & I & B6              & 18 28 47.86 & $+$00 08 39.8 & & 47800913 & B     & 08-Mar-1997 & &  5651200 &   & 01-Sep-2004 & & em  & ... & em  & sil. em.  & yes \\
OO Ser$^{\rm a}$  & L & \dots           & 18 29 49.13 & $+$01 16 20.6 & & 29000213 &       & 01-Sep-1996 & &          &   &             & & abs & abs & ... & sil. abs. & yes \\
                  &   &                 &             &               & & 34300318 &       & 24-Oct-1996 & &          &   &             & &     &     &     &           &     \\
                  &   &                 &             &               & & 47800223 & O     & 08-Mar-1997 & &          &   &             & &     &     &     &           &     \\
                  &   &                 &             &               & & 51301129 & O     & 12-Apr-1997 & &          &   &             & &     &     &     &           &     \\
                  &   &                 &             &               & & 67601734 &       & 22-Sep-1997 & &          &   &             & &     &     &     &           &     \\
CK 1              & L & M4              & 18 29 57.74 & $+$01 14 05.7 & & 10803126 & MBO   & 04-Mar-1996 & &  9828352 &   & 01-Sep-2004 & & ... & abs & ... & ...       & no  \\
$[$SVS76$]$ Ser 4$^{\rm c}$ & L & \dots & 18 29 57.89 & $+$01 12 46.3 & & 10803027 & MB!O  & 04-Mar-1996 & &  9407232 &   & 27-Mar-2004 & & abs & abs & ... & sil. abs. & ... \\
CK 2              & L & K5-M0           & 18 30 00.62 & $+$01 15 20.1 & & 10803228 & M!B!O & 04-Mar-1996 & & 11828224 &   & 01-Sep-2004 & & abs & abs & ... & sil. abs. & ... \\
S CrA$^{\rm d}$   & L & K6              & 19 01 08.70 & $-$36 57 19.8 & & 52301333 & O     & 22-Apr-1997 & & 11197952 &   & 20-Apr-2005 & & em  & ... & ... & sil. em.  & yes \\
                  &   &                 &             &               & &          &       &             & & 14920960 &   & 20-Oct-2005 & &     &     &     &           &     \\
HH 100 IRS        & L & K7              & 19 01 50.68 & $-$36 58 09.6 & & 11501029 & BO    & 11-Mar-1996 & &          &   &             & & abs & abs & ... & sil. abs. & yes \\
                  &   &                 &             &               & & 70400629 & BO    & 19-Oct-1997 & &          &   &             & &     &     &     &           &     \\
R CrA             & I & B5              & 19 01 53.68 & $-$36 57 08.2 & & 11501230 & BO    & 11-Mar-1996 & &          &   &             & & em  & ... & ... & sil. em.  & ... \\
T CrA             & I & F0              & 19 01 58.78 & $-$36 57 49.9 & & 14100562 & BO    & 06-Apr-1996 & &          &   &             & & em  & ... & ... & sil. em.  & ... \\
VV CrA            & L & K7              & 19 03 06.75 & $-$37 12 49.5 & & 52301015 & O     & 22-Apr-1997 & & 25680896 &   & 17-Nov-2008 & & abs & abs & ... & sil. abs. & ... \\
WW Vul            & I & A3              & 19 25 58.74 & $+$21 12 31.4 & & 17600465 & MBO!  & 11-May-1996 & & 16828672 &   & 20-Jun-2006 & & em  & ... & ... & sil. em.  & yes \\
                  &   &                 &             &               & & 51300108 & BO    & 12-Apr-1997 & &          &   &             & &     &     &     &           &     \\
BD +40 4124       & I & B3              & 20 20 28.25 & $+$41 21 51.5 & & 15900568 & B     & 24-Apr-1996 & &          &   &             & & ... & ... & em  & PAH dom.  & ... \\
LkHa 224          & I & F9              & 20 20 29.36 & $+$41 21 28.5 & & 14201271 & BO    & 07-Apr-1996 & &          &   &             & & ... & ... & em  & PAH dom.  & ... \\
PV Cep            & L & G0-K0$^{\rm e}$ & 20 45 53.94 & $+$67 57 38.7 & & 14302274 & BO    & 08-Apr-1996 & & 12287488 &   & 23-Oct-2004 & & abs & abs & ... & sil. abs. & yes \\
SV Cep            & I & A0              & 22 21 33.20 & $+$73 40 27.1 & & 56201203 & B     & 31-May-1997 & &          &   &             & & em  & ... & ... & sil. em.  & ... \\
\enddata
\tablecomments{For a detailed description of the columns of this table, see Sec.~\ref{sec:catalog}.}
\tablenotetext{(a)}{ISO observations for LDN\,1489\,IRS and OO\,Ser were taken in mapping mode.}
\tablenotetext{(b)}{Coordinates for Ced\,111\,IRS\,5 come from a Spitzer/IRAC 3.6$\,\mu$m image.}
\tablenotetext{(c)}{2MASS coordinates for the member of the SVS4 cluster that is brightest at $\ge$2.2$\,\mu$m, SVS4/9 ([EC92] 95).}
\tablenotetext{(d)}{Coordinates for S\,CrA is from \citet{prato2003}.}
\tablenotetext{(e)}{SIMBAD gives a spectral type of A5, but the star is probably a low-mass object of G0-K0 \citep{magakian2001}.}
\end{deluxetable}

\newpage

\begin{figure*}
\epsscale{0.95}
\plotone{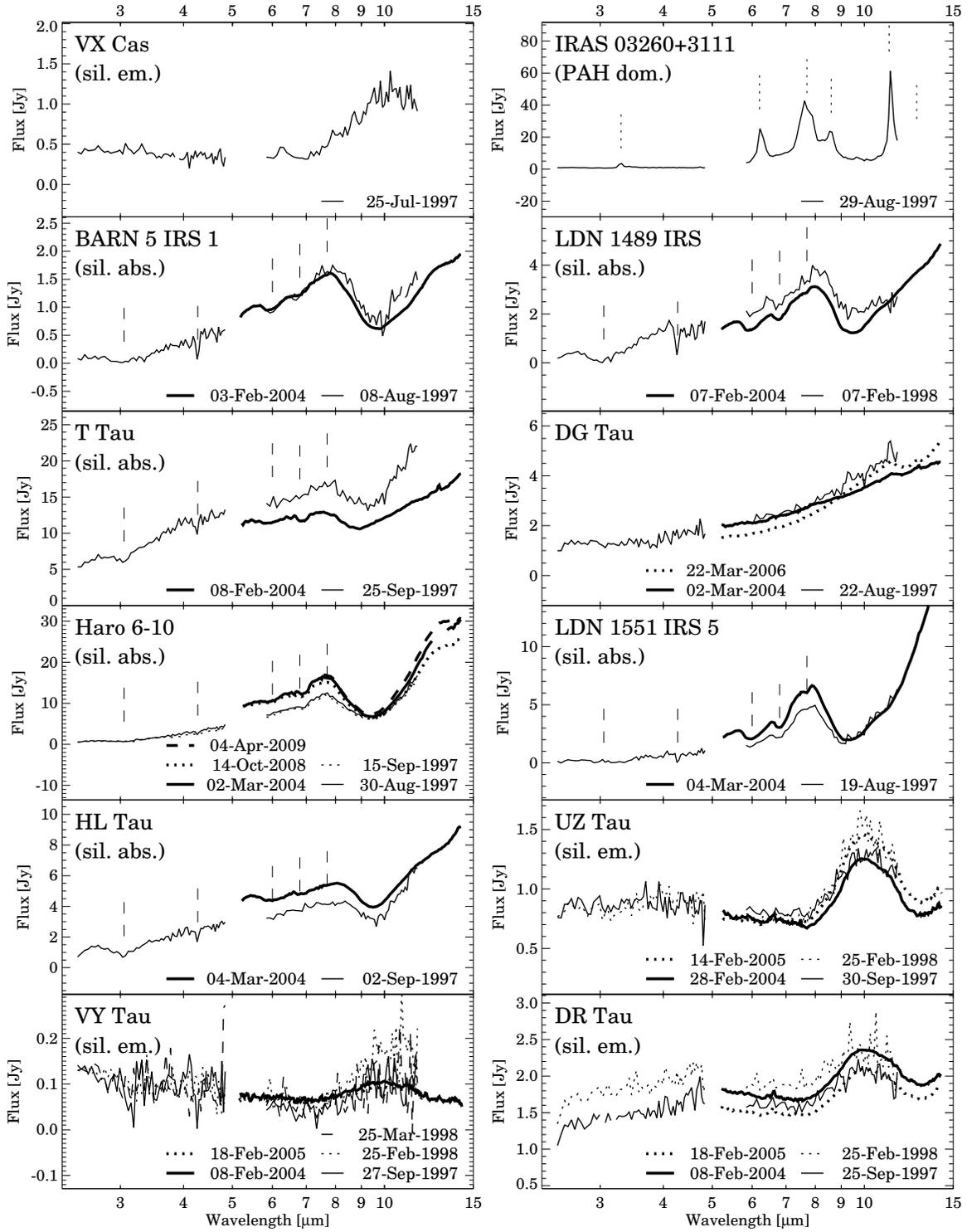}
\caption{ISOPHOT-S and Spitzer/IRS spectra for our object. Where
  clearly detected, vertical dashed lines mark the wavelengths of
  molecular ice bands and vertical dotted lines indicate the PAH
  features.
\label{map_var}}
\end{figure*}

\setcounter{figure}{1}
\begin{figure*}
\epsscale{1.0}
\plotone{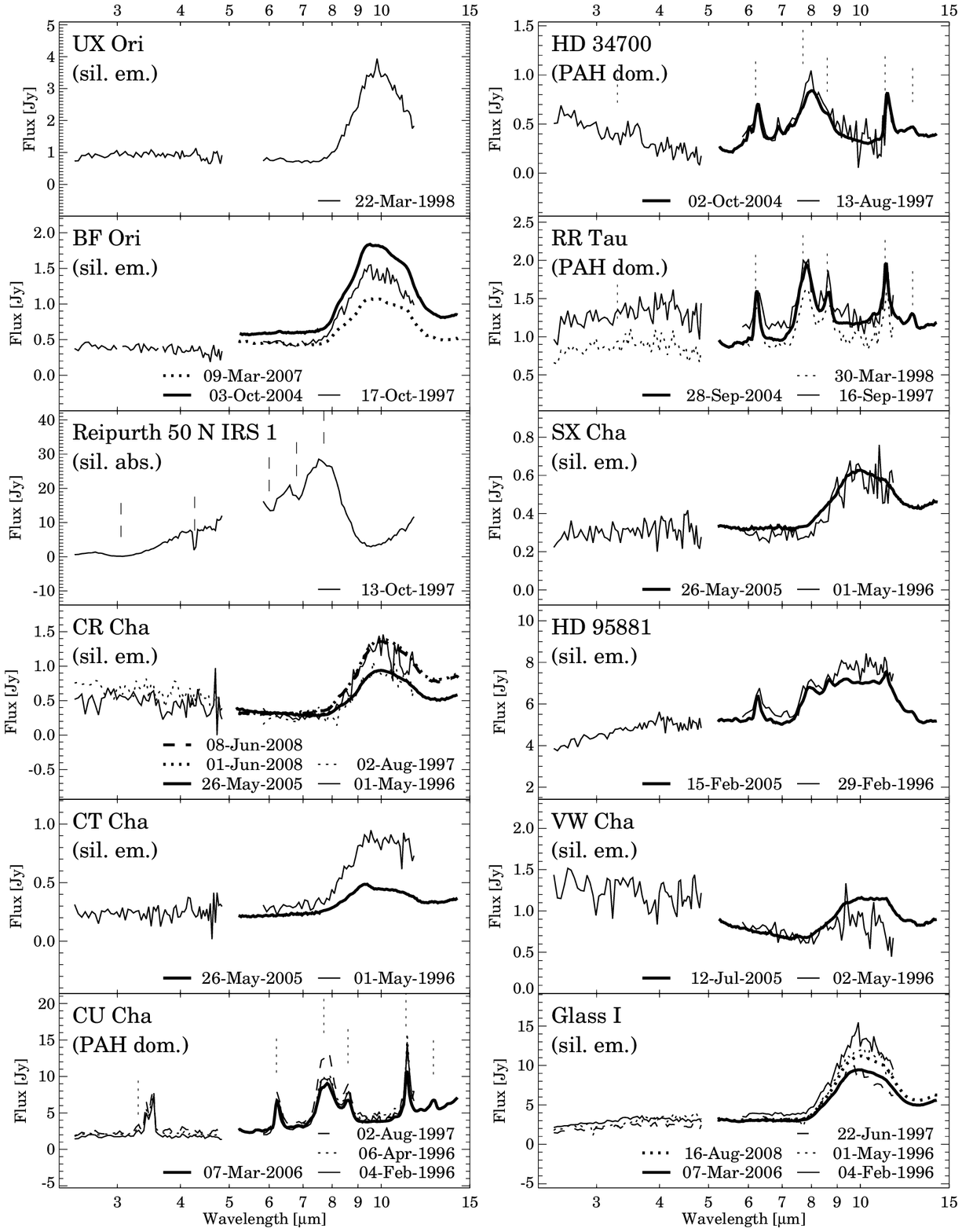}
\caption{(Continued)}
\end{figure*}

\setcounter{figure}{1}
\begin{figure*}
\epsscale{1.0}
\plotone{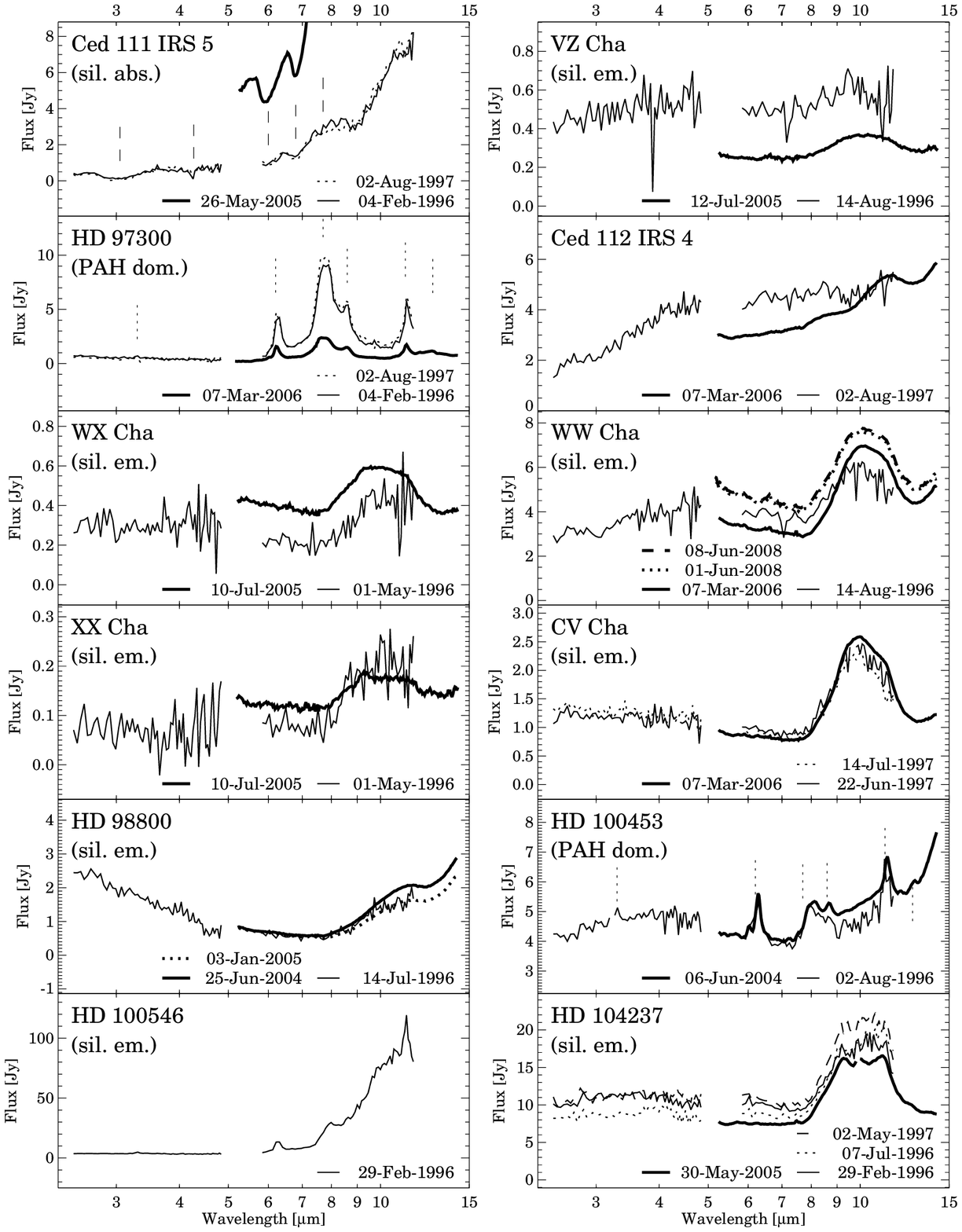}
\caption{(Continued)}
\end{figure*}

\setcounter{figure}{1}
\begin{figure*}
\epsscale{1.0}
\plotone{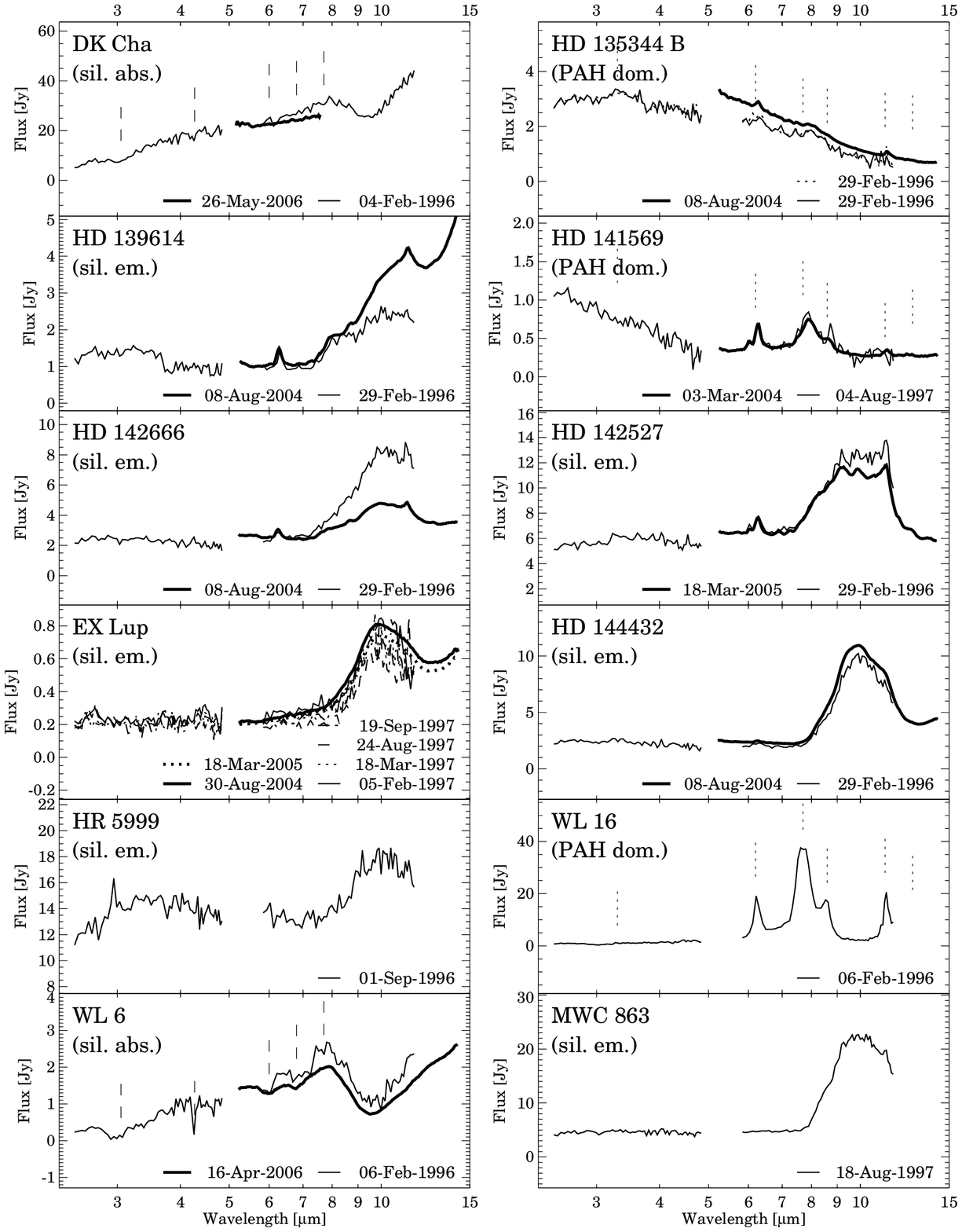}
\caption{(Continued)}
\end{figure*}

\setcounter{figure}{1}
\begin{figure*}
\epsscale{1.0}
\plotone{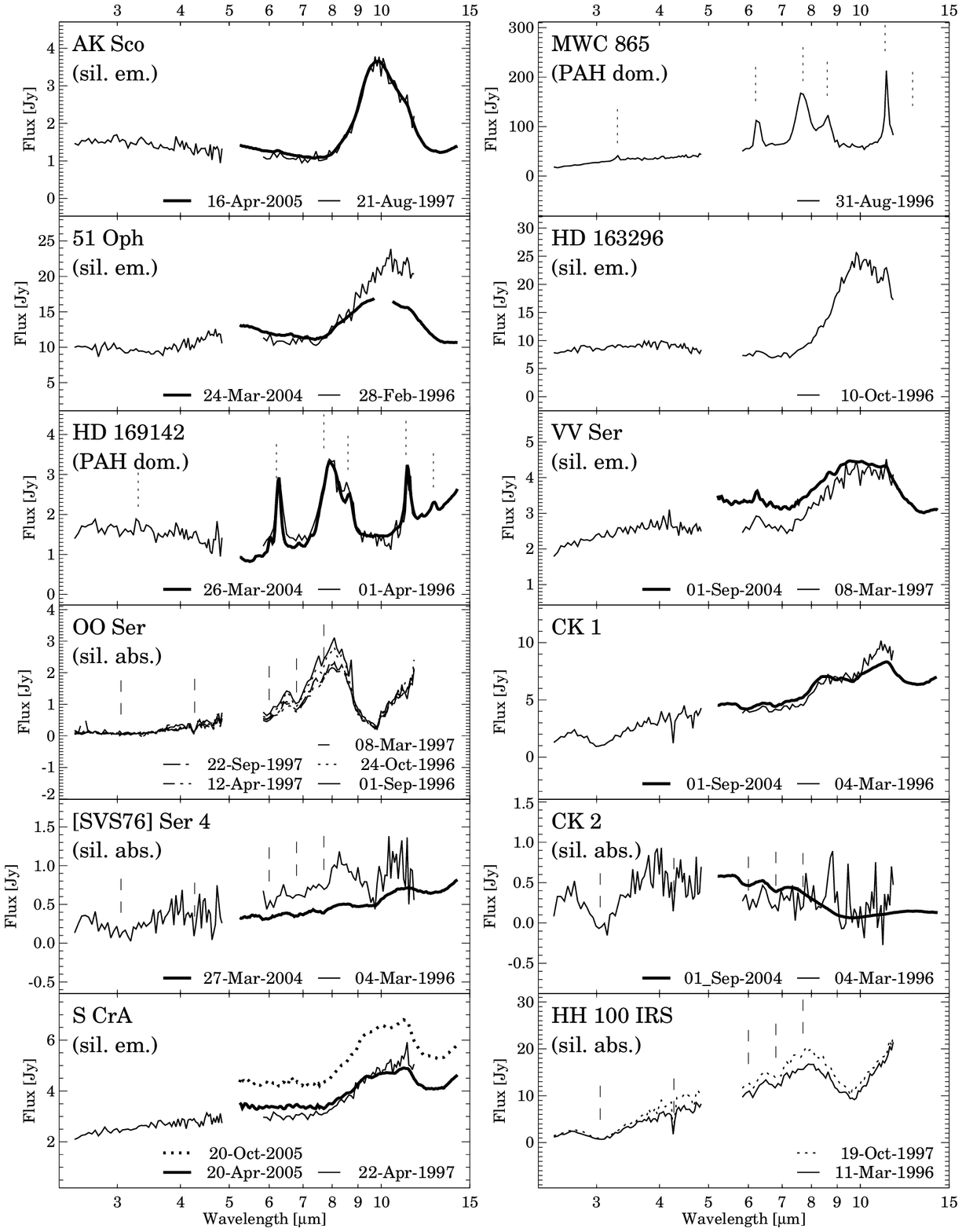}
\caption{(Continued)}
\end{figure*}

\setcounter{figure}{1}
\begin{figure*}
\epsscale{1.0}
\plotone{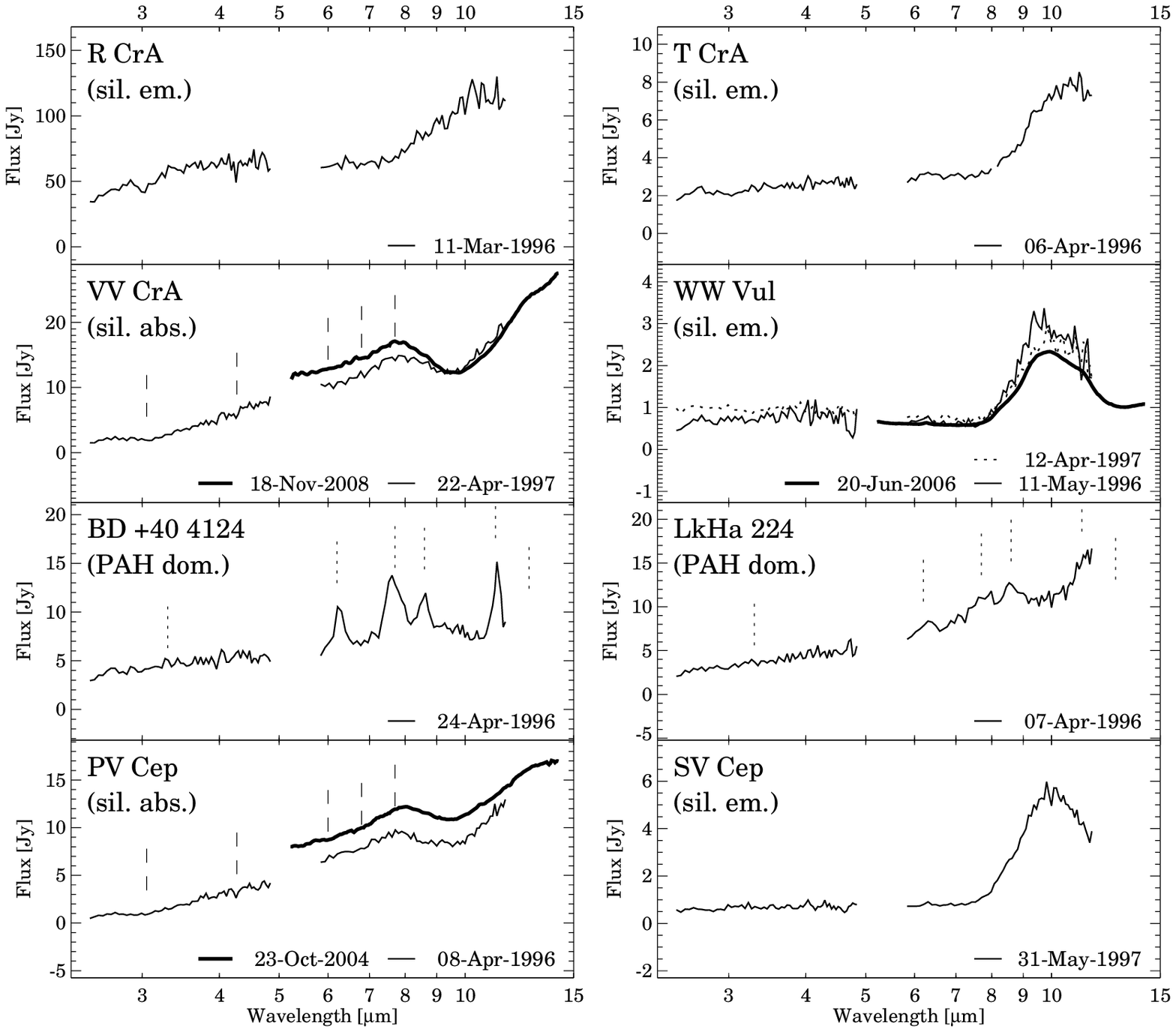}
\caption{(Continued)}
\end{figure*}

\clearpage

\begin{figure*}
\epsscale{0.95}
\plotone{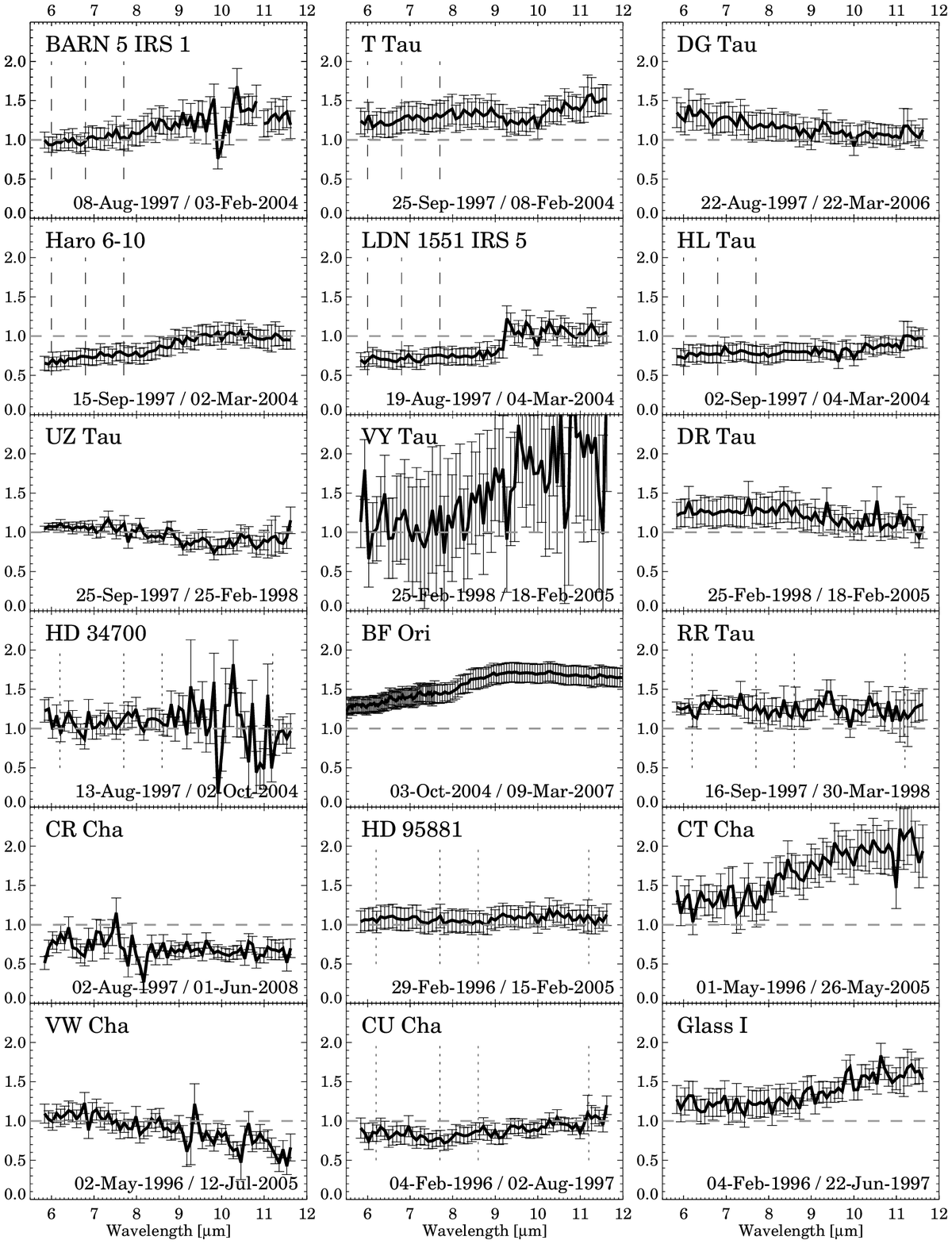}
\caption{Ratios of spectra obtained at different epochs. Where several
  spectra existed, we took the two most extremes. Error bars represent
  the quadratic sum of the errors of each spectrum.
\label{compare_var}}
\end{figure*}

\setcounter{figure}{2}
\begin{figure*}
\epsscale{1.0}
\plotone{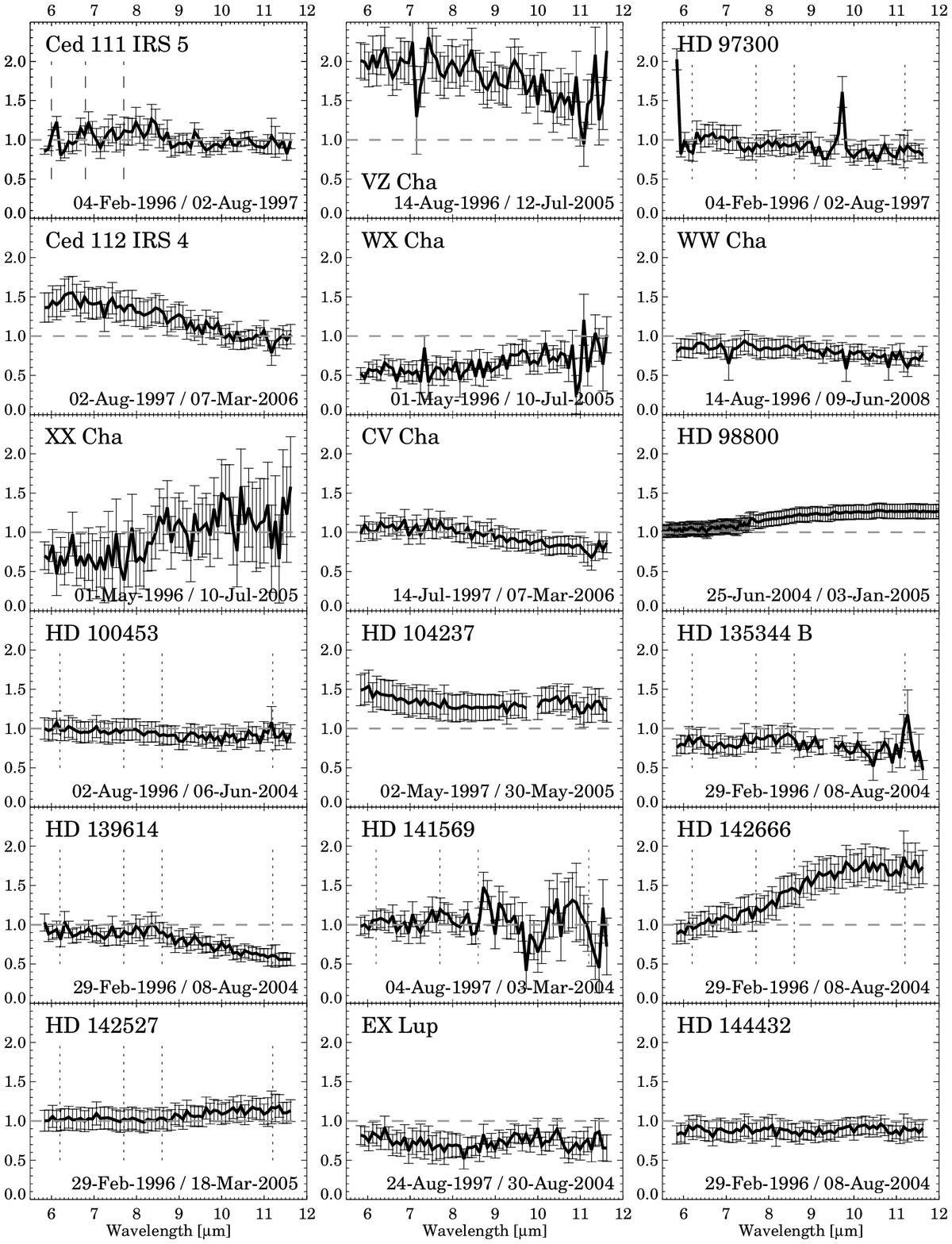}
\caption{(Continued)}
\end{figure*}

\setcounter{figure}{2}
\begin{figure*}
\epsscale{1.0}
\plotone{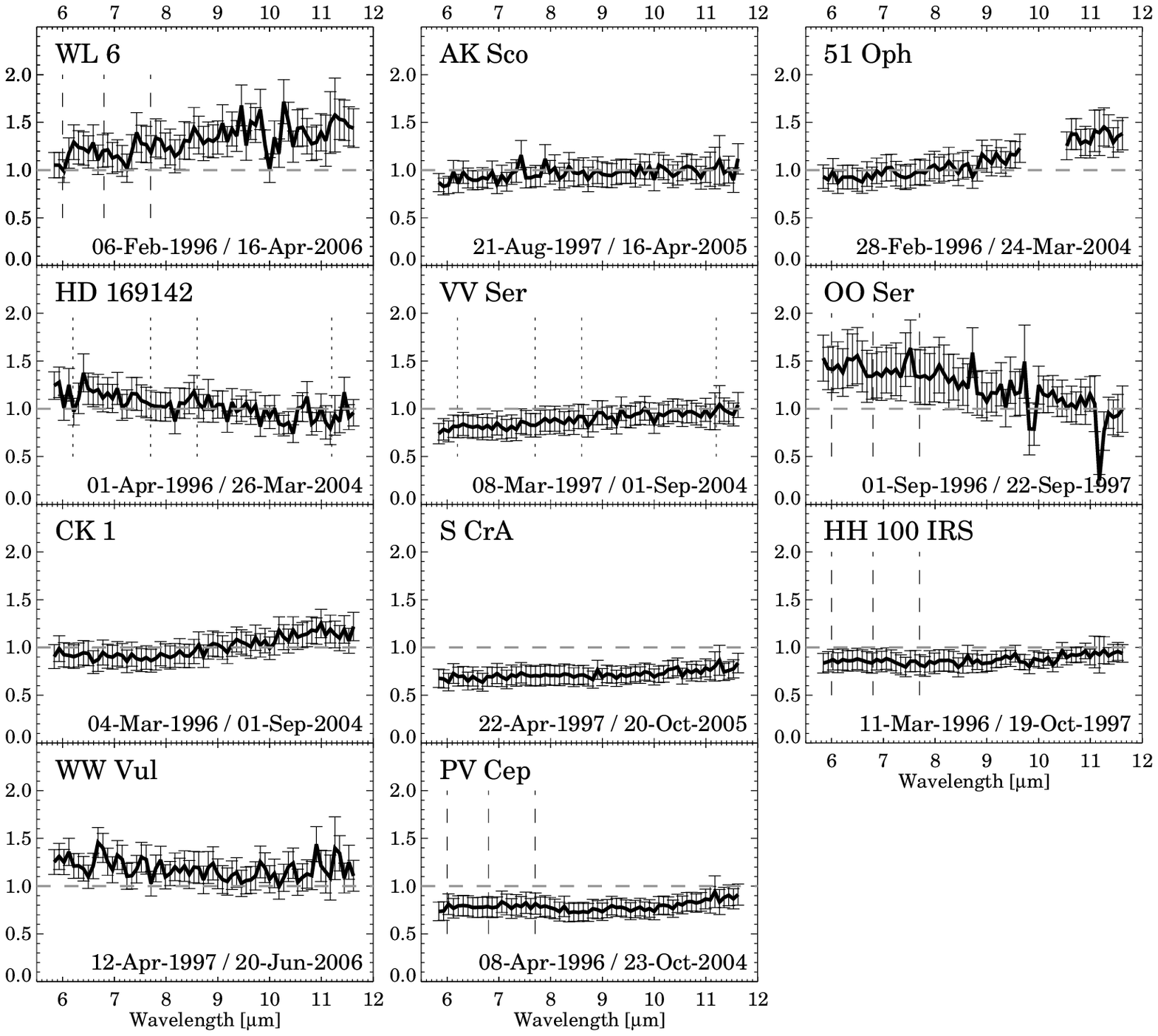}
\caption{(Continued)}
\end{figure*}

\begin{figure*}
\epsscale{0.55}
\plotone{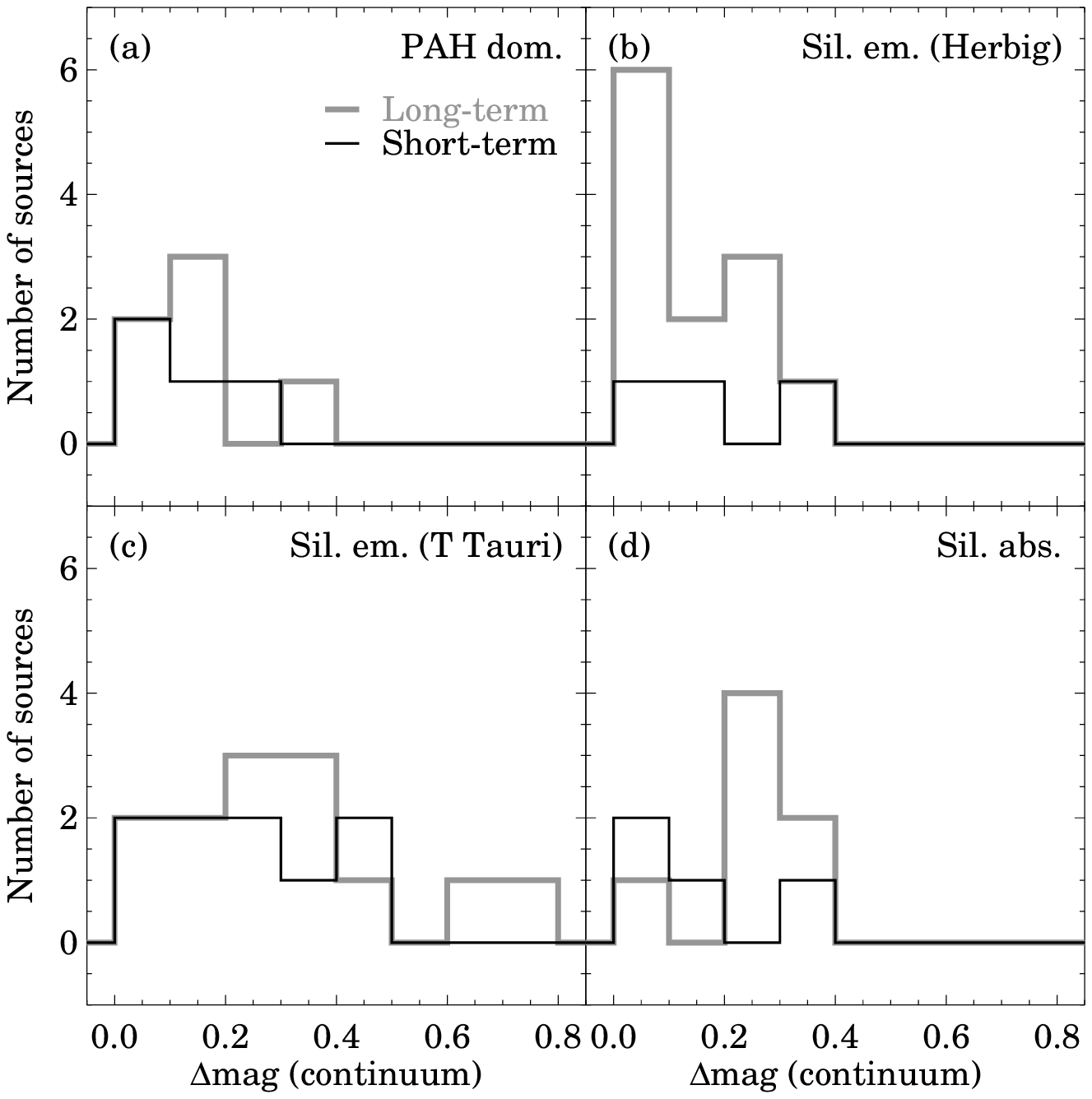}
\plotone{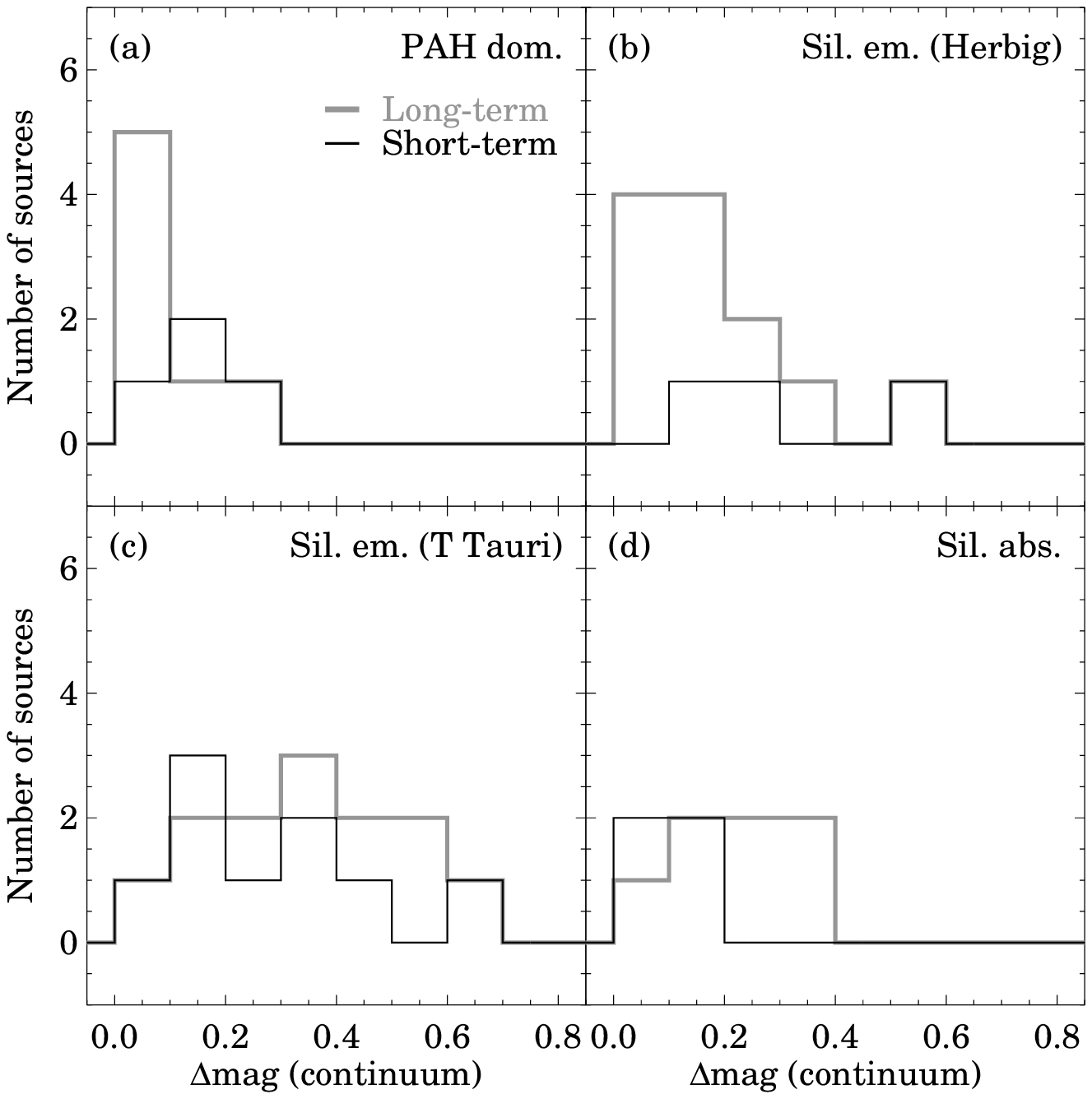}
\caption{Histograms showing the distribution of magnitude changes for
  our sources. {\it (a)--(d):} Magnitude changes in the continuum;
  {\it (e)--(h):} Magnitude changes in the feature (PAH features for
  PAH dominated objects, silicate feature for objects showing silicate
  emission or absorption). Different lines are plotted for magnitude
  changes calculated between the ISOPHOT-S and Spitzer/IRS spectra
  (long-term variability) and for magnitude changes calculated for
  spectra taken with the same instrument (short-term variability).
\label{timescales}}
\end{figure*}

\begin{figure*}
\epsscale{1.03}
\plotone{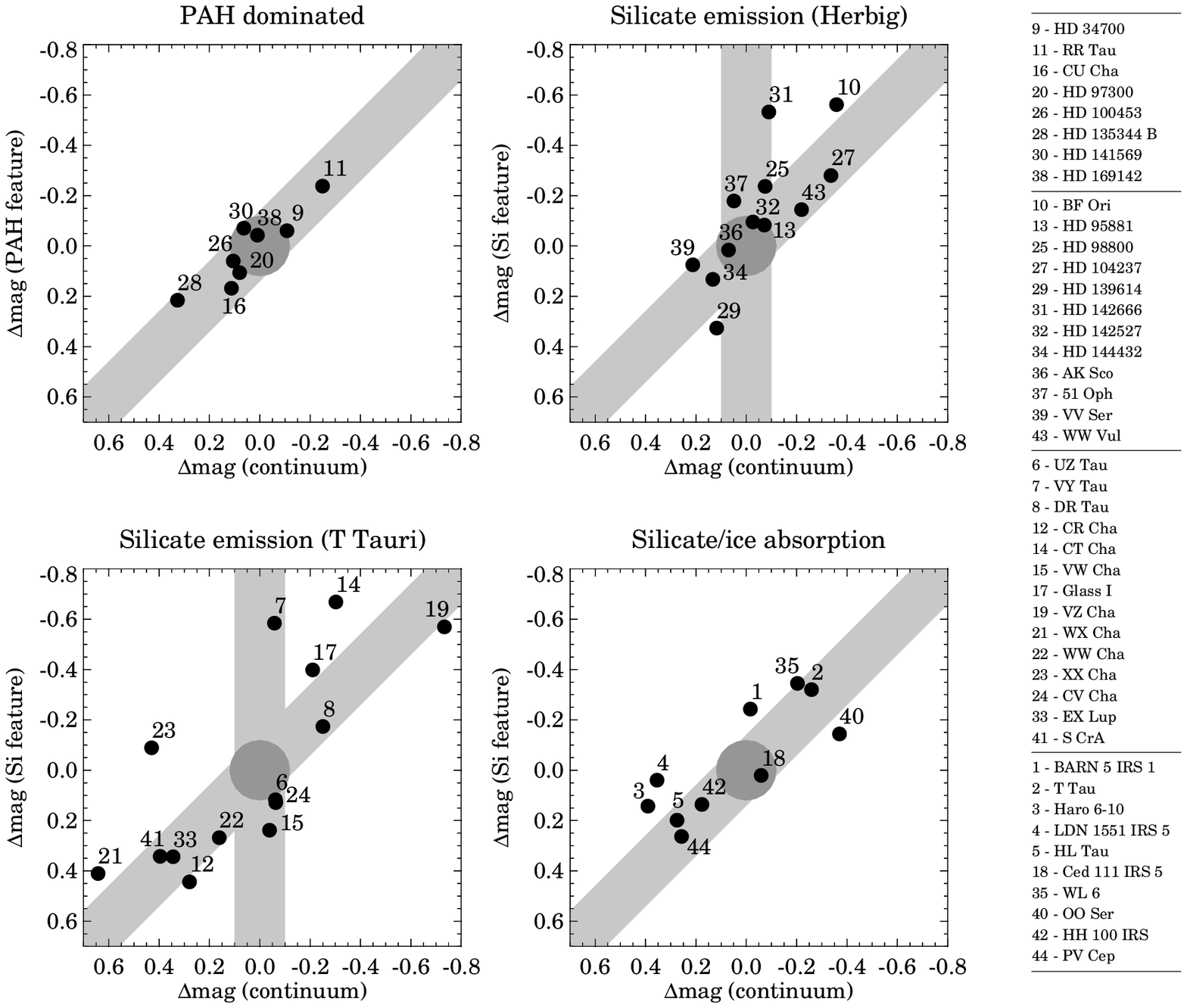}
\caption{Synthetic magnitude changes. The graphs display the
  brightness changes of our targets in the continuum and in the PAH or
  silicate band. The typical uncertainty of 0.1\,mag is represented by
  the dark gray circle in the middle of each panel. The vertical light
  gray stripe marks the area where sources with constant continuum but
  variable silicate feature fall. The diagonal light gray stripe marks
  the location of sources whose continuum and feature changed with
  similar amplitudes.\label{graph}}
\end{figure*}

\clearpage

\begin{figure}
\epsscale{1.0}
\plotone{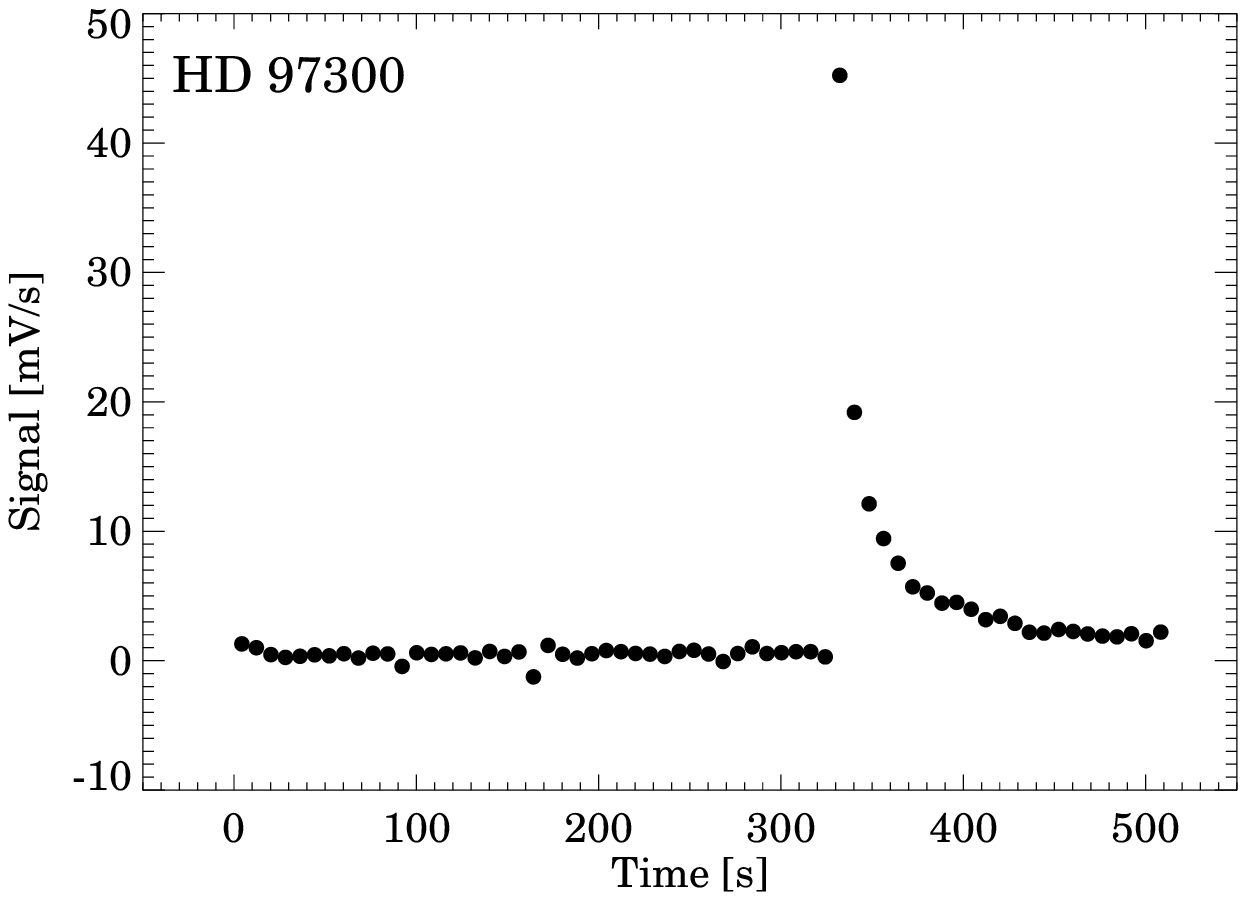}
\caption{Example for a cosmic glitch which was only partly corrected
  by the standard PIA deglitching algorithm, producing a spike in the
  spectrum (HD\,97300, TDT 62501316, Pixel 60). In our interactive
  processing all data points after t=330\,s are manually discarded.
\label{fig_glitch}}
\end{figure}

\begin{figure}
\epsscale{1.0}
\plotone{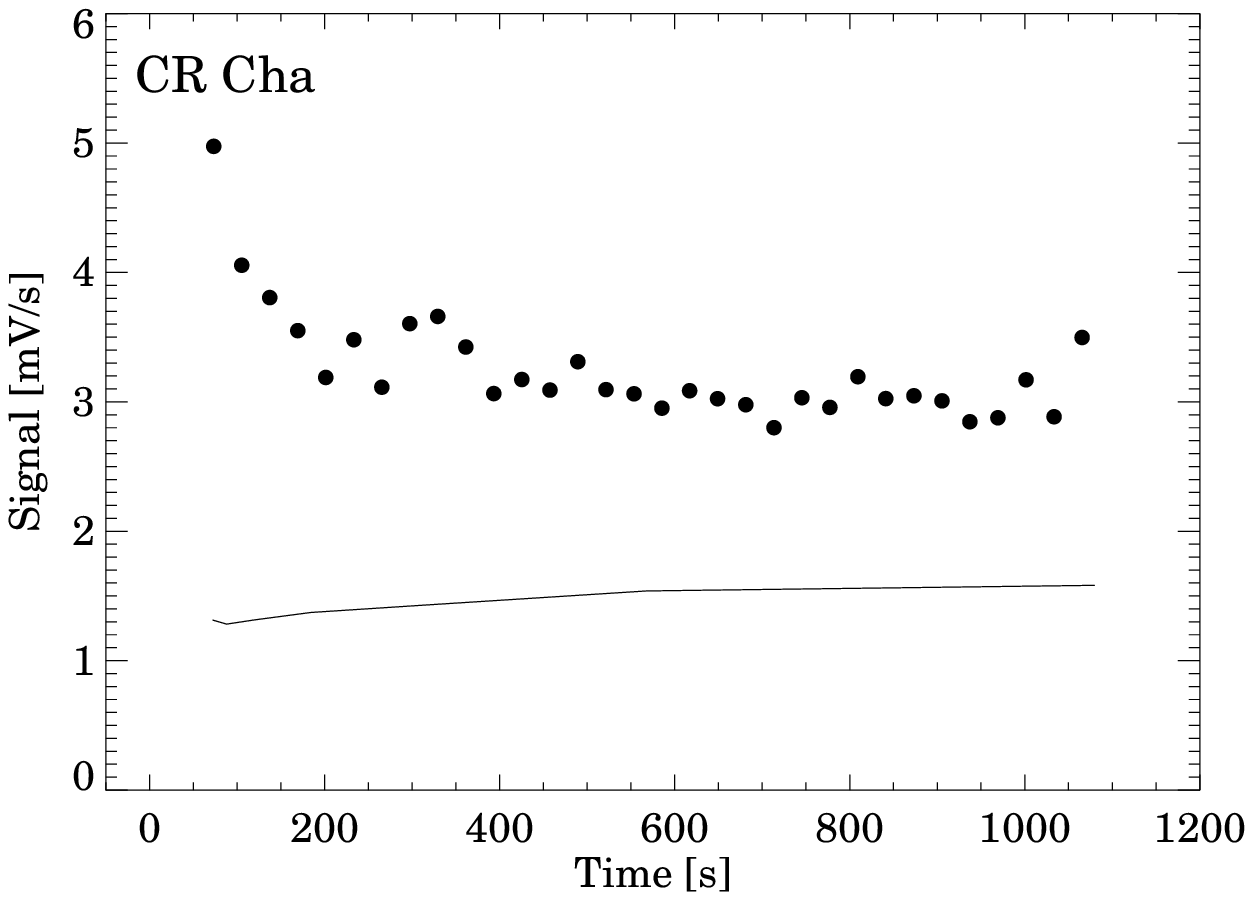}
\caption{Example for memory effect during the observation of CR\,Cha
  (TDT 62501703). The measured signal of Pixel\,93
  ($\lambda$=8.45\,$\mu$m) {\it decreases} rather than increases with
  time, due to recent illumination of the pixel by a bright
  source. The measured data points lie above the expected transient
  curve (solid line), leading to an overestimation of the derived
  flux.
\label{fig_memo_a}}
\end{figure}

\begin{figure}
\epsscale{1.0}
\plotone{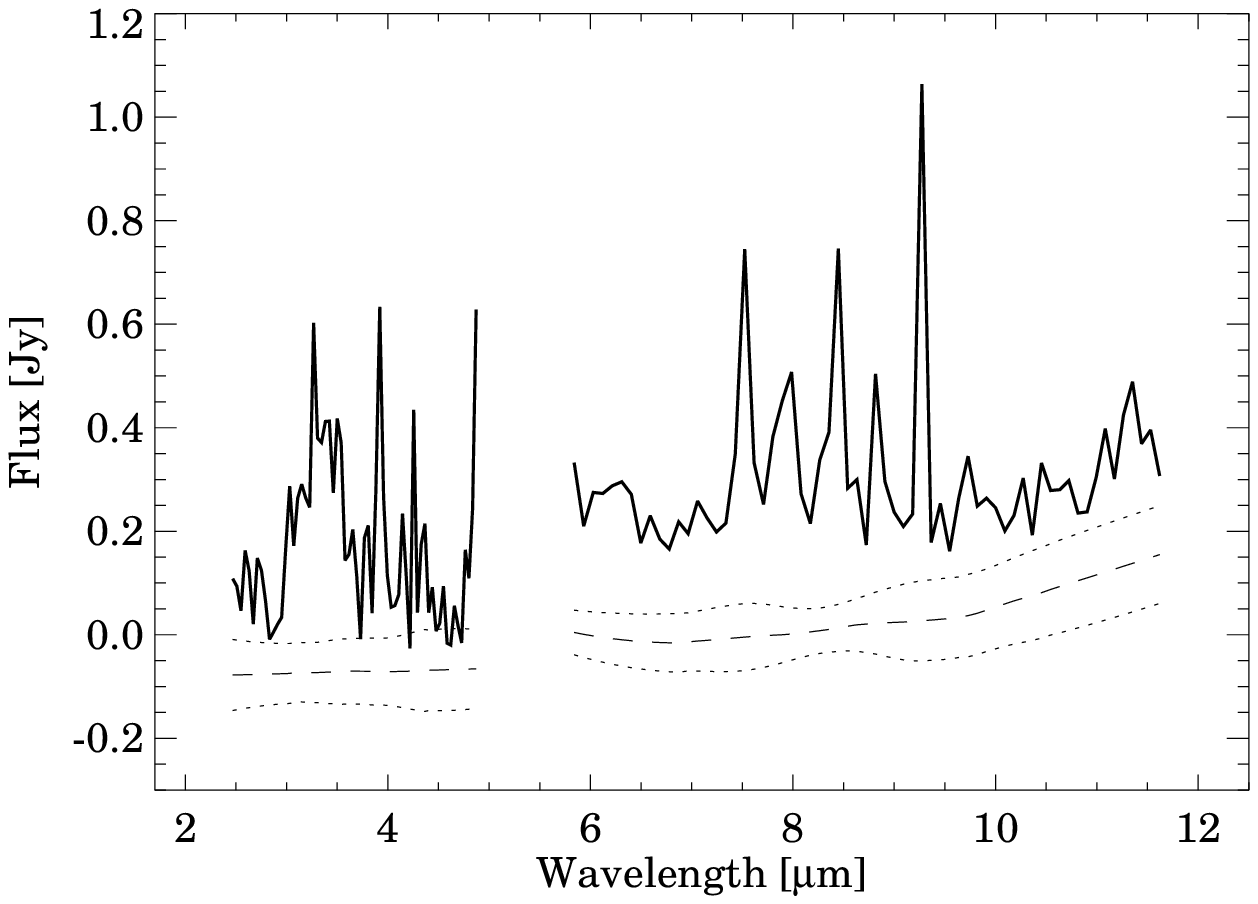}
\caption{Dark current observation before the observation of CR\,Cha
  (TDT 62501703). The observed signal (thick solid line) is well above
  the mission average value (dashed line, with 1$\sigma$ errors
  indicated by dotted lines).
\label{fig_memo_b}}
\end{figure}

\begin{figure}
\epsscale{1.0}
\plotone{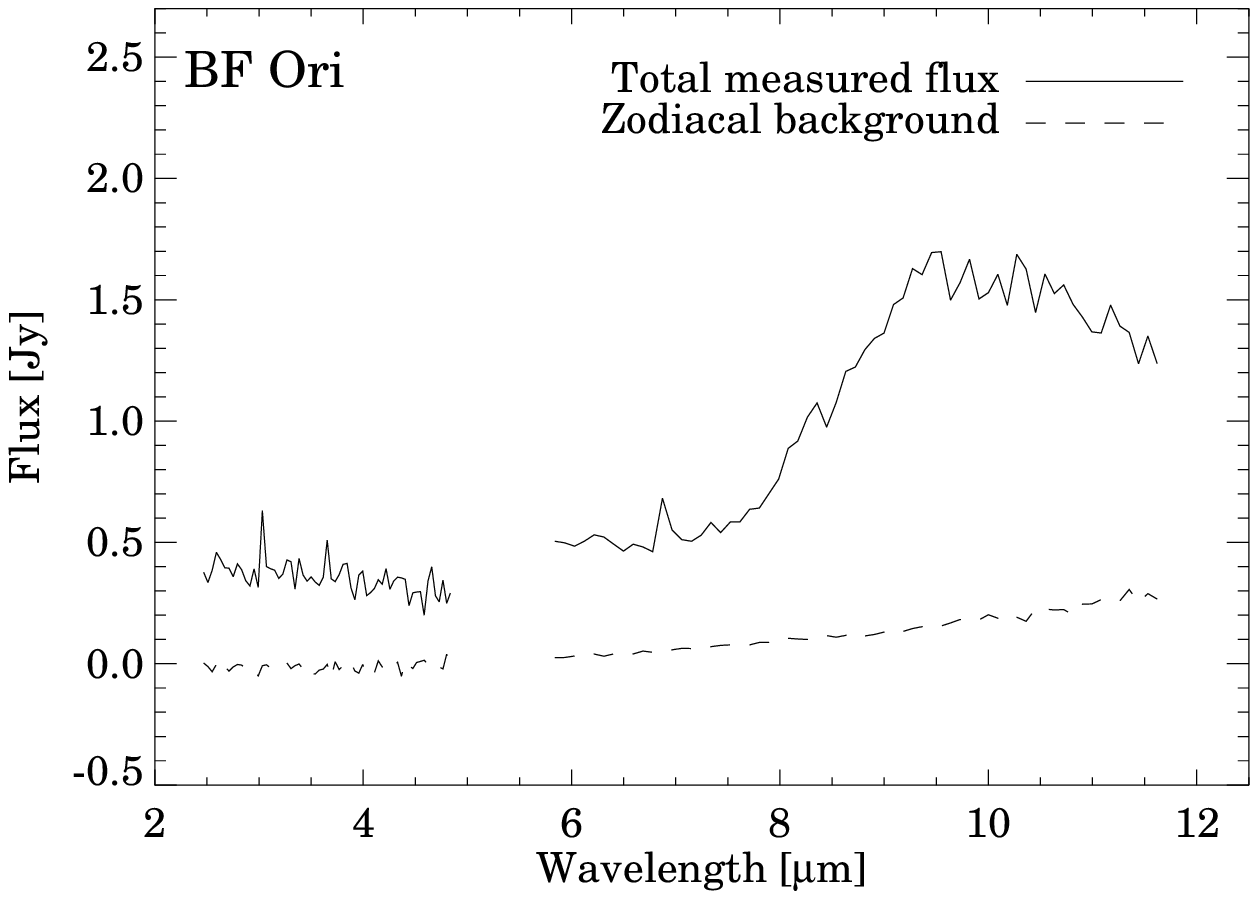}
\caption{Contribution of the zodiacal background (dashed line) to the
  measured flux of BF\,Ori (solid line). Since no dedicated background
  measurement was performed, the zodiacal spectrum was estimated from
  the DIRBE 4.9 and 12\,$\mu$m data points extracted for the position
  and date of the BF\,Ori observation.
       \label{fig_bgd}}
\end{figure}

\begin{figure}
\epsscale{1.0}
\plotone{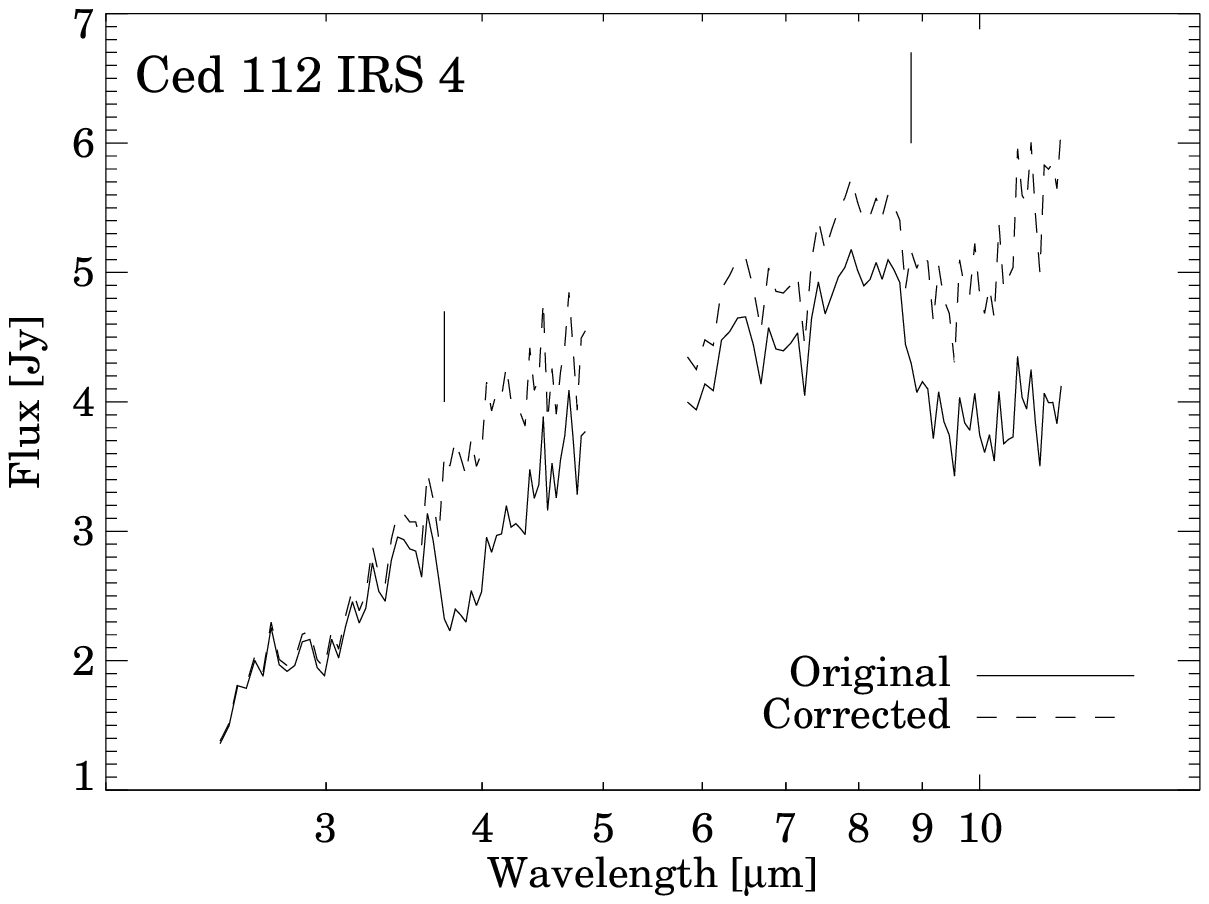}
\caption{Example for spectral artifacts introduced by off-center
  location of a point source in the ISOPHOT-S beam. The object
  (Ced\,112\,IRS\,4, TDT 62501217) was offset by $+4.8''$ and $-0.6''$
  in satellite Y- and Z-direction, respectively. Vertical marks show
  the wavelengths where the most obvious artifacts, in the forms of
  discontinuities, appear.
\label{fig_off}}
\end{figure}

\begin{figure}
\epsscale{1.0}
\plotone{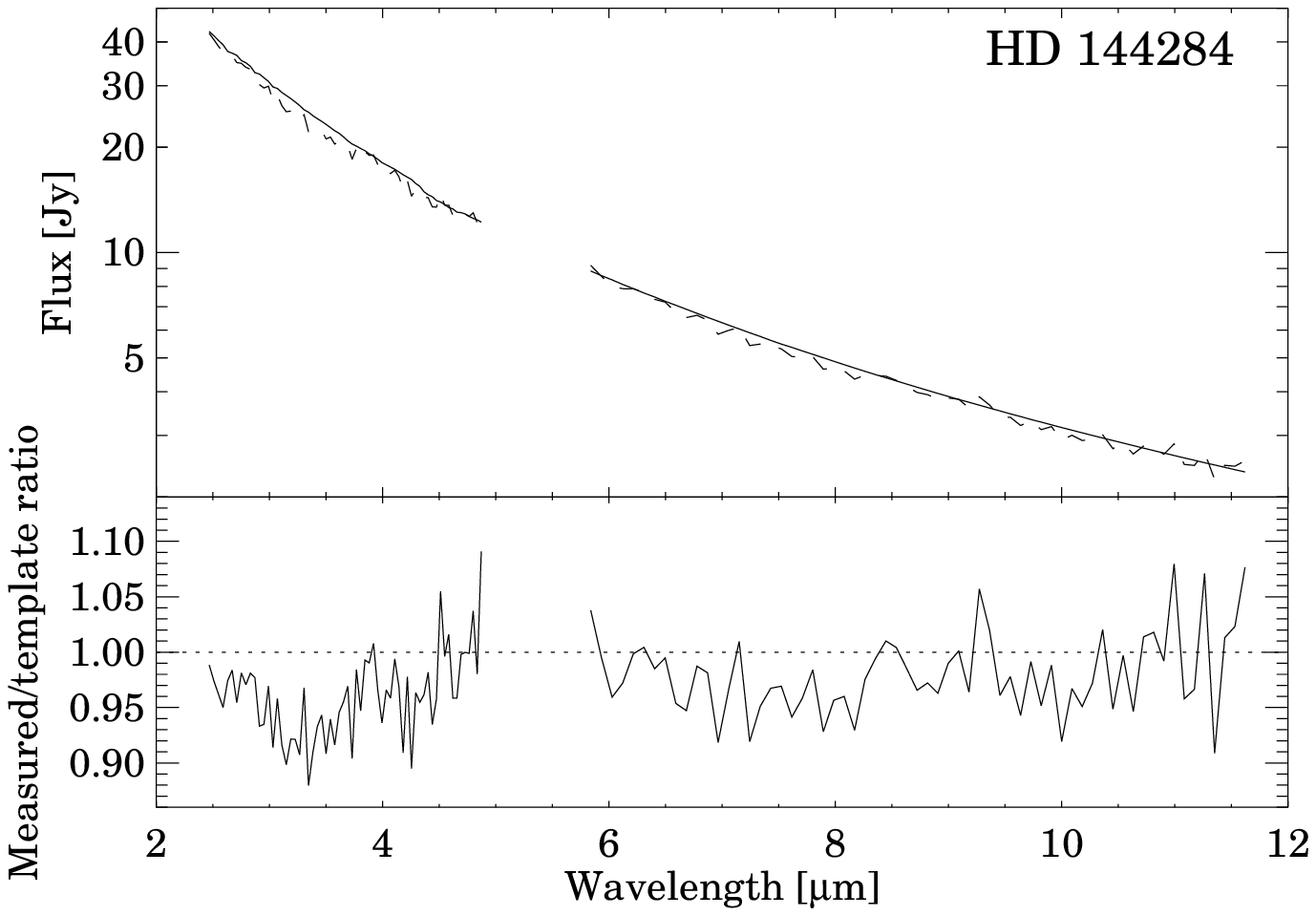}
\caption{{\it Upper panel}: Spectrum of the standard star HD\,144284
  (dashed line, TDT 52401702) compared with a stellar template
  prediction (solid line). {\it Lower panel}: Ratio of the measured
  flux values to the stellar template.
\label{fig_std}}
\end{figure}

\begin{figure}
\epsscale{1.0}
\plotone{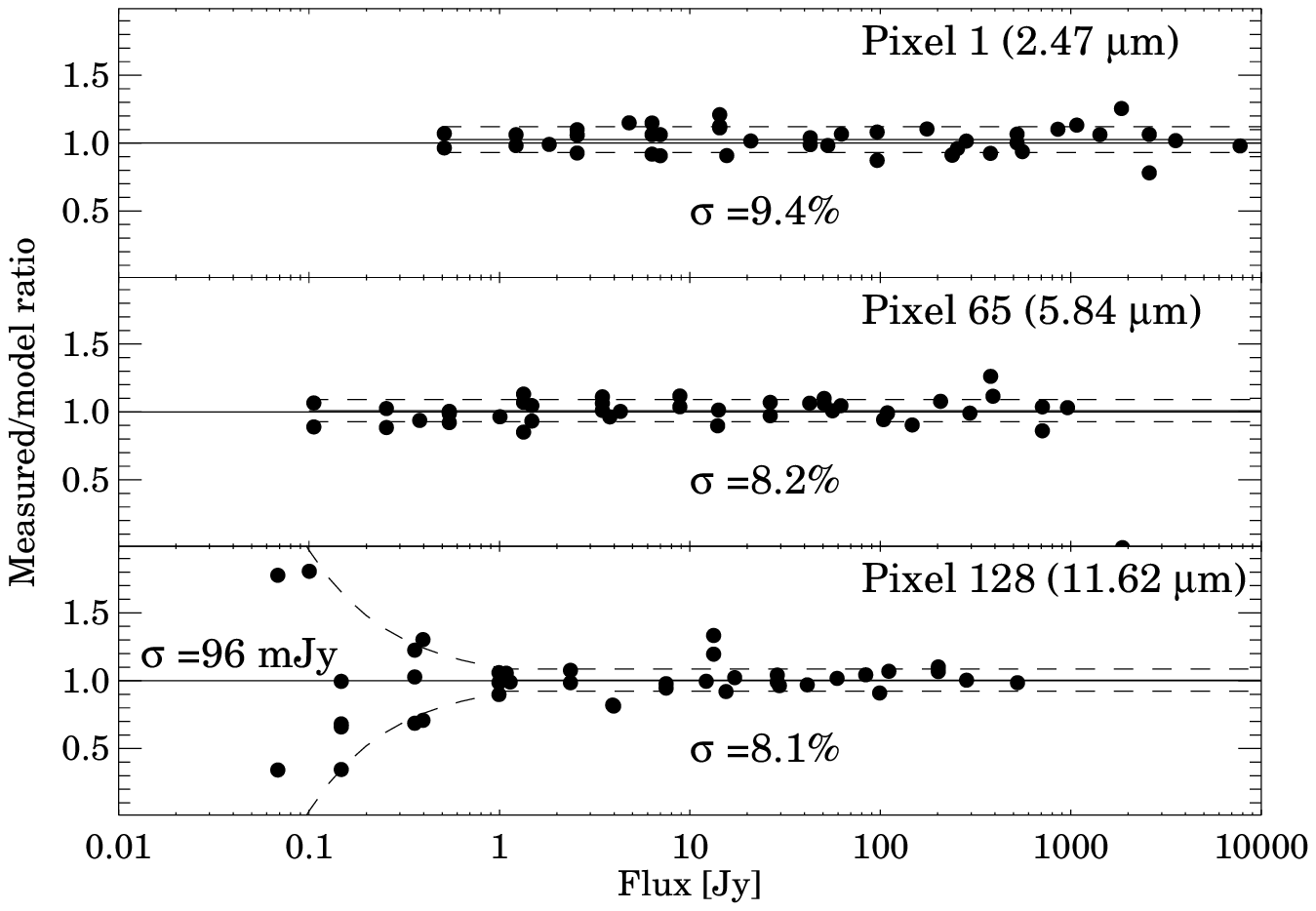}
\caption{Ratio of measured flux values to stellar template prediction
  for 3 wavelengths (all normal star observations are
  included). Uncertainty values representative of the whole ensemble
  are computed separately at high flux level (expressed in \%) and at
  low flux level (in mJy).
\label{fig_photcheck}}
\end{figure}

\begin{figure}
\epsscale{1.0}
\plotone{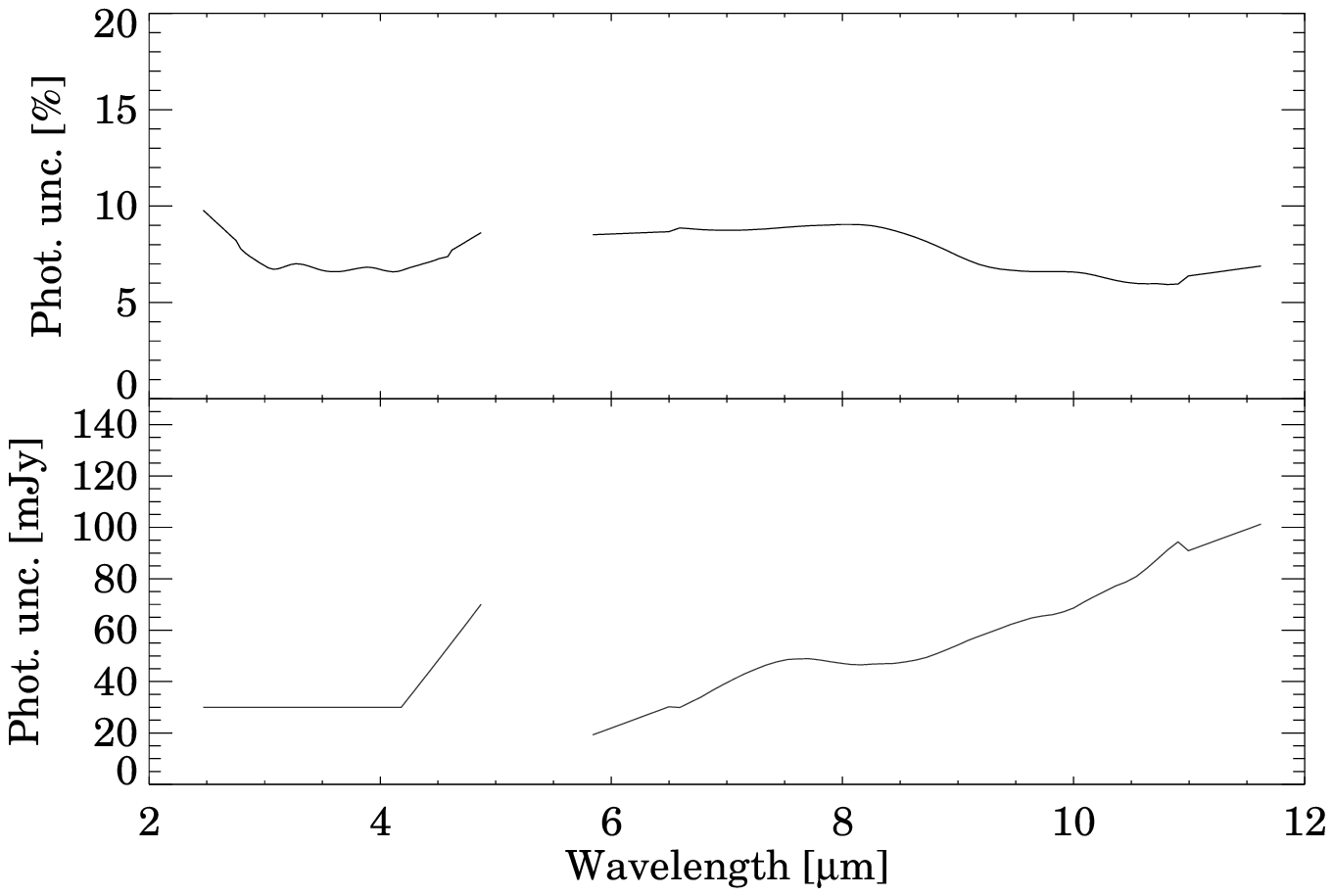}
\caption{Typical photometric uncertainties for the ISOPHOT-S
  wavelength range. Multiplicative errors (upper panel) are valid for
  brighter sources, while additive errors (lower panel) are for
  fainter sources ($<$1\,Jy).
\label{fig_unc}}
\end{figure}

\begin{figure}[h!]
\epsscale{0.75}
\plotone{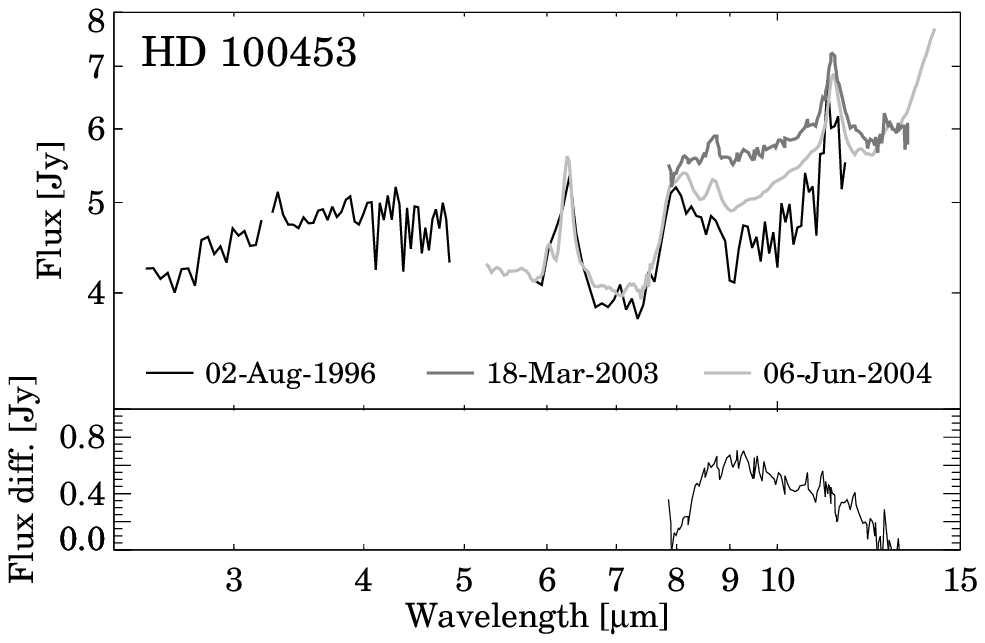}
\caption{Mid-IR spectra of HD\,100453. The ISOPHOT-S (black line) and
  Spitzer/IRS (light gray line) spectra are from this work, the TIMMI2
  (light gray) spectrum is from \citet{vanboekel2005}. The bottom
  panel shows the difference between the spectra obtained in 2003 and
  2004. \label{hd100453}}
\end{figure}

\begin{figure}[h!]
\epsscale{0.75}
\plotone{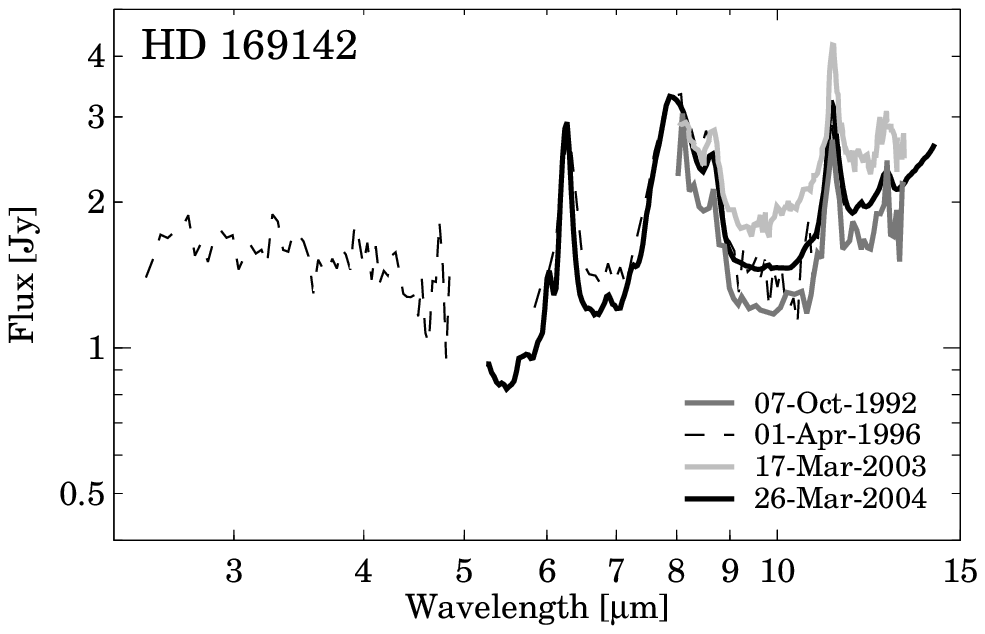}
\caption{Mid-IR spectra of HD\,169142. The ISOPHOT-S (dashed line) and
  Spitzer/IRS (black line) spectra are from this work, the TIMMI2
  (light gray) spectrum is from \citet{vanboekel2005}, the UKIRT/CGS3
  (dark gray) spectrum is from \citet{sylvester1996}. \label{hd169142}}
\end{figure}

\begin{figure}[h!]
\epsscale{0.75}
\plotone{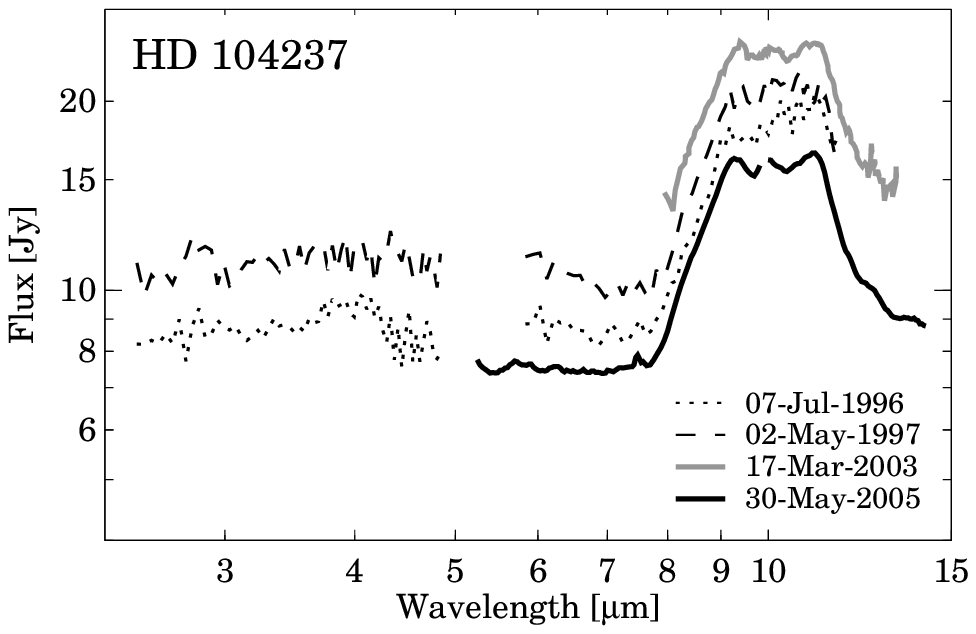}
\caption{Mid-IR spectra of HD\,104237. The ISOPHOT-S (dotted and
  dashed lines) and Spitzer/IRS (black line) spectra are from this
  work, the TIMMI2 (gray) spectrum is from \citet{vanboekel2005}.
  \label{hd104237}}
\end{figure}

\begin{figure}[h!]
\epsscale{0.75}
\plotone{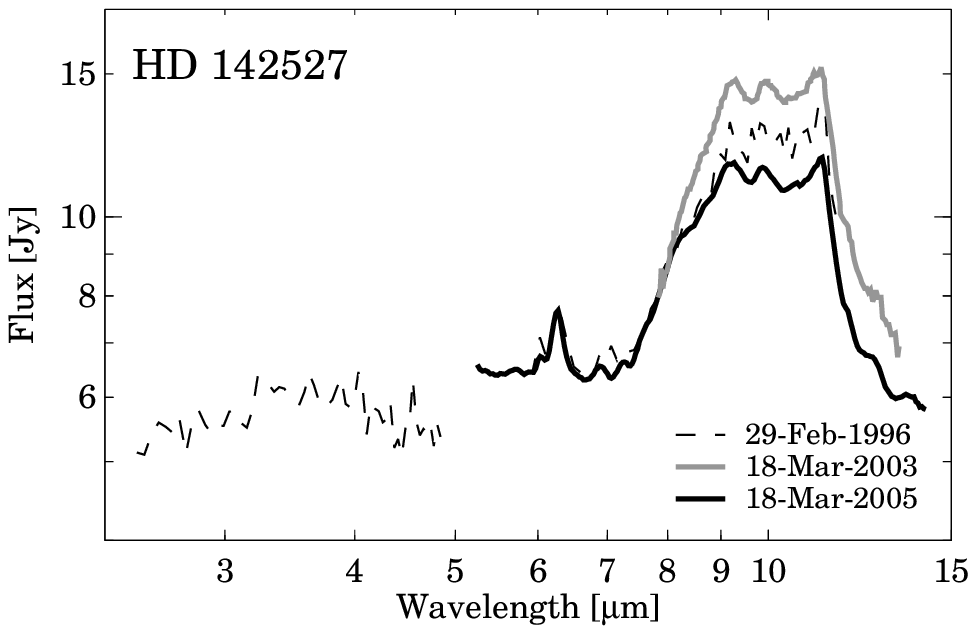}
\caption{Mid-IR spectra of HD\,142527. The ISOPHOT-S (dashed line)
  and Spitzer/IRS (black line) spectra are from this work, the TIMMI2
  (gray) spectrum is from \citet{vanboekel2005}.
  \label{hd142527}}
\end{figure}

\end{document}